\documentclass[11pt]{article}
\usepackage{caption}
\usepackage{relsize}
\usepackage{slashed}
\usepackage{float}
\restylefloat{table}
\usepackage{boldline,multirow}
\usepackage{multirow,array}
\usepackage{arydshln}
\usepackage{makecell}
\usepackage{longtable}
\usepackage{amssymb}
\usepackage{cancel}
\usepackage{amsmath}
\numberwithin{equation}{section}
\usepackage[dvips]{graphicx}
\usepackage{bm}
\usepackage{dsfont}
\usepackage[dvipsnames]{xcolor}
\usepackage{setspace}
\usepackage{amsfonts}
\usepackage{fancyhdr}
\usepackage{xcolor}
\usepackage{graphicx}
\usepackage{rotating}
\usepackage{comment}
\usepackage{color}
\usepackage{cite}
\usepackage{braket}
\usepackage{amssymb,amsmath,amsthm,mathrsfs,latexsym,amsxtra,graphicx,appendix,epstopdf,feynmf,subfig,setspace,fix-cm,color,cite}
\usepackage{pgfplots}
\usepgfplotslibrary{polar}
\usepgfplotslibrary{fillbetween}
\pgfplotsset{compat=newest}
\usepackage{mathtools}
\usepackage{tensor}
\usepackage{siunitx}
\usepackage{tikz}
\usetikzlibrary{trees}
\usetikzlibrary{decorations.pathmorphing}
\usetikzlibrary{decorations.markings}
\usetikzlibrary{decorations.markings}
\usepackage{float}
\usepackage{rotating}
\usepackage{simplewick}
\usepackage{amssymb}
\usepackage{amsmath}
\usepackage{setspace}
\usepackage{soul}
\usepackage{amsfonts}
\usepackage{fancyhdr}
\usepackage{xcolor}
\usepackage{graphicx}
\usepackage{rotating}
\usepackage{comment}
\usepackage{color}
\usepackage{cite}
\usepackage{braket}
\usepackage[colorlinks=true, linkcolor  = blue,urlcolor=blue,citecolor=violet]{hyperref}
\usepackage{simpler-wick}
\allowdisplaybreaks

\definecolor{darkgreen}{rgb}{0,0.5,0}
\definecolor{darkblue}{rgb}{0,0,0.6}
\definecolor{purple}{rgb}{0.4,0.15,0.21}
\definecolor{black}{rgb}{.2,.2,.2}

\DeclareSymbolFont{myletters}{OML}{ztmcm}{m}{it}
\DeclareMathSymbol{\uplambda}{\mathord}{myletters}{"15}

\newcommand{\bi}{\begin{itemize}}
\newcommand{\ei}{\end{itemize}}
\newcommand{\bea}{\begin{eqnarray}}
\newcommand{\eea}{\end{eqnarray}}
\newcommand{\be}{\begin{equation}}
\newcommand{\ee}{\end{equation}}

\newcommand{\dd}{\mathrm{d}}

\newcommand{\tpsi}{\widetilde{\psi}}

\newcommand{\tepsilon}{\widetilde{\epsilon}}
\newcommand{\bepsilon}{\overline{\epsilon}}
\newcommand{\btepsilon}{\overline{\widetilde{\epsilon}}}

\newcommand{\tvarphi}{\widetilde{\varphi}}

\newcommand{\bpsi}{\overline{\psi}}
\newcommand{\btpsi}{\overline{\widetilde{\psi}}}

\newcommand{\talpha}{\tilde{\alpha}}

\newcommand{\tF}{\widetilde{F}}

\newcommand{\ii}{{\rm i}}
\newcommand{\rme}{{\rm e}}
\newcommand{\rd}{{\rm d}}

\newcommand{\cC}{\mathcal{C}}
\newcommand{\cL}{\mathcal{L}}
\newcommand{\cN}{\mathcal{N}}
\newcommand{\cQ}{\mathcal{Q}}
\newcommand{\ph}[1]{\phantom{#1}}

\newcommand{\mb}[1]{\mathbf{#1}}
\newcommand{\mc}[1]{\mathcal{#1}}

\newcommand{\mf}[1]{\mathfrak{#1}}

\newcommand{\Z}{\mathbb{Z}}
\newcommand{\C}{\mathbb{C}}
\newcommand{\R}{\mathbb{R}}

\numberwithin{equation}{section}
\allowdisplaybreaks  

\begin{document}



\onehalfspacing

\begin{center}

~
\vskip4mm
{{\huge {dS$_2$ Supergravity
\quad 
 }
  }}
\vskip5mm

Dionysios Anninos,$^1$ Pietro Benetti Genolini,$^1$ and ~Beatrix M\"uhlmann$^2$ \\ 

\end{center}
\vskip2mm
\vskip4mm
\begin{center}
{
\footnotesize
{$^1$Department of Mathematics, King's College London, Strand, London WC2R 2LS, UK \newline\newline
$^2$ Department of Physics, McGill University, Montreal, QC H3A 2T8, Canada  \\
}}
\end{center}
\begin{center}
{\textsf{\footnotesize{
dionysios.anninos@kcl.ac.uk, pietro.benettigenolini@unige.ch, 
beatrix.muehlmann@mcgill.ca}} } 
\end{center}
\vskip5mm

\vspace{4mm}
 
\vspace*{0.6cm}

\vspace*{1.5cm}
\begin{abstract}
\noindent 
We construct two-dimensional supergravity theories  endowed with a positive cosmological constant, that admit de Sitter vacua. We consider the cases of  $\mathcal{N}=1$ as well as $\mathcal{N}=2$ supersymmetry, and couple the supergravity to a superconformal field theory with the same amount of supersymmetry. Upon fixing a supersymmetric extension of the Weyl gauge, the theories are captured, at the quantum level, by supersymmetric extensions of timelike Liouville theory with $\mathcal{N}=1$ and $\mathcal{N}=2$ supersymmetry respectively. The theories exhibit good ultraviolet properties and are amenable to a variety of techniques such as  systematic loop expansions and, in the $\mathcal{N}=2$ case, supersymmetric localization. Our constructions offer a novel path toward a precise treatment of the Euclidean gravitational path integral for de Sitter, and in turn, the Gibbons--Hawking entropy of the de Sitter horizon. We argue that the supersymmetric localization method applied to the $\mathcal{N}=2$ theory must receive contributions from boundary terms in configuration space. We also discuss how these theories overcome several  obstructions that appear upon combining de Sitter space with supersymmetry.

\end{abstract}

\newpage
\setcounter{page}{1}
\pagenumbering{arabic}

\tableofcontents

\section{Introduction}

A concrete framework to explore quantum properties of spacetime while retaining significant computational control is that of two-dimensional quantum gravity \cite{Polchinski:1989fn}. Along this vein, in this paper we study theories of two-dimensional quantum gravity with a positive cosmological constant \cite{Martinec:2014uva,Bautista:2015wqy,Anninos:2017hhn,Anninos:2021ene,Muhlmann:2022duj,Anninos:2021eit,Maldacena:2019cbz,Cotler:2019nbi} and construct models of two-dimensional de Sitter space (dS$_2$) at the quantum level which exhibit substantial theoretical control.\footnote{For the more studied case of negatively curved spacetimes, two-dimensional quantum gravity techniques have recently led to  insights regarding near-extremal black holes \cite{Saad:2019lba}, and novel applications of Euclidean gravity methodology  to black hole information release \cite{Almheiri:2019qdq, Penington:2019kki}.  } 
\vspace{0.2cm}\newline
Many of the challenges  associated to the problem of quantum de Sitter space  already appear in two-dimensions. Underscoring these is the difficulty in constructing  quantum observables in cosmological spacetimes,  reviewed for example  in \cite{Witten:2001kn,Anninos:2012qw}, as compared to observables in asymptotically anti-de Sitter or Minkowski spacetimes. Relatedly, the Cauchy surface is often compact for cosmological spacetimes and observers are surrounded by cosmological horizons shielding away any asymptotic portions of the spacetime. 
In light of these challenges, it is perhaps of some consolation that   gravitational theories with $\Lambda>0$, including theories in two dimensions, permit  a semiclassically well-defined gauge and field redefinition invariant Euclidean path integral $\mathcal{Z}_{\text{grav}}$ \cite{Gibbons:1976ue,Anninos:2020hfj}. The schematic structure of $\mathcal{Z}_{\text{grav}}$ is a path integral over all fields, including the metric field, placed on a topological sphere, or more generally over some compact manifold. Although the physical meaning of $\mathcal{Z}_{\text{grav}}$ remains somewhat obscure, it potentially constitutes the sharpest  gauge-invariant calculable quantity in quantum cosmology. Moreover, it has been hypothesised by Gibbons and Hawking \cite{Gibbons:1976ue,Gibbons:1977mu} that the simplest form of $\mathcal{Z}_{\text{grav}}$ computes the quantum entropy of the de Sitter cosmological event horizon.  As such, a sharper understanding of $\mathcal{Z}_{\text{grav}}$ may lead to a sharper understanding of {the} quantum properties of theories with $\Lambda>0$. 
\vspace{0.2cm}\newline
A novel ingredient incorporated into this work is supersymmetry. It has long been appreciated that supersymmetry and de Sitter space are difficult to reconcile  \cite{Lukierski:1984it,Pilch:1984aw}. In two dimensions, the story is somewhat different. The $\mathfrak{so}(1,2)$ isometry algebra of de Sitter and anti-de Sitter happen to coincide in two dimensions, and the group theoretic considerations forbidding a unitary representation of the supersymmetric extension of $\mathfrak{so}(1,2)$ are thus alleviated. Moreover, in two dimensions, the supergravity multiplet does not encode locally propagating degrees of freedom and the cosmological term in the gravitational action can be rendered supersymmetric independently of the kinetic term, which is topological. Thus, one can have two-dimensional supergravity with a positive cosmological constant, something that has at least implicitly long been known \cite{Polyakov:1981re}. To get semiclassically controlled de Sitter solutions, we must further add matter fields and one might be concerned that these might spoil supersymmetry when propagating in the dS$_2$ background. However, the matter theory can be arranged to be  superconformal, so that it can be supersymmetrically placed on a dS$_2$ background \cite{Hristov:2013spa, Anous:2014lia}. Given that we are coupling a superconformal matter field theory to a non-conformal supergravity with non-vanishing cosmological term, the theory can still physically discern Weyl equivalent spacetimes.\footnote{Tangentially, recent considerations of complexified saddles in Euclidean supergravity naturally lead to complexified notions of the supersymmetry transformations obeying non-standard reality conditions \cite{Freedman:2013oja}. Such generalizations might be adaptable to contexts, such as de Sitter, in which supersymmetry has not ordinarily played a role.}
\vspace{0.2cm}\newline
We thus find ourselves with the unusual task of constructing and exploring dS$_2$ supergravity theories at the quantum level. Throughout this work we will address this question by employing semiclassical Euclidean methods. The parameter governing the semiclassical limit is the central charge $c_m$ of the superconformal matter theory, which is taken to be large. This gives rise to a $1/c_m$ expansion controlling the gravitational fluctuations. An important technical ingredient is that in the supersymmetric extension of the Weyl gauge there are convincing arguments \cite{Antoniadis:1990mx,Distler:1989nt} indicating that an important portion of the problem reduces to supersymmetric Liouville theory. Supersymmetric Liouville theory, which is a superconformal field theory, can in turn be studied through a variety of techniques.
\vspace{0.2cm}\newline
While the restriction to two dimensions might seem like an enormous simplification of the physically relevant four-dimensional problem, the two- and four-dimensional theories share several common features. 
The group theoretic structure of the de Sitter group (reviewed for example in \cite{Dobrev:1977qv,Sun:2021thf,Anninos:2023lin})  in two and four spacetime dimensions, namely $SO(1,2)$ and $SO(1,4)$, share the Harish--Chandra discrete series unitary irreducible representations (UIR) in addition to the more generic principal and complementary series UIRs. Notably, the single-particle Hilbert space of the linearised gravitational field in dS$_4$ furnishes a discrete series UIR of $SO(1,4)$, and we will observe  that the fluctuations of the two-dimensional super-Weyl factor are themselves related to the discrete series UIR. The two-dimensional counterpart of the Euclidean gravitational path integral  $\mathcal{Z}_{\text{grav}}$ conjecturally capturing the de Sitter entropy retains a significantly rich structure, sharing  commonalities with the four-dimensional case \cite{Anninos:2020hfj}, but now in a setting where ultraviolet divergences are under substantial control \cite{Anninos:2021ene,Muhlmann:2022duj}. Interestingly, as in the four-dimensional case,  $\mathcal{Z}_{\text{grav}}$ exhibits \cite{Polchinski:1989fn} a conformal mode with unbounded kinetic term that can be treated by a careful choice of contour \cite{Gibbons:1978ac,Polchinski:1988ua}. Two-dimensional de Sitter space also appears as a factor of the  dS$_2 \times S^2$ Nariai solution of four-dimensional general relativity with $\Lambda>0$. 
\vspace{0.2cm}\newline
Although we mainly work in a semiclassical limit around an $S^2$ saddle, we provide indications that the theories under consideration admit good ultraviolet properties more generally. As such, the quantum completion of these models, if it exists, may not require a fully fledged superstring embedding -- two-dimensional gravity is renormalizable in and of itself. Whether or not this is an advantage or a drawback of our simplified two-dimensional setting is open to debate. {In this vein, it may be worth noting that the only smooth Einstein metric supported on $S^2$ and $S^3$ is the round one, whereas the situation on $S^D$ with $D\geq 5$ is quite opposite: an abundance of non-round Einstein metrics are known on odd-dimensional spheres, and non-round Einstein metrics have also been constructed on $S^6$, $S^8$, and $S^{10}$ as overviewed in \cite{nienhaus2023einstein}.
The borderline case of $S^4$ is still open, as it is not yet known whether there exists any Einstein metric other than the round four-sphere. (Depending on this one might place the four-dimensional theory closer or further away from its lower dimensional counterparts.)} 
\vspace{0.2cm}\newline
We note that the two-dimensional theory we consider is a supersymmetric version of `timelike' Liouville theory. Non-supersymmetric timelike Liouville theory is a non-unitary counterpart to the more ordinary `spacelike' Liouville theory. It has been proposed that upon coupling Liouville theory to a timelike Liouville matter theory \cite{Mertens:2020hbs, Collier:2023cyw} and taking a suitable limit it reduces to Jackiw-Teitelboim (JT) gravity \cite{Jackiw:1984je,Teitelboim:1983ux}. Attempts to realise quantum dS$_2$ in non-supersymmetric JT gravity face certain obstructions including a non-normalizable Hartle--Hawking wavefunction \cite{Maldacena:2019cbz,Cotler:2019nbi} and  a potentially divergent sphere partition function $\mathcal{Z}_{\text{grav}}$. The question of whether supersymmetric JT gravity in dS$_2$ overcomes these problems is left to future work. 
\vspace{0.2cm}\newline
As a final remark, ideas regarding the finite dimensional Hilbert space of a quantum de Sitter universe \cite{fischler,Banks:2006rx,Bousso:2000nf,Parikh:2004wh,Dong:2018cuv,Chandrasekaran:2022cip} do not seem orthogonal to the type of phenomena observed in the context of supersymmetry. Indeed, restricting to a BPS subsector of the Hilbert space has the effect of reducing a theory to a finite dimensional closed subset of states,  potentially allowing for precise supergravity computations of microscopic phenomena without the need for an ultraviolet completion. This perhaps creates room for an imperfect analogy to de Sitter space \cite{Anninos:2022ujl,Lin:2022rzw}.\footnote{More generally, connections between supersymmetry and de Sitter space have appeared in other higher dimensional contexts. These include the possibility of a twisted superstring/supergravity theory admitting de Sitter vacua \cite{Hull:1998vg},  supersymmetric extensions of higher spin dS/CFT \cite{Anninos:2011ui,Hertog:2017ymy}, supersymmetric edge modes at the de Sitter horizon \cite{Anninos:2021ihe}, supersymmetric localization methods for lower dimensional de Sitter \cite{Anninos:2022ujl,Castro:2023dxp}, and a supersymmetric organisation of de Sitter quasinormal modes \cite{Anninos:2011af}.} 
\vspace{0.2cm}\newline
{\textbf{Structure of paper.} The structure of the paper is the following. In section \ref{subsec:N1_SUGRA} we review $\mathcal{N}=1$ supergravity in two dimensions and its relation to $\mathcal{N}=1$ super-Liouville theory in the super-Weyl gauge. We emphasize the positivity of the cosmological constant, the nature of the residual gauge supersymmetries, and the incorporation of supersymmetric matter fields in the form of an $\mathcal{N}=1$ superconformal field theory. In section \ref{subsec:N1_TimelikeLiouville}, we study the super-Weyl gauge-fixed $\mathcal{N}=1$ supergravity in the semiclassical limit of large central charge for the matter theory. We discuss how an important part of the problem reduces to an $\mathcal{N}=1$ supersymmetric version of timelike Liouville theory, whose properties we study while emphasising its relation to the underlying $\mathcal{N}=1$ supergravity. A one-loop calculation of the central charge is provided. It is shown that the theory admits Euclidean dS$_2$ as a semiclassical solution about which systematic loop corrections can be computed. The resulting $\mathcal{N}=1$ supergravity path integral takes the form (\ref{eq:Z_grav_N1_final}). Part of the answer takes a form suggestive of an entanglement entropy. 
In section \ref{subsec:SUSYdS2} we review the obstructions that appear upon supersymmetrizing a theory placed in de Sitter space, and the form in which our two-dimensional setup evades them. 
In section \ref{sec:N2_SUGRA} we discuss $\mathcal{N}=2$ dS$_2$ supergravity, which has a richer gravity multiplet whose content includes a graviphoton $U(1)$ gauge field. In section \ref{tSLN2} the super-Weyl gauge of $\mathcal{N}=2$ dS$_2$ supergravity is studied. An important component of the problem reduces to an $\mathcal{N}=2$ super-Liouville theory. Again, a one-loop calculation of the central charge is provided, and it is shown that the theory admits Euclidean dS$_2$ as a semiclassical solution and a systematic loop expansion. The resulting form of the $\mathcal{N}=2$ supergravity path integral is (\ref{eq:Z_grav_N2_final}). In section \ref{N2Loc}, we discuss supersymmetric localization as applied to $\mathcal{N}=2$ super-Liouville and conclude that boundary terms in the space of configurations must contribute to the path integral. Details of the various calculations and notational choices are discussed in appendix \ref{app:ConventionsEuclidean} and \ref{app:Fermions}. The spacelike counterparts of the $\mathcal{N}=1$ and $\mathcal{N}=2$ timelike super-Liouville theories in the main text are briefly discussed in appendix \ref{spacelikeL}. 

\section{\texorpdfstring{$\mathcal{N}=1$ dS$_2$}{N1 dS2} supergravity}
\label{subsec:N1_SUGRA}

We begin by discussing the first concrete model, namely $\mathcal{N}=1$  supergravity in two spacetime dimensions,  introduced in \cite{Howe:1978ia}. For our purposes, it will prove important to further couple the $\mathcal{N}=1$ supergravity to an $\mathcal{N}=1$ superconformal field theory. We will show that the theory admits a positive cosmological constant and two-dimensional de Sitter solutions (dS$_2$), for which reason we refer to the theory as $\mathcal{N}=1$ dS$_2$ supergravity.  To analyse the theory, we will resort to a supersymmetric extension of the Weyl gauge. In this gauge, the relation of  $\mathcal{N}=1$ supergravity to $\cN=1$ super-Liouville theory will be uncovered. Our considerations lie mostly in Euclidean signature.

\subsection{\texorpdfstring{$\mathcal{N}=1$}{N1} supergravity multiplet \& action}\label{N1sugra}

In Lorentzian signature, the two-dimensional off-shell ${\cN=(1,1)}$ gravity multiplet contains the zweibein $\rme^a_\mu$, a spin-3/2 Majorana gravitino $\chi_\mu$, and a real scalar field $A$.\footnote{{\label{footnote:EuclideanCounting}In two dimensions, a Lorentzian theory with $(\cN, \cN)$ supersymmetry admits $\cN$ Majorana--Weyl supercharges with either chirality. In Euclidean signature, though, the Majorana condition is not consistent with the Weyl condition, so a $(\cN, \cN)$ theory is invariant under $\cN$ (complex) Weyl spinors with either chirality, or $2\cN$ Majorana spinors (Majorana spinors being equivalently minimal), and thus $\cN$ Dirac spinors. We use the notation $\cN$ to indicate the number of Dirac spinor supercharges preserved by the theory, corresponding to $4\cN$ real supercharges. Occasionally, the same number of supercharges is also denoted $(\cN, \cN)$, in keeping with the Lorentzian naming.
}} The supersymmetry variations  are given by \cite{Howe:1978ia, Uematsu:1984zy}
\begin{equation}
\label{eq:N1_SUGRA_Variations}
	\delta \rme^a_\mu = \overline{\epsilon}\gamma^a\chi_\mu ~, \qquad \delta\chi_\mu = \left( D_\mu - \frac{A}{2} \gamma_\mu \right) \epsilon ~, \qquad \delta A = - \frac{1}{2}\overline{\epsilon}\gamma^{\mu\nu} \left( D_\mu - \frac{A}{2} \gamma_\mu \right) \chi_\nu ~.
\end{equation}
Here $\epsilon$ is a Majorana spinor parameter, $\bepsilon$ its Majorana conjugate, $\gamma^a$ generate Cliff$(1,1)$, and we are suppressing the spinor indices. The covariant derivative acting on spinors is denoted by $D_\mu$. Our conventions for spinors are summarized in appendix \ref{app:ConventionsEuclidean}. 
In terms of the Majorana fermionic coordinate $\theta$, the Berezinian superfield, the superspace analogue of the metric determinant $\rme$, is defined as follows
\begin{equation}
	{\rm E} = \rme \left[ 1 + \frac{1}{2} \bar{\theta} \gamma^\mu \chi_\mu + {\frac{1}{2}} \bar{\theta}\theta \left( - A + \frac{1}{4}\overline{\chi}_\mu \gamma^{\mu\nu} \chi_\nu \right) \right] ~,
\end{equation}
and the scalar curvature superfield is
\begin{equation}
\label{eq:N1ScalarCurvature}
	\mc{R} = - A + \bar{\theta}\gamma^{\mu\nu}\left( D_\mu - \frac{A}{2} \gamma_\mu \right) \chi_\nu + \frac{1}{2}\bar{\theta}\theta \left( \frac{R}{2} - A^2 - \overline{\chi}^\mu \gamma^\nu D_{[\mu}\chi_{\nu]} - \frac{A}{4} \overline{\chi}^\mu \chi_\mu \right) ~,
\end{equation}
where $R$ is the Ricci scalar of the two-dimensional metric. With these two superfields, one can write the simplest supersymmetrization of the Einstein-Hilbert Lagrangian.
\vspace{0.2cm}\newline
{In the remainder of what follows we will be mostly considering Euclidean signature. Due to the absence of Majorana--Weyl spinors in Euclidean signature in two spacetime dimensions,  the content of the Lorentzian multiplet $(\rme^a_\mu,\chi_\mu,A)$ must be complexified, thereby doubling the physical number of degrees of freedom and complexifying the metric. In order to compute physical quantities, we must subsequently consider half-dimensional contours in the complexified configuration space, as  will be specified throughout our analysis.} Our {$\cN=1$} Euclidean action is given by  the Euclidean superfields integrated over a genus $h$ compact surface $\Sigma_h$. As a  superspace integral it reads
\begin{equation}\label{SN1}
S^{(h)}_{\mathcal{N}=1} =  	\int_{\Sigma_h} \rd^2x \rd^2\theta \, {\rm E} \, \left(-\frac{ \vartheta }{\pi}\mc{R} + {2} \ii {\boldsymbol{\mu}} \right) ~,
\end{equation}
where $\boldsymbol{\mu}\in \R$ is a  parameter with dimensions of inverse length, and $\vartheta$ is a dimensionless coupling. The first term in $S^{(h)}_{\mathcal{N}=1} $ reduces to a topological invariant. Using the expressions above one shows that
\begin{equation}
\label{eq:SuperEulerNumber}
	\frac{\vartheta}{\pi} \int_{\Sigma_h} \rd^2x \, \rd^2\theta \, {\rm E} \, \mc{R} = \frac{\vartheta}{\pi} \int_{\Sigma_h} \rd^2 x \, \rme \, \frac{R}{4} =  \vartheta  \chi_h ~,
\end{equation}
where $\chi_h = 2-2h$ is the Euler number of $\Sigma_h$.
This is due to the fact that two-dimensional gravity does not have any propagating degrees of freedom: the Ricci scalar is the density of a topological number, and the canonical gravitino kinetic term, which has the form $\overline{\chi}_\mu \gamma^{\mu\nu\rho} D_\nu \chi_\rho$, vanishes, as in two dimensions there is no rank-3 Clifford algebra element $\gamma^{\mu\nu\rho} \equiv \gamma^{[\mu}\gamma^{\nu} \gamma^{\rho]} \equiv 0$. In other words, there is no local kinetic term for the supermultiplet. The second term, which is going to play the role of a cosmological constant, can be expressed in component form as
\begin{equation}
\label{eq:IntegralBerezinian}
	{2} \ii \boldsymbol{\mu} \int_{\Sigma_h} \rd^2x\,  \rd^2\theta \, {\rm E} = \boldsymbol{\mu} \int_{\Sigma_h} \rd^2x \, \rme \left( -  {\ii}  A + \frac{ {\ii} }{4} \overline{\chi}_\mu \gamma^{\mu\nu}\chi_\nu \right) ~.
\end{equation}
 \vspace{0.2cm}\newline
\textbf{$\mathcal{N}=1$ matter content.} In order to add matter fields to $S^{(h)}_{\mathcal{N}=1}$ while preserving the same amount of supersymmetry, we consider an $\mathcal{N}=1$ supersymmetric quantum field theory in the presence of  a background $\mathcal{N}=1$ gravity multiplet. For the sake of concreteness and computational control, we will further take the matter theory to be an  $\cN=1$ superconformal field theory (SCFT) with central charge $c_m$. Apart from the value of $c_m$, the details of the $\mathcal{N}=1$ matter theory will not play an important role in what follows, as we will be focusing mostly on an $S^2$ topology and study properties of the undecorated sphere path integral.
\newline\newline
All in all, the undecorated gravitational path integral over the $\mathcal{N}=1$ supergravity action (\ref{eq:IntegralBerezinian}) coupled to an $\mathcal{N}=1$ SCFT takes the following form 
\begin{equation}
\label{eq:ZSUGRA_General}
\mathcal{Z}_{\text{grav}}^{\mathcal{N}=1} = \sum_{h=0}^\infty e^{\vartheta \chi_h}\int [\mathcal{D} \rme^a_\mu] [ \mc{D} \chi_\mu ] [\mc{D}A] \, e^{\boldsymbol{\mu} \int_{\Sigma_h} \rd^2x \, \rme \left( \ii A - \frac{ \ii}{4} \overline{\chi}_\mu \gamma^{\mu\nu}\chi_\nu \right) }\times Z_{\text{SCFT}}^{(h)}[\rme^a_\mu, \chi_\mu, A ]~.
\end{equation}
Here, $Z_{\text{SCFT}}^{(h)}$ is the partition function of the $\mathcal{N}=1$ superconformal matter field theory at genus $h$. In addition to (\ref{eq:ZSUGRA_General}), one can further consider diffeomorphism invariant insertions of operators. Remarkably, due to the conjectural relationship of $\mathcal{N}=1$ supergravity coupled to  $\mathcal{N}=1$ superconformal minimal models with $c_m<3/2$ to matrix integrals \cite{Seiberg:2003nm}, there is significant evidence that $\mathcal{Z}_{\text{grav}}^{\mathcal{N}=1}$ exists at the quantum level, at least in a restricted sense. Where it exists, and upon further resorting to the supersymmetric extension of the Weyl gauge, one has a concrete proposal for the measure of the path integral \cite{Distler:1989nt}, rendering the gravitational path integral at hand {\color{teal}c}onsiderably {\color{teal}c}oncrete.

\subsection{\texorpdfstring{$\mathcal{N}=1$}{N1} Super-Weyl gauge}

The path integration measure over the components of the supermultiplet in \eqref{eq:ZSUGRA_General} does not have a known precise  definition prior to gauge fixing. Nonetheless, using the redundancy under super-diffeomorphisms  one can impose a supersymmetric extension of the Weyl gauge \cite{Howe:1978ia}, namely
\begin{equation}
\label{eq:SuperWeylGauge}
	\rme^a_\mu = e^{b \varphi} \tilde{\rme}^a_\mu ~, \qquad \chi_\mu = e^{\frac{1}{2} b \varphi} \gamma_\mu \psi ~, \qquad A = {\frac{\ii}{b}} e^{ - b \varphi} F~.
\end{equation}
Here, $\tilde{\rme}^a_\mu$ is the zweibein for the reference metric which we take to be the round sphere of radius $r$. The fields $\varphi$, $F$, and $\psi$ are the Euclidean continuation of a super-multiplet with two real scalars and a Majorana spinor.
We can view the super-multiplet $(\varphi,\psi,F)$ stemming from (\ref{eq:SuperWeylGauge}) as the supersymmetric extension of the Weyl factor in ordinary two-dimensional gravity. The corresponding superfield in superspace notation is given by $\Phi = \varphi + \bar{\theta} \psi + \frac{1}{2} \bar{\theta}\theta F$. {As already mentioned, in Euclidean signature all fields are {a priori} complexified, thus the minimal $\cN=1$ multiplet contains two {complex} scalars and a {Dirac} spinor. One then imposes a reality condition on the complexified multiplet to reflect the physical content of the underlying supergravity configurations.} For the scalar content, it is natural to consider the contour along real-valued $\varphi$ and $F$.  The Euclidean fermion $\psi$ requires more care, as there is no Majorana--Weyl representation in Euclidean signature. Instead, we impose the Euclidean Majorana condition overviewed, for example, in section 3 of \cite{VanProeyen:1999ni} and discussed around (\ref{eq:EuclideanReality}) of appendix \ref{app:ConventionsEuclidean}, namely $\psi = \gamma_* \psi^C$, where $\psi^C$ is the charge conjugate of $\psi$.
\vspace{0.2cm}\newline
Returning to our path integral (\ref{eq:IntegralBerezinian}), we note that the partition function of an $\mathcal{N}=1$ SCFT on $S^2$ in the super-Weyl gauge is governed by the superconformal anomaly
\begin{equation}
Z_{\text{SCFT}}^{(0)}[\varphi, \psi, F]= \mathcal{C}e^{-S_{\text{s-anomaly}}[\varphi, \psi, F]}~,
\end{equation}
where $\mathcal{C}$ is a normalization constant independent of $c_m$.  The anomaly theory, given a suitable normalization of the super-Weyl field, is given by
\begin{equation}\label{Sanom}
S_{\text{s-anomaly}}[\varphi,\psi, F]= -\frac{c_m}{48\pi}\int_{S^2} \dd^2 x \, \tilde{\rme} \left(\frac{1}{2}\tilde{g}^{\mu\nu}\partial_\mu \varphi \partial_\nu \varphi - \frac{\ii}{2} \bpsi \slashed{D} \psi - \frac{1}{2} F^2 + \widetilde{R} \varphi\right)~.
\end{equation}
In addition, one must incorporate the contribution from the supergravity sector. 
\vspace{0.2cm}\newline
It has been argued in \cite{Distler:1989nt}  that the resulting action governing the super-Weyl field, upon including the contributions from the Jacobian of the gravitational path integration measure, as well as the SCFT anomaly action stemming from the ghost and matter sectors, is given by the  $\cN=1$ superconformal super-Liouville action \cite{Polyakov:1981re}. Resorting to a standard normalization for the fields, the action of  $\cN=1$ superconformal super-Liouville is given by
\begin{equation}
\label{eq:N1_SL_Action}
\mathcal{S}^{\mathcal{N}=1}_{\text{L}}=\frac{1}{4\pi} \int_{S^2} \dd^2 x\, \tilde{\rme}\,\left(\frac{1}{2} \partial^\mu \varphi \partial_\mu {\varphi} - \frac{\ii}{2}  \bpsi \slashed{D} \psi -\frac{1}{2} F^2 + \frac{1}{2} Q \widetilde{R} \varphi + \mu e^{b\varphi} F - \frac{\ii}{2}  \mu b e^{b\varphi} \overline{\psi} \psi \right) ~,
\end{equation}
where $\widetilde{R}=2/r^2$ is the Ricci scalar of the round sphere, $Q=b+b^{-1}$. The central charge of the theory is given by $c_{\text{L}}^{\cN=1}=\frac{3}{2}+3Q^2$, as we will confirm through a one-loop analysis in a later section.\footnote{\label{footnote:Normalization}A technical point useful for comparing with the literature is that the normalization of the bosonic part in the above $\mathcal{N}=1$ action differs from the non-supersymmetric theory by a factor of $1/2$. This convention, commonly used starting from \cite{Polyakov:1981re}, is chosen to obtain the relation $Q=b+b^{-1}$. If on the other side we were to rescale $\varphi \rightarrow \sqrt{2}\varphi$, $Q\rightarrow \sqrt{2}Q$ such that the normalization of the action agrees with the non-supersymmetric case, then we would obtain the rescaled relation $Q= b/2+ b^{-1}$.} 
The path integration measure is given by the standard flat measure on the space of $\varphi$, $\psi$, $F$. 
Finally,  we must consider the contribution to the path integral from the ghost system associated to the imposition of the super-Weyl gauge \eqref{eq:SuperWeylGauge}: we need both a fermionic $\mf{bc}$-ghost system with central charge $c_{\mathfrak{bc}}=-26$ and a bosonic $\beta\gamma$-ghost system with $c_{\beta\gamma}=11$.
\vspace{0.2cm}\newline
It is worth noting that the interaction terms in (\ref{eq:N1_SL_Action}) can be directly related to the integral of the Berezinian \eqref{eq:IntegralBerezinian} in the super-Weyl gauge. With the superfield normalized as in (\ref{eq:SuperWeylGauge}), it reads 
\begin{equation}
\label{eq:Berezinian_SCGauge_N1} 
		\frac{\boldsymbol{\mu}}{b} \int \dd^2 x \,  \tilde{\rme} \left( e^{b\varphi} F - \frac{\ii b}{2} \, e^{b\varphi} \overline{\psi} \psi \right) \, .
\end{equation}
The parameter $\boldsymbol{\mu}$ in \eqref{SN1} is related to $\mu$ in \eqref{eq:N1_SL_Action} as $\boldsymbol{\mu} = b\mu/(4\pi)$.
We note that in Euclidean signature with the fermions subject to our Euclidean reality conditions (\ref{eq:EuclideanReality}), the fermionic interaction term is pure imaginary (\ref{eq:N1_SL_Action}). In Lorentzian signature this leads to the standard reality conditions for a unitary theory. 
\vspace{0.2cm}\newline
Interestingly, integrating out $F$ in \eqref{eq:N1_SL_Action} leads to the following term $\tilde{\rme} \mu^2 e^{2b\varphi}/2$ in the action. Recalling \eqref{eq:SuperWeylGauge}, this term can be viewed as a gauge-fixed form of the cosmological term with {positive} cosmological constant. More generally, the Euclidean Einstein-Hilbert action in $d$ dimensions has the form $S_{\rm EH} = \frac{1}{16\pi G}\int \rd^d x  \sqrt{g} (-R + 2\Lambda)$, such that $\Lambda>0$  damps contributions with large volume in the path integral.
\vspace{0.2cm}\newline
Given that the net Weyl anomaly stemming from the gravity, matter, and ghost theories must vanish \cite{Distler:1988jt,David:1988hj}, one can relate the central charge of the $\mathcal{N}=1$ super-Liouville theory to the matter and ghost central charges. In particular,
\begin{equation}
\label{charge_conservation}
	c_{\text{L}}^{\cN=1} + c_m + c_{\mathfrak{bc}} + c_{\beta\gamma} = 0  \quad \Leftrightarrow \quad Q= \sqrt{\frac{9}{2} - \frac{c_m}{3} } ~,  \quad \ b = \frac{1}{2}\left(\sqrt{\frac{9}{2} - \frac{c_m}{3}} - \sqrt{\frac{1}{2} - \frac{c_m}{3}}\right)~.
\end{equation}
These relations imply that in the range $c_m\leq 3/2$ both $Q$ and $b$ are real-valued. The critical value $c_m=3/2$ is the $p\rightarrow \infty$ limit of the unitary $\mathcal{N}=1$ superconformal minimal  models, which are labelled by an integer $p\geq 3$ such that 
\begin{equation}\label{cmMM}
c_m = \frac{3}{2} - \frac{12}{p(p+2)}~.
\end{equation}
Upon coupling the $\mathcal{N}=1$ supergravity theory to one of these unitary superconformal minimal models \cite{Fredenhagen:2007tk} a conjectured duality exists relating the quantum $\mathcal{N}=1$ supergravity theory coupled to superconformal matter to a complex matrix integral (see e.g. \cite{Seiberg:2003nm}). When the matter central charge exceeds $c_m\geq 27/2$, both $b$ and $Q$ are pure imaginary. In this regime we can render the Liouville action real-valued by analytically continuing $(b,Q) \rightarrow \ii (\beta, -q)$ alongside an analytic continuation of the superfields $(\varphi, \psi,F)$. We call the resulting theory $\mathcal{N}=1$ timelike Liouville theory. In figure \ref{fig:centralchargesN1} we depict these two regimes.  The division between spacelike and timelike Liouville theory along with the `forbidden' regime organised in terms of the matter central charge is structurally analogous to the non supersymmetric case for which the boundary values are $c_m=1$ and $c_m =25$. In that case, for $c_m \leq 1$ we have non-supersymmetric spacelike Liouville theory, whereas the case $c_m \geq 25$ corresponds to the non-supersymmetric timelike Liouville theory. 

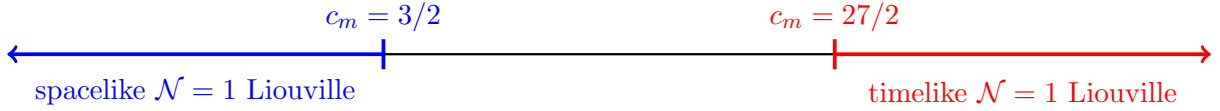
\begin{figure}
\begin{center}
\begin{tikzpicture}
\draw[line width=0.3mm] (-5,0) -- (6,0);
\draw[line width=0.5mm, red,->] (3,0) -- (8,0);
\draw[line width=0.5mm, red] (3,.2) -- (3,-.2);
\draw[line width=0.5mm,blue,<-] (-8,0) -- (-3,0);
\draw[line width=0.5mm,blue] (-3,.2) -- (-3,-.2);
\node[scale=1,blue] at (-3,.5)   {$c_m=3/2$ ~ };
\node[scale=1,blue] at (-5.5,-.5)   {spacelike $\mathcal{N}=1$ Liouville ~ };
\node[scale=1,red] at (3,.5)   {$c_m=27/2$ ~ };
\node[scale=1,red] at (5.5,-.5)   {timelike $\mathcal{N}=1$ Liouville ~ };
\end{tikzpicture}
\end{center}
\caption{The line shows the central charge of the superconformal field theory coupled to $\mathcal{N}=1$ super-Liouville theory. For $c_m\leq 3/2$ we the Liouville parameters $b$ and $Q$ are real-valued. This is the spacelike Liouville regime. For $c_m\geq 27/2$ both $b$ and $Q$ become imaginary. In this regime we can continue the parameters and fields to obtain timelike $\mathcal{N}=1$ Liouville theory. In the regime $c_m \in (3/2,27/2)$ the bosonic part of the $\mathcal{N}=1$ super-Liouville action  is complex. To our knowledge this theory has not been studied.}
\label{fig:centralchargesN1}
\end{figure}

\subsection{Residual gauge supersymmetries}

The theory governed by the action $\mathcal{S}^{\mathcal{N}=1}_{\text{L}}$ (\ref{eq:N1_SL_Action}) is invariant under superconformal transformations. The detailed form of these will be discussed in the next section. 
From the perspective of quantum field theory, these are ordinary physical symmetries of the theory. From the underlying gravitational perspective, these symmetries are in fact redundancies associated to the imposition of the super-Weyl gauge (\ref{eq:SuperWeylGauge}). Physical observables must be invariant under them. This is the underlying reason that the net central charge of the combined matter, ghost, and Liouville SCFT must vanish, as noted in (\ref{charge_conservation}).
\vspace{0.2cm}\newline
The subgroup of transformations leaving the super-Weyl gauge condition \eqref{eq:SuperWeylGauge} invariant, which are moreover well-defined over the whole $S^2$, is given by $OSp(1|2,\C)$. This is the global sub-group of the two-dimensional superconformal group with two real Poincaré supercharges. Its Lie superalgebra is $\mf{osp}(1|2,\C)$, which is the minimal graded extension of the global conformal algebra in two dimensions, $\mf{sl}(2,\C)$. It consists of the six real bosonic generators of $\mf{sl}(2,\C)$ represented by three complex $\mc{J}_m$, and four fermionic generators grouped in a Dirac spinor $\mc{Q}^\alpha$ with commutation relations
\begin{equation}
\label{eq:osp12C_v1}
\begin{aligned}
    [\mc{J}_m,\mc{J}_n] &= \ii \epsilon_{mnr}\mc{J}_r ~, &\qquad   [\mc{J}_m, \cQ^\alpha] &= - \frac{1}{2}(\gamma_m)^\alpha_{\ph{\alpha}\beta} \cQ^\beta ~, &\qquad  \{ \cQ^\alpha, \cQ^\beta\} &= (\gamma^m)^{\alpha\beta}\mc{J}_m ~.
\end{aligned}
\end{equation}
In terms of real generators, we can split $\mc{J}_m$ into the $J_m$ generating the isometries of $S^2$, and the $K_m$ generating the conformal symmetries, while the $\cQ^\alpha$ can be split into Poincaré and conformal supercharges $Q^\alpha$ and $S^\alpha$, obeying  the commutation relations
\begin{equation}
\label{eq:osp12C_v2}
\begin{aligned}
	[J_m, J_n] &= \ii \epsilon_{mnr} J_r ~, &\qquad [J_m, K_n] &= \ii \epsilon_{nmr}K_r ~, \\
	 [K_m, K_n] &= - \ii \epsilon_{mnr}J_r ~, &\qquad \{ Q^\alpha, Q^\beta\} &= (\gamma^m)^{\alpha\beta}J_m ~, \\ 
	 \{ S^\alpha, S^\beta \} &= - (\gamma^m)^{\alpha\beta}J_m ~, &\qquad \{Q^\alpha, S^\beta \} &= (\gamma^m)^{\alpha\beta} K_m ~, \\
	[J_m, Q^\alpha] &= - \frac{1}{2}(\gamma^m)^\alpha_{\ph{\alpha}\beta} Q^\beta ~, &\qquad [K_m, Q^\alpha] &= - \frac{1}{2}(\gamma^m)^\alpha_{\ph{\alpha}\beta} S^\beta\, \\
		[J_m, S^\alpha] &= - \frac{1}{2}(\gamma^m)^\alpha_{\ph{\alpha}\beta} S^\beta ~, &\qquad [K_m, S^\alpha] &= - \frac{1}{2}(\gamma^m)^\alpha_{\ph{\alpha}\beta} Q^\beta ~.
\end{aligned}
\end{equation}
Due to the superconformal invariance of the theory, the above describe the symmetries of $\mathcal{N}=1$ super-Liouville formulated on any space which is related to the flat metric on the complex plane by a Weyl transformation.
\vspace{0.2cm}\newline
{\textbf{$\mathcal{N}=1$ super-Moebius group.}} For the sake of completeness, we can also express $OSp(1|2,\C)$ in the superspace picture. The {$OSp(1|2,\mathbb{C})$ supergroup}  generalises the $SL(2,\mathbb{C})$  group of ordinary Moebius transformations, which appears as its maximal bosonic subgroup. It can be defined by the set of  supermatrices subject to supermatrix multiplication 
\begin{equation}\label{SM}
M = 
\begin{pmatrix}
a & b  & b\alpha-a \rho  \\
c & d   & d\alpha-c \rho \\ 
\alpha & \rho    & 1 - \alpha \rho
\end{pmatrix} \equiv \begin{pmatrix}
A   & B \\ 
C      & D
\end{pmatrix} ~.
\end{equation}
Here $a,b,c,d \in \mathbb{C}$, and  $\alpha$, $\rho$ are complex-valued Grassman numbers, subject to the condition $a d-b c = 1+\alpha\rho$. The matrix A is the bosonic $2\times 2$ submatrix. The final condition imposes a unit Berezinian
\begin{equation}
\text{ber}(M)\equiv \det A \, \det\left( D -  C A^{-1} B\right)^{-1} =1~.
\end{equation}
Denoting the holomorphic superspace coordinate as $Z \equiv (z|\theta)$, the super-Moebius transformations are given by \cite{Gieres:1992sc, Levy:2018xpu}
\begin{eqnarray}\label{st}
z({Z}') &=& \frac{a {z}' + b}{c {z}' + d} + {\theta}' \frac{\alpha {z}' + \rho}{(c {z}' + d)^2}~, \\
\theta({Z}') &=& \frac{\alpha {z}' + \rho}{c {z}' + d} +  \frac{{\theta'}}{c {z}' + d}~,
\end{eqnarray}
where $a,b,c,d$ are complex numbers and $\alpha,\rho$ are complex Grassmann numbers such that $ad-bc=1+\alpha\rho$ yielding a total of six bosonic generators and four fermionic generators, as in (\ref{eq:osp12C_v2}). 
The elements of our holomorphic Jacobian matrix are
\begin{eqnarray}
 \partial_{{z}'} z&=&\frac{1}{(c{z}'+d)^2}\left(1+\frac{{ \theta'}}{(c{z'}+d)} ({\alpha(d- c{z'})-2\rho c})\right)~, \\
\partial_{{z'}} {\theta} &=&\frac{1}{(c{z'}+d)^2} (\alpha d- \rho c) -{\theta'}\frac{c}{(c{z'}+d)^2}~, \\
 \partial_{{\theta'}} {z} &=& \frac{\alpha {z'} + \rho}{(c {z'} + d)^2}~, \quad  \partial_{{\theta'}} {\theta} = \frac{1}{c {z'} + d}~.
\end{eqnarray}
The superdeteminant of interest is
\begin{equation}
\sigma \equiv \text{ber} \begin{pmatrix} \partial_{{z'}} z & \partial_{{z'}} {\theta} \\ \partial_{{\theta'}} {z} & \partial_{{\theta'}} {\theta} \end{pmatrix}  = \frac{1}{c {z'} + d} + \frac{{\theta'}}{(c {z'} + d)^2}(\alpha d+\rho c)  ~.
\end{equation}
When placing $\mathcal{N}=1$ super-Liouville theory on the complex plane, the holomorphic super-Moebius transformation of the superfield $\Phi$ is then given by
\begin{equation}\label{SMPhi}
{\Phi}'[{Z}'] = \Phi[z({Z'})|\theta({Z'})]+ \frac{1}{\beta} \log \sigma^{-1}~,
\end{equation}
and similarly for the anti-holomorphic part. Upon setting $\alpha$ and $\rho$ to zero we retrieve the ordinary $SL(2,\mathbb{C})$ transformations, while setting all parameters except $\rho$ to zero we retrieve the $\mathcal{N}=1$ supersymmetry transformations.

\subsection{Precise \texorpdfstring{$\mathcal{N}=1$}{N1} supergravity path integral}

Once the dust settles, and upon path integration of the ghost and matter fields, the genus zero contribution to the $\mathcal{N}=1$ supergravity path integral \eqref{eq:ZSUGRA_General} takes the form
\begin{equation}
\label{eq:Z_grav_N1_0}
	\mc{Z}^{\cN=1}_{{\rm grav}, (0)} = e^{2\vartheta} \times {\mc{A}} \times \left( \frac{r}{\ell_{\text{uv}}}\right)^{\frac{c_m+c_{\mf{bc}} + c_{\beta\gamma}}{3}} \times \int \frac{[\mc{D}\varphi][\mc{D}\psi][\mc{D}F]}{{\rm vol}_{OSp(1|2;\C)}} e^{- \mc{S}_{\text{L}}^{\cN=1}} ~,
\end{equation}
where $\mc{A}$ is a $\mu$-independent prefactor stemming from the partition functions of the SCFT and the ghost systems on the round sphere with unit radius and any additional normalization choices from the path integral measure. We can absorb $\mc{A}$ into the definition of the physical coupling $\vartheta$. The length scale $\ell_{\text{uv}}$ is a reference scale, such as the ultraviolet cutoff scale, necessary to render the expression dimensionless.  In light of (\ref{charge_conservation}) we note that $\mc{Z}^{\cN=1}_{{\rm grav}, (0)}$ is independent of $r$ as it should be. Given that the leading ultraviolet divergences can be arranged to cancel in supersymmetric path integrals due to the boson/fermion degeneracy, the supergravity theory at hand stands a chance to be ultraviolet finite altogether, at least when placed on an $S^2$ topology.
\vspace{0.2cm}\newline
At this point, we are invited to direct our attention to the super-Liouville undecorated path integral
\begin{equation}
\label{eq:Z_L_N1}
	\mathcal{Z}_{\text{L}}^{\mathcal{N}=1}[\mu] \equiv \int \frac{[\mc{D}\varphi][\mc{D}\psi] [\mc{D}F]}{{\rm vol}_{OSp(1|2;\C)}} e^{- \mc{S}_{\text{L}}^{\cN=1}} ~,
\end{equation}
and related insertions, always bearing in mind the relationship to the underlying $\mathcal{N}=1$ supergravity theory and ensuring that the volume ${\rm vol}_{OSp(1|2;\C)}$ in the measure is properly accounted for.

\section{Timelike \texorpdfstring{$\mathcal{N}=1$}{N1} super-Liouville theory}
\label{subsec:N1_TimelikeLiouville}

Motivated by our considerations of  $\mathcal{N}=1$ supergravity in  two dimensions coupled to $\mathcal{N}=1$ superconformal matter, in this section we consider $\cN=1$ super-Liouville theory. For the sake of concreteness, and especially due to the physical motivation of having real-valued dS$_2$ solutions, we will take our SCFT to have large and positive central charge $c_m$. As we will explore in detail, the limit of large $c_m$ governs a semi-classical limit governing small geometric fluctuations around a classical saddle point solution.

\subsection{Action and symmetries}

We are interested in the $\mathcal{N}=1$ super-Liouville action \eqref{eq:N1_SL_Action} in the regime $c_m\gg1$. In light of (\ref{charge_conservation}), to render the bosonic part of the action real (upon placing $F$ on-shell)  we must analytically continue the fields as $\varphi \to \ii\varphi$, $\psi \to \ii \psi$, and $F\to \ii F$ while keeping $\mu$ fixed. Further introducing $\beta = \ii b$, and $q = -\ii Q = 1/\beta - \beta$, we are led to the following action
\begin{equation}
\label{eq:N1_TimelikeLiouville}
\mathcal{S}^{\mathcal{N}=1}_{\text{tL}} =\frac{1}{4\pi} \int_{S^2} \dd^2 x \, \tilde{\rme}\bigg( - \frac{1}{2}\tilde{g}^{\mu\nu}\partial_\mu \varphi \partial_\nu {\varphi} + \frac{\ii}{2}  \bpsi \slashed{D} \psi +   \frac{1}{2}F^2 - \frac{1}{2}q \widetilde{R} \varphi + \ii  \mu e^{\beta\varphi} F {+}\frac{1}{2} \mu \beta  e^{\beta\varphi} \overline{\psi} \psi \bigg) ~.
\end{equation}
Although the spacelike and timelike actions are related by analytic continuation, the timelike Liouville SCFT is not the analytic continuation of the spacelike Liouville SCFT. Rather, at least in the non-supersymmetric case, the timelike theory is a distinct non-unitary solution to the conformal crossing equations. 
\vspace{0.2cm}\newline
After the analytic continuation of the action, the superpotential term in (\ref{eq:N1_TimelikeLiouville}) differs by an overall $\ii$ as compared to \eqref{eq:N1_SL_Action}. The theory preserves the $OSp(1|2;\mathbb{C})$ superconformal group in its original form. Due to this $\ii$ the Euclidean action no longer has the standard reality properties. This is the supersymmetric counterpart of the non-unitary nature of ordinary timelike Liouville theory. From now onwards, we denote the  theory dictated by the above action as $\mathcal{N}=1$ timelike super-Liouville theory.  This mirrors the terminology used for the non-supersymmetric counterpart. 
\vspace{0.2cm}\newline
It is natural to hypothesise that (\ref{eq:N1_TimelikeLiouville}) is a non-unitary $\mathcal{N}=1$ superconformal field theory with central charge $c_{\text{tL}}^{\mathcal{N}=1} =3/2-3q^2$, whose value we will verify via a one-loop analysis in a later section, such that
\begin{equation}\label{qbeta_N1}
q = \sqrt{\frac{c_m}{3}-\frac{9}{2}}~,\quad\quad \beta = \frac{1}{2} \left({\sqrt{\frac{c_m}{3}-\frac{1}{2}} - \sqrt{\frac{c_m}{3}-\frac{9}{2}}} \right) ~.
\end{equation}
Here, as above, we choose a sign for $q$ and derive $\beta$  from the relation $q = \beta^{-1}-\beta$. The semi-classical limit  $c_m\to \infty$ corresponds to $\beta \to 0^+$. We also recall that for $c_m \geq 27/2$, both $q$ and $\beta$ are real. For $c_m \leq 3/2$, both $q$ and $\beta$ are purely imaginary and we can recover the spacelike super-Liouville action \eqref{eq:N1_SL_Action}, as summarized in figure \ref{fig:centralchargesN1}.
\vspace{0.2cm}\newline
The field content of $\mathcal{S}^{\mathcal{N}=1}_{\text{tL}} $ in \eqref{eq:N1_TimelikeLiouville} can be organised in terms of a Euclidean $\cN=1$ multiplet consisting of a {complex} scalar $\varphi$, a {Dirac} spinor $\psi$ and an auxiliary {complex} scalar $F$.
Moreover, $\mathcal{S}^{\mathcal{N}=1}_{\text{tL}} $ can be written in canonical form, in terms of a superpotential $W(\varphi) = \mu e^{\beta\varphi}/\beta$, as follows
\begin{equation}
\label{N1Liouville}
\mathcal{S}_{\text{tL}} ^{\mathcal{N}=1} =\frac{1}{4\pi} \int_{S^2} \dd^2 x \, \tilde{\rme}\bigg[- \frac{1}{2}\tilde{g}^{\mu\nu}\partial_\mu\varphi \partial_\nu {\varphi} +\frac{\ii}{2}  \bpsi \slashed{D} \psi+ \frac{1}{2}F^2 - \frac{1}{2}q \widetilde{R} \varphi + \ii \left(F \, W'(\varphi) - \frac{\ii}{2}  W''(\varphi) \bpsi \psi \right)\bigg]~.
\end{equation} 
This action is invariant under the  $\mathcal{N}=1$ supersymmetry transformations
\begin{equation}
\label{eq:N1_SCFT_Transformations}
\delta \varphi = \overline{\epsilon}\psi ~, \quad\quad
\delta\psi = \ii \slashed{\partial}\varphi \, \epsilon + q\ii \slashed{D}\epsilon - \ii F \, \epsilon ~, \quad\quad
\delta F = - \overline{\epsilon}\slashed{D}\psi ~,
\end{equation}
where $\epsilon$ is a Dirac conformal Killing spinor on the sphere. In addition to these, the theory enjoys an infinite set of superconformal transformations, altogether generating the full $\mathcal{N}=1$ superconformal algebra.\footnote{The infinite superconformal algebra is more readily written using the generators $L_n$, $G_\alpha^s$ on the plane, rather than the basis used in \eqref{eq:osp12C_v2}, which is adapted to the sphere. A map between the two can be found by specializing to $\mc{N}=1$ the map in \cite{Hori:2013ika, Doroud:2013pka}.} Focusing on the global subalgebra  \eqref{eq:osp12C_v2}, the transformations \eqref{eq:N1_SCFT_Transformations} (for a given  independent conformal Killing spinor -- see \textit{e.g.} \cite{Doroud:2012xw}) constitute the representation on the fields of each of the fermionic generators, and the action of the bosonic generators matches the action obtained computing commutators of the supersymmetry transformations\footnote{\label{footnote:SpinorialLieDerivative}The spinorial Lie derivative is defined by 
\[
\cL_\xi \psi \equiv \xi^\mu D_\mu \psi + \frac{1}{4}\nabla_\mu \xi_\nu \gamma^{\mu\nu}\psi ~.
\]
There is an additional term $\overline{\epsilon}_{[1}D^2 \epsilon_{2]}$ in the commutator on $F$, but it vanishes for the Killing spinors on the sphere.}
\begin{equation}
\label{eq:N1SCA_Fields}
\begin{split}
	[\delta_{\epsilon_1}, \delta_{\epsilon_2}] \varphi &= \cL_\xi \varphi + \Delta \, q ~, \\
	[\delta_{\epsilon_1}, \delta_{\epsilon_2}] \psi &= \cL_\xi \psi + \frac{1}{2} \Delta \, \psi ~, \\
	[\delta_{\epsilon_1}, \delta_{\epsilon_2}] F &= \cL_\xi F + \Delta \, F ~.
\end{split}
\end{equation}
Here
\begin{equation}
	\xi^\mu \equiv 2\ii \overline{\epsilon}_2 \gamma^{\mu} \epsilon_1 ~, \qquad \Delta \equiv \frac{1}{2} \nabla_\mu \xi^\mu ~,
\end{equation}
are, respectively, the parameter of a spacetime translation and a dilation. Notice that if both $\epsilon_{1,2}$ are positive or negative Killing spinors on the sphere, then $\xi$ is a Killing vector and thus $\Delta =0$, whereas if one supersymmetry parameter is positive and the other negative, then $\xi$ is a conformal Killing vector and $\Delta \neq 0$ (see appendix \ref{app:ConventionsEuclidean} for the conventions). This clearly matches the structure of the anti-commutators of the fermionic generators in \eqref{eq:osp12C_v2}, and the bosonic subalgebra $\mf{sl}(2,\C)$ is generated by isometries and conformal isometries of the sphere. 
\vspace{0.2cm}\newline
From \eqref{eq:N1SCA_Fields}, we notice that the infinitesimal dilation action is not linear in $\varphi$, leading to a finite transformation by a shift $\varphi \to \varphi + q$. This leaves invariant the action \eqref{N1Liouville} only in the semi-classical limit, whereas showing the invariance for non-infinitesimal $\beta$ requires taking into account quantum corrections. Because of this peculiar representation of the scale transformation on the scalar field of the multiplet $(\varphi, \psi, F)$, characteristic of the Liouville field, the latter does not behave as a canonical  supermultiplet.
\vspace{0.2cm}\newline
Upon regularization of the theory one must introduce a cutoff procedure that breaks superconformal invariance. This leads to a theory that is invariant under a massive subalgebra of the superconformal transformations above. This is concretely done by choosing $\epsilon$ to be a positive Killing spinor on the sphere \cite{Gerchkovitz:2014gta}, so that \eqref{eq:N1_SCFT_Transformations} become
\begin{equation}
\label{eq:N1_SUSY_Transformations}
\delta \varphi = \overline{\epsilon}\psi ~,  \quad\quad
\delta\psi = \ii \slashed{\partial}\varphi \, \epsilon - \frac{q}{r} \epsilon - \ii F \, \epsilon ~,  \quad\quad
\delta F = - \overline{\epsilon}\slashed{D}\psi ~.
\end{equation}
These transformations obey the $\cN=1$ supersymmetry algebra on the sphere, which is $\mf{osp}(1|2)$ sitting as a real algebra, generated by $J_m$ and $Q^\alpha$, inside the complexified $\mf{osp}(1|2,\C)$ of \eqref{eq:osp12C_v2}.
\vspace{0.2cm}\newline
{As previously emphasized, the action \eqref{eq:N1_TimelikeLiouville} includes twice the degrees of freedom needed to describe the real slice of $\cN=1$ supergravity. In order to avoid this doubling, when performing the path integration we impose a reality condition, specifically $\varphi$ and $F$ are required to be real, and $\psi$ to satisfies the Euclidean Majorana condition (\ref{eq:EuclideanReality}). With this choice, }
the  action \eqref{eq:N1_TimelikeLiouville} encodes the dynamical content  of two-dimensional  $\mathcal{N}=1$ supergravity coupled to an  $\mathcal{N}=1$ superconformal field theory when placed on an $S^2$. Upon integrating out the auxiliary $F$ field, we are led to the undecorated path integral
\begin{equation}
\label{N1Liouville_outF}
\mathcal{Z}_{\text{tL}}^{\mathcal{N}=1}[\mu]  = \int \frac{[\mathcal{D} \varphi][\mathcal{D} \psi]}{{\rm vol}_{OSp(1|2;\C)}}\, e^{-\frac{1}{4\pi}\int_{S^2} \dd^2 x\, \tilde{\rme}\,\left(- \frac{1}{2}\partial^\mu\varphi \partial_\mu {\varphi} +\frac{\ii}{2}  \bpsi \slashed{D} \psi - \frac{q}{r^2} \varphi +\frac{1}{2}\mu^2  e^{2\beta \varphi} + \frac{1}{2}\mu \beta e^{\beta \varphi}\bpsi \psi \right)}~.
\end{equation}
We note, again, that the corresponding cosmological term in the underlying $\mathcal{N}=1$ supergravity has a positive cosmological constant. As it stands, the above path integral still requires a careful choice of contour for $\varphi$, since the na\"ive real contour leads to a divergent expression. This is analogous to the conformal mode problem in Euclidean gravity \cite{Gibbons:1978ac}, and a reflection of the non-unitary nature of the timelike super-Liouville theory. It is worth emphasizing that the  non-unitarity is of the same form as that of the non-supersymmetric timelike Liouville theory, and not a feature that appears due to the supersymmetrization. Rather than attempting to define the theory non-perturbatively, for example via the methods of the conformal bootstrap, we  proceed to explore the path integral through a saddle point approximation.

\subsection{\texorpdfstring{dS$_2$}{dS2} saddle point  \texorpdfstring{\&}{and} Gaussian fields}

In the limit $\beta \to 0^+$, the $\mathcal{N}=1$ super-Liouville theory lends itself to a saddle point approximation.  To proceed with the saddle-point approximation, we start with the classical equations of motion stemming from \eqref{eq:N1_TimelikeLiouville}:
\begin{equation}\label{eom}
-	\nabla^2 \varphi = \mu^2 \beta e^{2\beta\varphi} - \frac{q}{r^2} + \frac{1}{2} \mu \beta^2 e^{\beta\varphi} \overline{\psi} \psi ~, \qquad \slashed{D}\psi = \ii \mu \beta e^{\beta \varphi} \psi ~. 
\end{equation}
To leading order in the  $\beta \to 0^+$ limit, the above equations are invariant under the $\mathcal{N}=1$ super-Moebius transformations. As such, given a solution  $(\varphi_*, \psi_*,F_*)$ to (\ref{eom}) we can generate an $OSp(1|2,\C)$ family of solutions. A classical solution has vanishing fermionic field and constant scalar field, so it has the form
\begin{equation}
\label{eq:N1_ClassicalSolution}
	\varphi_* = \frac{1}{2\beta} \log \frac{q}{r^2\mu^2\beta} ~, \qquad \psi_* = 0 ~.
\end{equation}
Interestingly, $\varphi_*$ is real-valued provided $\mu^2\beta>0$, indicating that the corresponding physical metric $g_{\mu\nu} = e^{2\beta \varphi_*}\tilde{g}_{\mu\nu}$ is the round metric on $S^2$ -- the Euclidean continuation of two-dimensional de Sitter space. The size of the two-sphere is parametrically large in the $\beta \to 0^+$ limit, its volume scales as $\sim 1/\mu^2\beta^2$. In contrast,  the classical solutions of the spacelike $\mathcal{N}=1$ super-Liouville theory, as explored in appendix \ref{spacelikeL} and specifically \eqref{eq:N1_Spacelike_Saddles}, are complex-valued. 
\vspace{0.2cm}\newline
The particular solution  \eqref{eq:N1_ClassicalSolution} breaks the $OSp(1|2,\C)$ super-Moebius group, as can be seen for instance by computing the variation $\delta\psi$. Nonetheless, we should recall that the theory has a $OSp(1|2,\C)$ family of saddle-point solutions which are all gauge equivalent from the perspective of the underlying $\mathcal{N}=1$ supergravity. As such, in this context there is no sense in which the $OSp(1|2,\C)$ is or can be broken. Instead, we path integrate over the whole classical solution space and divide by the residual volume of $OSp(1|2,\C)$.\footnote{This is a supersymmetric generalization of the situation for ordinary timelike Liouville theory in the semiclassical limit \cite{Anninos:2021ene}, whereby any saddle point solution breaks the $PSL(2,\C)$ Moebius group, which is restored upon integrating over the whole moduli space of solutions. Again, from the underlying gravitational perspective, the $PSL(2,\C)$ is a gauge redundancy which cannot be broken.}
\vspace{0.2cm}\newline
Given the classical solution \eqref{eq:N1_ClassicalSolution}, we can compute the on-shell action to obtain the leading contribution of the saddle to the undecorated $S^2$ path integral in the semiclassical limit $\beta\to 0^+$
\begin{equation}\label{ZtLs}
\mathcal{Z}_{\text{tL}}^{\mathcal{N}=1}[\mu]  \approx \mathcal{Z}^{\mathcal{N}=1}_{\text{saddle}}[\mu]=  \left(\frac{q}{e\mu^2 r^2 \beta}\right)^{\frac{q}{2\beta}}~.
\end{equation}
Inspection of the exact path integral (\ref{N1Liouville_outF}) shows that by resorting to a shift in $\varphi$ one can argue that the $\mu$ dependence in (\ref{ZtLs}) is in fact exact.
\vspace{0.2cm}\newline
{\textbf{Gaussian theory.}} Given the saddle point solution, we can expand the fields 
\begin{equation}
\varphi= \varphi_*+ \delta \varphi~,\quad \psi= \psi_*+ \delta \psi~,
\end{equation}
in order to obtain a Gaussian Lagrangian, which we split into a fermionic and bosonic piece 
\begin{equation}\label{GaussianL}
\mathcal{L}_{\text{bos}}^{(2)}[\delta\varphi] = -\frac{1}{8\pi}\,\delta \varphi\left(-\nabla^2 -\frac{2q\beta}{r^2}\right)\delta\varphi ~,\quad\quad
\mathcal{L}_{\text{fer}}^{(2)}[\delta\psi] = \frac{\ii}{8\pi}\,\delta\bpsi \left( \slashed{D} -\ii\frac{\sqrt{q\beta}}{r}\right)\delta\psi ~.
\end{equation}
The corresponding Gaussian path integral is then given by
\begin{equation}\label{Zfluc}
\mathcal{Z}_{\text{tL},\text{pert}}^{\mathcal{N}=1} = \int [\mathcal{D}\delta \varphi] [\mathcal{D}\delta \psi]\, e^{-\int_{S^2} \dd^2 x\, \tilde{\rme}\,\left( \mathcal{L}_{\text{bos}}^{(2)}[\delta\varphi]+ \mathcal{L}_{\text{fer}}^{(2)}[\delta\psi]\right) }~.
\end{equation}
Recalling the relation $q= \beta^{-1}-\beta$, we note that the bosonic fluctuations in (\ref{GaussianL}) have negative `mass' squared.\footnote{Generally speaking, when we use the term `mass' we mean the non-kinetic coefficient of the quadratic term in the fluctuation Lagrangian. A proper characterization is best expressed in terms of the appropriate irreducible representation of the dS$_2$ group $SO(2,1)$ organizing the single-particle Hilbert space rather than the more ambiguous notion of a mass in curved space.} This phenomenon parallels the non-supersymmetric timelike Liouville case. Relatedly, the fermionic `mass' term in (\ref{GaussianL}) is non-standard and would correspond to a pure imaginary `mass' in Lorentzian signature. {It is interesting to compare the situation with the four-dimensional case. The existence of a discrete series spin-$3/2$ unitary irreducible representation (UIR) of the four-dimensional de Sitter group $SO(1,4)$ was established in \cite{ottoson1968classification}. Attempts to realise this UIR in terms of a field theoretic Rarita--Schwinger Lagrangian in dS$_4$ suggest an imaginary `mass' for the Rarita--Schwinger field \cite{Deser:2003gw,Letsios:2023qzq}. What we are observing might be viewed as a two-dimensional counterpart of this.}
\vspace{0.2cm}\newline
The non-standard reality properties of the Gaussian fluctuations place some tension on the unitarity of the theory. On the other hand, we ought to remember that the fluctuations represent pieces of an underlying gauge field subject to gauge constraints that remove all propagating degrees of freedom, equally so in the supersymmetric and non-supersymmetric cases.   What is essential, in the least, is that upon taking these constraints into account one can make sense of the path integral. Along this vein, we now consider the path integral at the one-loop level. 

\subsection{One-loop contribution}\label{sec:oneLoopN1}

To compute the one-loop contribution $\mathcal{Z}_{\text{tL,pert}}^{\mathcal{N}=1}$ in (\ref{Zfluc}) it is convenient to expand the fields in a basis of spherical harmonics. We can expand $\varphi$ in a complete basis of real spherical harmonics $Y_{lm}(\Omega)$
\begin{equation}\label{phiexp}
\varphi(\Omega) = \sum_{l=0}^\infty \sum_{m=-l}^l \varphi_{lm} Y_{lm}(\Omega)~,\quad\quad  -\nabla^2 Y_{lm}(\Omega) = \frac{1}{r^2}(l+1)Y_{lm}(\Omega)~.
\end{equation}
Here $\Omega$ is a point on $S^2$ with unit area, the coefficients $\varphi_{lm}$ are real-valued, and $Y_{lm}(\Omega)$ are defined in appendix \ref{app:Fermions}.  We have normalized the $Y_{lm}(\Omega)$ such that 
\begin{equation}
\int_{S^2} \dd\Omega \,Y_{lm}(\Omega)Y_{l' m'}(\Omega) = \delta_{l l'} \delta_{m m'}~,
\end{equation}
where $\dd\Omega$ denotes the volume measure on the unit round sphere of area $4\pi$. We observe that all but the $l=0$ mode in (\ref{Zfluc}) are Gaussian unsuppressed. Following \cite{Gibbons:1978ac}, to render the path integral well defined we analytically continue $\delta \varphi_{lm} \rightarrow \pm \ii \delta \varphi_{lm}$ for all $l\geq 0$ and $-l \leq m \leq l$. The Jacobian of this transformation is a local term that can be absorbed in the measure. Upon continuing all the modes, the $l=0$ mode $\delta \varphi_{00}$ which was suppressed before the analytic continuation is now unsuppressed. Continuing it back, as in \cite{Polchinski:1988ua}, introduces an overall factor of $\pm \ii$.
\vspace{0.2cm}\newline
We can similarly expand the fermion $\psi$ in a complete basis of eigenfunctions of the Dirac operator on the two-sphere \cite{Camporesi:1995fb}, 
\begin{equation}
\psi(\Omega) =\frac{1}{\sqrt{r}} \sum_{l=0}^\infty \sum_{m=0}^l\left(\alpha_{+\,lm} \psi_{+\, lm}^{(+)}(\Omega) + \alpha_{-\,lm} \psi_{-\, lm}^{(+)}(\Omega) + \beta_{+\,lm} \psi_{+\, lm}^{(-)}(\Omega) + \beta_{-\,lm}  \psi_{-\, lm}^{(-)}(\Omega)\right)~.
\end{equation}
We must  subject $\psi$ to the Euclidean Majorana condition discussed in (\ref{eq:EuclideanReality}). As such,
\begin{equation}
\label{psiexp}
\psi(\Omega) =\frac{1}{\sqrt{r}} \sum_{l=0}^\infty \sum_{m=0}^l\left(\alpha_{+\,lm} \psi_{+\, lm}^{(+)}(\Omega) + \alpha_{-\,lm} \psi_{-\, lm}^{(+)}(\Omega) -\ii\alpha_{-\,lm}^* \psi_{+\, lm}^{(-)}(\Omega) -\ii \alpha^*_{+\,lm} \psi_{-\, lm}^{(-)}(\Omega)\right)~.
\end{equation}
The eigenspinors $\psi_{\pm\,lm}^{(\pm)}(\Omega)$, defined in appendix \ref{app:Fermions}, obey the eigenvalue equation 
\begin{equation}\label{snabla}
\slashed{D} \psi_{\pm\,lm}^{(s)}(\Omega) = \pm \frac{\ii}{r}(l+1)\psi_{\pm,lm}^{(s)}(\Omega)~,\quad s= \pm~,
\end{equation} 
and satisfy the orthogonality condition
\begin{equation}\label{ortho_fermions}
\int_{S^2} \dd \Omega\, \bpsi^{(+)}_{\pm \, lm}(\Omega) \psi^{(-)}_{\pm \, l'm'} (\Omega)= -  \ii \delta_{l,l'}\delta_{m,m'}~,\quad \int_{S^2} \dd \Omega\, \bpsi^{(-)}_{\pm \, lm}(\Omega) \psi^{(+)}_{\pm \, l'm'}(\Omega) =  +\ii \delta_{l,l'}\delta_{m,m'}~.
\end{equation}
The spectrum of the Dirac operator (\ref{snabla}) is purely imaginay, and there is no vanishing eigenvalue for $l\geq 0$ in  (\ref{snabla}).
The coefficients $\alpha_{\pm\,lm}$ and $\alpha^*_{\pm\,lm}$ in (\ref{psiexp}) are dimensionless  complex Grassmann numbers. Each fermionic eigenvalue $l\in \mathbb{Z}$ is $2(l+1)$-fold degenerate. In (\ref{snabla}) this is captured by the $\pm$ superscript, which accounts for eigenfunctions that would otherwise have negative $m$.
\vspace{0.2cm}\newline
The path integration measure on the  bosonic and fermionic eigenmodes (\ref{phiexp}) and (\ref{psiexp}) is defined as
\begin{equation}\label{fermionMeasure}
\quad [\mathcal{D}\psi] =  \prod_{l=0}^\infty \prod_{m=0}^l\left(\frac{2\pi}{\Lambda_{\text{uv}}r^2}\right)\prod_{\pm}\dd{\alpha}_{\pm\, lm}\dd{\alpha}^*_{\pm\, lm}~,\quad [\mathcal{D}\varphi] =  \prod_{l=0}^\infty \prod_{m=-l}^l\left(\frac{\Lambda_{\mathrm{uv}} r^2}{\pi}\right)^{\frac{1}{2}}{\text{d}   \varphi_{lm}}~,
\end{equation}
leading to
\begin{equation}
1=\int [\mathcal{D}\psi] e^{-\frac{1}{2\sqrt{2\pi}}\Lambda^{1/2}_{\text{uv}}\int_{S^2} \dd^2  x \, \tilde{\rme}\,\bpsi \psi}~,\quad\quad 1= \int [\mathcal{D} \varphi] e^{- \Lambda_{\mathrm{uv}} \int_{S^2} \dd^2 x \, \tilde{\rme}\, \varphi(x)^2}~.
\end{equation}
The parameter $\Lambda_{\text{uv}} \equiv 1/\ell^2_{\text{uv}}$ is an ultraviolet reference scale of units inverse length squared required to define the path integration measure. The reason for our choice of normalization for the fermionic measure will become evident in the evaluation of the one-loop determinant.
\vspace{0.2cm}\newline
\textbf{Fermionic one-loop determinant.}
 The fermionic Gaussian action evaluates to 
\begin{equation}\label{LF2}
S_{\text{tL},\text{fer}}^{(2)}[\delta\psi]  = \frac{1}{4\pi}   \sum_{l= 0}^\infty \sum_{m= 0}^l\left((l+1 - \sqrt{q\beta})\alpha_{+\, lm}\alpha_{-\,lm}^* - (l+1 + \sqrt{q\beta}) \alpha_{-\,lm}\alpha_{+\,lm}^*\right)~.
\end{equation}
Since the fermionic Gaussian part mixes the real mass term and the imaginary kinetic term, the action $S_{\text{tL,fer}}^{(2)}$ is inherently complex. Furthermore it is asymmetric in the coefficients for $\alpha_{+\, lm}\alpha_{-\,lm}^*$ and $ \alpha_{-\,lm}\alpha_{+\,lm}^*$. 
Since the eigenvalue $l$ is integer valued we observe that the second term in (\ref{LF2}) vanishes for $l=0$ up to order $\mathcal{O}(\beta^2)$ effects. As such we have two fermionic (almost) zero modes. 
The fermionic one-loop contribution stemming from (\ref{Zfluc}) thus reads
\begin{equation}\label{fermionGaussianN1}
\mathcal{Z}_{\text{tL,fer},1\text{-loop}}^{\mathcal{N}=1}\equiv \frac{2\pi}{\Lambda_{\text{uv}}r^2}\times \int [\mathcal{D}'\delta\psi]\, e^{-S_{\text{tL},\text{fer}}^{(2)}[\delta\psi]} = \frac{2\pi}{\Lambda_{\text{uv}}r^2}\times \prod_{l= 1}^\infty\left(\frac{(1+l)^2 - q\beta}{8\pi\Lambda_{\text{uv}}r^2}\right)^{l+1}~,
\end{equation}
where the prime in the measure indicates that we omitted the two $l=0$ zero modes. The zero modes must be treated separately, as their effect is to cancel part of the volume of $OSp(1|2,\C)$. We have nevertheless kept their contribution from the path integration measure, as they contribute to the ultraviolet divergent structure of the path integral.
\vspace{0.2cm}\newline
Using the same techniques as \cite{Anninos:2020hfj, Anninos:2021ene}, a heat kernel analysis can be employed to evaluate the infinite product. After some manipulations, we find 
\begin{multline}
\sum_{l=  1}^{\infty} (l+1) \log \frac{(1+l)^2- q\beta}{8 \pi\Lambda_{\text{uv}}r^2}  \cr
=-\int_{0}^\infty\frac{\dd t}{2t}\left[\frac{2e^{-\frac{t}{2}}}{(1-e^{-t})}\left(\frac{2e^{-\frac{t}{2}+\ii\nu t}}{(1-e^{-t})}-2e^{-\frac{t}{2}+\ii\nu t} - e^{-(\frac{5}{2}+\ii\nu)t}\frac{(1-2e^t +e^{2\ii\nu t})}{(1-e^{-t})}\right)\right]~,
\end{multline}
where $\nu \equiv \sqrt{-q\beta}$ and we take $\sqrt{-1}=\ii$. We can identify part of the above bracket (to leading order in $\beta$), namely
\begin{equation}\label{chi32}
\chi_{\Delta=3/2}(t) = \frac{e^{-\frac{3}{2}t}}{(1-e^{-t})}
\end{equation}
as the Harish--Chandra character of the double cover of $SO(1,2) \cong  PSL(2,\mathbb{R})$ discrete series representation with $\Delta=3/2$. The remaining terms exhibit subleading divergent structure near $t = 0$. This is reminiscent of the structure of the sphere path integral for gauge fields appearing in general dimensions, as studied in \cite{Anninos:2020hfj}.  Interestingly, although we have an imaginary Lorentzian mass for the fermionic fluctuations, the $SO(1,2)$ character $\chi_{\Delta=3/2}(t)$ is that of a unitary irreducible representation.
The heat kernel regularization amounts to the cutoff procedure
\begin{equation}\label{bessel}
-\frac{1}{2} \log \frac{{\color{black}\tilde{\lambda}}}{8\pi r^2\Lambda_{\mathrm{uv}}} = \int_0^\infty \frac{\dd\tau}{2\tau} e^{-\frac{\varepsilon^2}{4\tau}-{\color{black}\tilde{\lambda}}\tau} =  K_0\left(\sqrt{{\color{black}\tilde{\lambda}} } \varepsilon \right) \approx -\frac{1}{2}\log \frac{\varepsilon^2\, e^{2\gamma_E}{{\color{black}\tilde{\lambda}}}}{4}~,
\end{equation}
where $\varepsilon$ is a small dimensionless parameter given by $\varepsilon = {e^{-\gamma_E}}/{\sqrt{2\pi r^2 \Lambda_{\mathrm{uv}}}}$, and $\gamma_E$ denotes the Euler--Mascheroni constant. 
We then obtain 
\begin{equation}
\log \mathcal{Z}_{\text{tL,fer},1\text{-loop}}^{\mathcal{N}=1} = -\frac{2}{\varepsilon^2} +\left(\frac{5}{6}-\beta^2\right)\log \varepsilon  + \text{constants}~,
\end{equation}
where the constants are independent of the radius $r$,  $\Lambda_{\text{uv}}$, and consequently $\varepsilon$.
\vspace{0.2cm}\newline
\textbf{Bosonic one-loop determinant.}
The bosonic case can be treated similarly. Since the eigenvalues of the spherical Laplacian (\ref{phiexp}) are integer-valued and $q= \beta^{-1}-\beta$ the bosonic Gaussian in (\ref{GaussianL}) has three almost zero modes, i.e. zero modes that are lifted at order $\mathcal{O}(\beta^2)$.
In our one-loop analysis we treat them separately.
Following \cite{Anninos:2021ene} we obtain
\begin{multline}\label{Zbos_loop_N1}
\mathcal{Z}_{\text{tL,bos},1\text{-loop}}^{\mathcal{N}=1}\equiv {\color{black}\pm \ii} \left(\frac{4\pi r^2\Lambda_{\mathrm{uv}}}{\beta q}\right)^{\frac{1}{2}}  \left({\frac{r^2\Lambda_{\mathrm{uv}}}{\pi}}\right)^{\frac{3}{2}}\times  \int [\mathcal{D}'\delta\varphi]\, e^{-\frac{1}{4\pi} \int_{S^2} \dd x^2 \, \tilde{\rme}\,  \frac{1}{2}\delta\varphi\left( -\nabla^2- \frac{2}{r^2}{q\beta}\right)\delta\varphi } \cr
= {\color{black}\pm \ii}  \left(\frac{4\pi r^2\Lambda_{\mathrm{uv}}}{\beta q}\right)^{\frac{1}{2}}  \left({\frac{r^2\Lambda_{\mathrm{uv}}}{\pi}}\right)^{\frac{3}{2}}\times \prod_{l= 2}^\infty\left(\frac{8\pi r^2\Lambda_{\mathrm{uv}}}{l(l+1)-2\beta q}\right)^{l+\frac{1}{2}}~,
\end{multline}
where the $\pm \ii$ is the Jacobian arising when Wick rotating the $l=0$ mode of $\varphi$. 
The infinite product evaluates to 
\begin{multline}\label{heatkernel}
-\frac{1}{2} \sum_{l = 2}^\infty ({2l+1} ) \log \left( \frac{l(l+1) -2\beta q}{8\pi\Lambda_{\mathrm{uv}} r^2} \right)  =  \int_{0}^\infty\frac{\dd t}{2t}\bigg[ \frac{1+e^{-t}}{(1-e^{-t})}\frac{2e^{-\frac{t}{2}+\ii\nu t}}{(1-e^{-t})} \cr
+ e^{-(\frac{5}{2}-\ii\nu)t} \left(\frac{3-2e^t+5e^{3t}}{1-e^{-t}} - e^{2t}\frac{(5e^t-3)(1- e^{-(3+2\ii\nu) t})}{(1-e^{-t})^2}\right)\bigg]~,
\end{multline}
where $\nu = \sqrt{-1/4-2q\beta}$. With this choice the last term in the second line of (\ref{heatkernel}) contributes only at order $\mathcal{O}(\beta^2)$ and the leading divergence still comes from the first term. For the leading contribution $\nu = 3\ii/2 $ we can identify 
\begin{equation}
\chi_{\Delta=2}(t) = \frac{e^{-{2t}}}{(1-e^{-t})}
\end{equation}
as the Harish--Chandra character of the $SO(1,2)$ discrete series representation with $\Delta=2$. The corresponding quadratic Casimir $\Delta(1-\Delta) = -2$ is a negative integer.\footnote{Interestingly, the single-particle Hilbert space of a linearised graviton in dS$_4$, prior to implementing  the constraints stemming from the residual $SO(1,4)$ diffeomorphisms,  furnishes a discrete series representation in $SO(1,4)$. This does not occur for general dimensionality, in particular the discrete series are altogether absent in odd spacetime dimensions.} Due to the corrections in $\beta$, this is only an approximation. From (\ref{heatkernel}) we can read off the leading divergences
\begin{equation}
\log \mathcal{Z}_{\text{bos},1\text{-loop}}^{\mathcal{N}=1} = \frac{2}{\varepsilon^2} -\left(\frac{7}{3}-2\beta^2\right) \log \varepsilon+ \text{constants} 
\end{equation}
where $\varepsilon = {e^{-\gamma_E}}/{\sqrt{2\pi r^2 \Lambda_{\mathrm{uv}}}}$ follows from analogous steps as (\ref{bessel}).
\vspace{0.2cm}\newline\newline
Combining the fermionic and bosonic contribution we notice that the polynomial divergence cancels. This is the usual cancellation of the divergent contribution to the cosmological constant in supersymmetric theories. We expect a similar cancelation of the leading ultraviolet divergence to occur for the superconformal matter theory, as well as the combined $\mathfrak{bc}$-ghost and $\beta\gamma$-ghost theories. The latter cancelation follows from the structure of the combined ghost determinants \cite{Polyakov:1981re}.  
\vspace{0.2cm}\newline
Expanding the resulting path integral in the semiclassical limit $\beta \rightarrow 0^+$ limit
we obtain
\begin{equation}\label{ZtLN1}
\mathcal{Z}_{\text{tL}}^{\mathcal{N}=1} \approx  {\pm \ii} \times \text{const}\times  \left(\frac{\mu}{\beta}\right)^{-\frac{1}{\beta^2}+1}\Lambda_{\text{uv}}^{\frac{5}{4}-\beta^2} \, r^{c_{\text{tL}}^{\mathcal{N}=1}/3} \, e^{-\frac{1}{2\beta^2}- \left( \frac{1}{\beta^2}-1 \right)\log \beta^2 }  \times {\beta^{-1} } \times \left( 1 + \mathcal{O}(\beta^2) \right)~,
\end{equation}
where `const' denotes a real-valued constant independent of $\beta$,  $\mu$, and $r$. The $\mathcal{N}=1$ timelike Liouville central charge is $c_{\text{tL}}^{\mathcal{N}=1} = 3/2-3q^2$ as was declared below (\ref{eq:N1_TimelikeLiouville}). One no longer needs to divide by the volume of $OSp(1|2,\mathbb{C})$ in (\ref{ZtLN1}), as it has been used in fixing the three bosonic zero modes and the two fermionic zero modes. A proper treatment of this would require computing a Faddeev--Popov superdeterminant for the gauge fixing procedure, as in \cite{Anninos:2021ene}, which we leave to the future. Inspection of (\ref{SMPhi}) suggests that to leading order in the small $\beta$ expansion, the Faddeev--Popov superdeterminant contributes a factor of $1/\beta$ for each bosonic zero mode and a factor of $\beta$ for each fermionic zero mode yielding the prefactor of $\beta^{2-3} = \beta^{-1}$ in $\mathcal{Z}_{\text{tL}}^{\mathcal{N}=1}$. 

\subsection{Non-Gaussian interactions}
We now consider the leading effect due to the interactions, which is a two-loop contribution. Expanding the action to quadratic order in $\beta$, we  encounter interaction terms between the fermionic and bosonic fields. Concretely,
\begin{equation}\label{eq:LflucN1timelike}
4\pi \mathcal{L}_{\text{tL,int}}^{\mathcal{N}=1}[\delta\varphi,\delta\psi] = - \frac{2\ii}{3r^2}q\beta^2 \delta\varphi^3 + \frac{1}{3r^2}q\beta^3\delta\varphi^4 + \frac{\ii}{2r}\beta\sqrt{q\beta}\,\delta\varphi\delta\overline{\psi}\delta\psi - \frac{1}{4r}\beta^2\sqrt{q\beta}\,\delta\varphi^2\delta\overline{\psi}\delta\psi~,
\end{equation}
where we recall that we have Wick rotated $\delta \varphi \rightarrow + \ii\delta \varphi$ to render the fluctuations (up to zero modes) Gaussian suppressed. Choosing the positive sign in this rotation is a choice, and none of our results will rely on it. 
To proceed with the two-loop expressions, we obtain the relevant propagators. The appearance of $r$ in (\ref{eq:LflucN1timelike}) is immaterial. Upon integrating over the two-sphere of size $r$, any remaining $r$ dependence can be removed by a simple rescaling of $\delta\psi$. Thus, provided there are no ultraviolet divergences in the loop diagrams, the fluctuating theory produces no further $r$-dependent contributions to $\mathcal{Z}_{\text{tL}}^{\mathcal{N}=1}$ in (\ref{ZtLN1}). 
\vspace{0.2cm}\newline
{\textbf{Fermionic propagator.}} The fermionic propagator is given by\footnote{Here and below we define a momentum space expectation value as
\begin{equation}
\langle \mathcal{H}(\alpha_{\pm lm},\alpha^*_{\pm l'm'})\rangle  \equiv \left(\mathcal{Z}_{\text{fer},1\text{-loop}}^{\mathcal{N}=1} \right)^{-1} \int \left[\mathcal{D}\delta \psi \right] e^{-S_{\text{L},\text{fer}}^{(2)}[\delta\psi]  }   \mathcal{H}(\alpha_{\pm lm},\alpha^*_{\pm l'm'})~,
\end{equation}
implementing the momentum space action (\ref{LF2}) and momentum space measure (\ref{fermionMeasure}). Here $\mathcal{H}(\alpha_{\pm lm},\alpha^*_{\pm l'm'})$ is an arbitrary polynomial in $\alpha_{\pm lm}$, $\alpha^*_{\pm l'm'}$.  The integration variables $\alpha^*_{\pm lm}$ and $\alpha_{\pm lm}$ are treated as independent Grassmann variables with the convention that
\begin{equation}
\int \dd \alpha_{+lm}\dd\alpha_{+lm}^* \dd \alpha_{-lm}\dd\alpha_{-lm} \, \alpha_{+lm}\alpha_{+lm}^*\alpha_{-lm}\alpha_{-lm}^*=1~,
\end{equation}
for all possible combinations of $l,m$.
}
\begin{multline}\label{FN1}
H(\Omega,\Omega') \equiv  \frac{1}{\mathcal{Z}_{\text{fer},1\text{-loop}}^{\mathcal{N}=1}}  \frac{2\pi}{\Lambda_{\text{uv}}r^2} \int [\mathcal{D}'\delta\psi] \, e^{-S_{\mathrm{pert}}^{(2)}[\delta\psi]} \delta\bpsi(\Omega) \delta\psi(\Omega') \cr
=\frac{4\pi}{r}\sum_{l\geq 1}\sum_{0\leq m\leq l}\frac{1}{(l+1)^2-q\beta} \Big[ \ii (l+1+\sqrt{q\beta})\left(\bpsi^{(+)}_{+\,lm}(\Omega) \psi_{+\,lm}^{(-)}(\Omega')-\bpsi^{(-)}_{+\,lm}(\Omega) \psi_{+\,lm}^{(+)}(\Omega')\right)\cr
+ \ii (l+1-\sqrt{q\beta})\left(\bpsi^{(-)}_{-\,lm}(\Omega) \psi_{-\,lm}^{(+)}(\Omega')-\bpsi^{(+)}_{-,lm}(\Omega) \psi_{-\,lm}^{(-)}(\Omega')\right)\Big]~,
\end{multline}
where $S_{\mathrm{pert}}^{(2)}[\delta\psi]$ is the action associated to the Lagrangian in (\ref{GaussianL}). 
For the discussion that follows, the behaviour of (\ref{FN1}) in the coincident point limit is of importance and combining (\ref{ortho_fermions}) with (\ref{FN1}) we readily conclude that this is given by
\begin{equation}\label{eq:F00N1}
H(\Omega_0,\Omega_0) = \frac{1}{4\pi}\int_{S^2} \dd \Omega \, H(\Omega,\Omega)  = + \frac{4\sqrt{q \beta}}{r}\sum_{l\geq 1}\frac{l+1}{(l+1)^2-q\beta}~,
\end{equation}
which is logarithmically divergent. 
\newline\newline
{\textbf{Bosonic propagator.}}
For the bosonic propagator $G(\Omega,\Omega')$ one similarly obtains
\begin{equation}\label{propagatorN1}
G(\Omega,\Omega') = 4\pi \sum_{l \neq 1} \sum_{-l\leq m\leq l} \frac{{Y}_{l m}(\Omega) {Y}_{l m}(\Omega')}{l(l+1)-2q\beta}  ~.
\end{equation}
Given that we have removed all three $l=1$ modes, which form a three-dimensional representation of $SO(3)$, the two-point function $G(\Omega;\Omega')$ remains $SO(3)$ invariant.
Similar to the fermionic coincident point limit, the bosonic propagator at coincident points diverges logarithmically. It is given by
\begin{equation}\label{divloop_N1}
G(\Omega_0,\Omega_0) = \frac{1}{4\pi}\int_{S^2}\dd\Omega \,G(\Omega,\Omega) = \sum_{l \neq 1 } \frac{2l+1}{l(l+1)-2q\beta} ~.
\end{equation}
\newline
{\textbf{Two-loop contributions.}} Given the bosonic and fermionic propagators, as well as the perturbative interactions in (\ref{eq:LflucN1timelike}), one can systematically study quantum corrections near the saddle point solution. The simplest of these is the two-loop contribution to the undecorated path integral of timelike super-Liouviolle theory about the Euclidean dS$_2$ saddle. 
$\,$\vspace{0.2cm}\newline
Expanding the fluctuations in (\ref{eq:LflucN1timelike}) and performing Gaussian integrals, we obtain purely bosonic diagrams and diagrams that combine fermionic and bosonic propagators. The purely bosonic two-loop contributions are 
\begin{multline}\label{loops_b_N1}
\text{loops}_b = -\frac{\beta^2}{8\pi^2} \int_{S^2}\dd\Omega\dd\Omega'G(\Omega,\Omega) \, G(\Omega',\Omega')G(\Omega,\Omega') -\frac{\beta^2}{12\pi^2} \int_{S^2} \dd\Omega\dd\Omega' \, G(\Omega,\Omega')^3
\\ - \frac{\beta^2}{4\pi}\int_{S^2} \dd{\Omega} \, G(\Omega,\Omega)^2~.
\end{multline}
Diagrammatically, the above correspond to double tadpoles, melonic and cactus diagrams as shown in figure \ref{fig:diagramsValpha}.
\begin{figure}[H]
\begin{center}
\begin{tikzpicture}[scale=.5]

\draw[black,thick] (-2.5,0) --(-1.5,0);
\draw[thick] (-0.5,0) circle (1cm);
\draw[thick] (-3.5,0) circle (1cm);

\draw[thick] (6.5,0) circle (1cm);
\draw[black,thick] (7.5,0) --(5.5,0);

\draw[thick] (12.5,0) circle (1cm);
\draw[thick] (14.5,0) circle (1cm);

\end{tikzpicture}
\end{center}
\caption{Double-tadpoles, melons and cactus diagrams.}
\label{fig:diagramsValpha}
\end{figure}
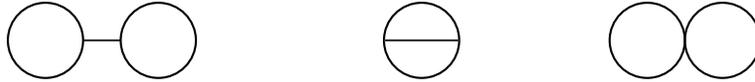
\noindent
These diagrams are shared by the non-supersymmetric timelike Liouville theory studied in \cite{Anninos:2021ene}, which can be obtained from the bosonic subsector of the fluctuation theory (\ref{eq:LflucN1timelike}) by rescaling $\delta\varphi\to \sqrt{2} \delta\varphi$ and $\beta\to \beta/\sqrt{2}$ to leading order at small $\beta$ (see footnote \ref{footnote:Normalization}).  In \cite{Anninos:2021ene} it was established that although the double tadpole and cactus diagrams are logarithmically divergent, they exhibit an exact cancellation.
The melonic diagrams, instead, are finite and evaluate to 
\begin{equation}\label{melons_N1}
\mathlarger{\mathlarger{\ominus}} \equiv -\frac{1}{6}\sum_{l_1,l_2,l_3\neq 1} \frac{(2l_1+1)(2l_2+1)(2l_3+1)}{(l_1(l_1+1)-2)(l_2(l_2+1)-2)(l_3(l_3+1)-2)} \begin{pmatrix} l_1 & l_2 & l_3 \\ 0 & 0 & 0\end{pmatrix}^2~,
\end{equation}
where the last term denotes the 3j symbol.
The novel contributions arise from diagrams involving also fermion loops in the $\mathcal{N}=1$ timelike super-Liouville theory. These are given by 
\begin{multline}\label{2loop_exp_N1}
 \text{loops}_f = \frac{r}{16\pi^2}\beta^2\int_{S^2} \dd\Omega\dd\Omega' \, H(\Omega,\Omega)G(\Omega',\Omega') G(\Omega,\Omega') - \frac{r^2}{128\pi^2}\beta^2\int_{S^2} \dd\Omega\dd\Omega' \, H(\Omega,\Omega) H(\Omega',\Omega')G(\Omega,\Omega') \cr
 + \frac{r}{16\pi}\beta^2\int_{S^2}\dd{\Omega} \, H(\Omega,\Omega)G(\Omega,\Omega)
 - \frac{r^2}{128\pi^2}\beta^2 \int_{S^2} \dd\Omega\dd\Omega' \, H(\Omega,\Omega')^2 G(\Omega,\Omega')  ~.
\end{multline}
and are diagrammatically expressed in figure \ref{fig:twoloopN1a}.
\begin{figure}[H]
\begin{center}
\begin{tikzpicture}[scale=.5]
\draw[thick] (0,0) circle (1cm);
\draw[thick] (1,0)--(2,0);
\draw[thick, magenta, densely dashed] (3,0) circle (1cm);

\draw[thick, magenta, densely dashed] (7,0) circle (1cm);
\draw[thick] (8,0)--(9,0);
\draw[thick, magenta, densely dashed] (10,0) circle (1cm);

\draw[thick] (14,0) circle (1cm);
\draw[thick, magenta, densely dashed] (16,0) circle (1cm);

\draw[thick, magenta, densely dashed] (20,0) circle (1cm);
\draw[thick] (19,0)--(21,0);

\end{tikzpicture}
\end{center}
\caption{Two-loop contribution for $\mathcal{N}=1$ timelike super-Liouville theory. The magenta dashed lines correspond to fermions, whereas black lines are bosonic propagators.}
\label{fig:twoloopN1a}
\end{figure}
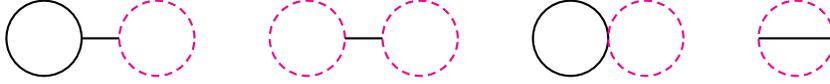
\noindent
We adopt the convention that bold dashed magenta lines are fermion propagators, whereas black lines correspond to the propagation of $\delta\varphi$. The first three diagrams in (\ref{2loop_exp_N1}) are easily evaluated and yield schematically
\begin{align}\label{eq:diagrams_N1_loopf}
\frac{r}{16\pi^2}\int_{S^2} \dd\Omega\dd\Omega' \, H(\Omega,\Omega)G(\Omega',\Omega') G(\Omega,\Omega') &= \frac{r}{(-2)}H(\Omega_0,\Omega_0)G(\Omega_0,\Omega_0) \approx  -4 \log^2 (\Lambda_{\text{uv}}r^2) + \ldots ~,\cr
 - \frac{r^2}{128\pi^2}\int_{S^2} \dd\Omega\dd\Omega' \, H(\Omega,\Omega) F(\Omega',\Omega')G(\Omega,\Omega')  &= - \frac{r^2}{8 } \frac{1}{(-2)} H(\Omega_0,\Omega_0)^2 \approx \log^2 (\Lambda_{\text{uv}}r^2) + \ldots ~,\cr
 \frac{r}{16\pi}\int_{S^2}\dd{\Omega} \, H(\Omega,\Omega)G(\Omega,\Omega) &=\frac{r}{4}H(\Omega_0,\Omega_0)G(\Omega_0,\Omega_0) \approx 2\log^2 (\Lambda_{\text{uv}}r^2) + \ldots ~,
\end{align}
where the minus two is the propagator for the bosonic $l=0$ mode (\ref{propagatorN1}) and the coincident point limit of the bosonic and fermionic propagator is given in (\ref{divloop_N1}) and (\ref{eq:F00N1}) respectively.  
The melonic-type diagram is more involved, but numerically we infer
\begin{equation}
- \frac{r^2}{128\pi^2}\beta^2   \int_{S^2} \dd\Omega\dd\Omega' \, H(\Omega,\Omega')^2 G(\Omega,\Omega') \approx \log^2 (\Lambda_{\text{uv}}r^2)  + \ldots~,
\end{equation}
and thus the leading $\log^2$ divergence of the combined diagrams in (\ref{2loop_exp_N1}) vanishes.
\vspace{0.2cm}\newline
Finally, we note that the spacelike counterpart of the $\mathcal{N}=1$ timelike super-Liouville theory, which we discuss in appendix \ref{spacelikeL}, exhibits a very similar perturbative loop expansion. In particular, the cancellation of the two- and higher-loop ultraviolet divergences in the small $b$ expansion of one theory imply cancellation in the small $\beta$ expansion of the other.

\subsection{\texorpdfstring{$\mathcal{N}=1$}{N1} summary} \label{subsec:N1summary}

To summarize,  the structure of the $\mathcal{N}=1$ timelike Liouville theory $S^2$ path integral is given by
\begin{equation}\label{eq:N1timelikeFinal}
\mathcal{Z}_{\text{tL}}^{\mathcal{N}=1} \approx   \pm \ii \, \text{const}\times  \left(\frac{\mu}{\beta}\right)^{-\frac{1}{\beta^2}+1}\Lambda_{\text{uv}}^{\frac{5}{4}-\beta^2}r^{c_{\text{tL}}^{\mathcal{N}=1}/3} e^{-\frac{1}{2\beta^2}- \left(\frac{1}{\beta^2} -1 \right) \log \beta^2} \times {\beta^{-1}}\times (1 + \text{loops}\,\beta^2+\ldots)~,
\end{equation}
where `loops' denotes order $\mathcal{O}(\beta^2)$ contributions arising in the one-loop and two-loop effects as well as potential contributions from the Faddeev--Popov gauge fixing procedure. Upon fixing our path integration measure, the overall constant is independent of $\beta$ and $\mu$. 
Several remarks concerning (\ref{eq:N1timelikeFinal}) are in order:
\begin{itemize}
\item Due to the unboundedness of the $l=0$ scalar mode, and upon implementing the Gibbons--Hawking--Perry contour prescription \cite{Gibbons:1978ac}, the partition function of $\mathcal{N}=1$ timelike super-Liouville theory on the two-sphere is purely imaginary. This feature is shared by the corresponding non-supersymmetric path integral computed in \cite{Anninos:2021ene}. 
\item The conformal anomaly of the sphere partition function (\ref{eq:N1timelikeFinal}) provides a first explicit check of the central charge of $\mathcal{N}=1$ timelike Liouville theory, namely $c_{\text{tL}}^{\mathcal{N}=1} = 3/2-3q^2$.
\item Higher-loop contributions can be systematically evaluated. At two-loop order, ultraviolet divergences cancel. 
\item Following the hypothesis that timelike $\mathcal{N}=1$ super-Liouville theory is a (non-unitary) two-dimensional $\mathcal{N}=1$ SCFT, one is led to explore quantum properties of the theory via more general conformal bootstrap methodology. Of particular interest is the OPE coefficient of the operator
\begin{equation}
\mathcal{W}_\beta \equiv  F  e^{\beta \varphi}  - \frac{\ii}{2}\beta e^{\beta \varphi} \overline{\psi} \psi~
\end{equation}
which encodes the $S^2$ path integral. For spacelike $\mathcal{N}=1$ super-Liouville theory $\mathcal{W}_\beta$ can be obtained by acting with  superconformal generators on the NS primary $\mathcal{V}_\alpha \equiv e^{\alpha \varphi}$ \cite{Belavin:2007gz} with $\alpha=\beta$. Alternatively, the OPE coefficients can be obtained by differentiating the $S^2$ path integral with respect to $\mu$. Generally, we do not expect the timelike structure constants to be the analytic continuation of the spacelike structure constants. The case $\alpha = \beta$ is special, as it corresponds to an operator that is not  part of the physical spectrum of the theory.  In the non-supersymmetric case this justifies the analytic continuation of the structure constant \cite{Anninos:2021ene}. We briefly discuss the $\mathcal{N}=1$ counterpart  in appendix \ref{sec:spacelikeN1}.
\item From a purely field theory perspective, the sphere partition function of an $\cN=1$ SCFT is ambiguous, as it depends on the regularization scheme \cite{Gerchkovitz:2014gta}. Here, on the other hand, we are only viewing it as one of the building blocks in the two-dimensional supergravity \eqref{eq:Z_grav_N1_0}. Both leading and logarithmic ultraviolet divergences cancel in any SUSY-preserving regularization scheme for the entire supergravity, so the overall two-dimensional gravity leads to less ambiguous results.
\end{itemize}
$\,$\newline
Returning to the gravitational perspective, the structure of the undecorated $\mathcal{N}=1$ supergravity path integral (\ref{eq:ZSUGRA_General}) on an $S^2$ takes the following form
\begin{equation}
\label{eq:Z_grav_N1_final}
	\log \mc{Z}^{\cN=1}_{{\rm grav}, (0)} = 2\vartheta -\left(\frac{12}{\left(\sqrt{2 c_m-27}-\sqrt{2 c_m -3}\,\right)^2}- \frac{1}{2} \right) \log \frac{\mu^2}{\Lambda_{\text{uv}}}+ f_0(c_m)~.
\end{equation}
Here $f_0(c_m)$ is a series in $1/c_m$ encoding the quantum corrections from loop corrections stemming from the super-Liouville path integral. At large $c_m$, the coefficient of the logarithm goes as $c_m/6 - 7/4 + \ldots$, whose leading term takes the form \cite{Holzhey:1994we, Calabrese:2004eu, Casini:2011kv} appearing in the entanglement entropy of two-dimensional conformal field theory across a finite interval. One can compare the first subleading coefficient $-7/4$ to the one appearing in the non-supersymmetric case \cite{Anninos:2021ene} which is $-19/6$. Perhaps the difference can be attributed to an  entanglement entropy contribution from the differing gravity and ghost sectors.

\section{de Sitter \& supersymmetry}

\label{subsec:SUSYdS2}

Throughout our discussion, we observe two phenomena that are somewhat unusual from the perspective of supergravity. The first is the presence of a positive value for the cosmological constant, and the second is the appearance of de Sitter vacua in a theory with linearly realised supergravity transformations. In this brief section, we reflect on these features in light of the general relation between de Sitter space and supersymmetry.

\subsection{Supersymmetric quantum fields in de Sitter}

Before delving into the question of de Sitter supergravity, we review some general obstructions that arise upon constructing a supersymmetric quantum field theory on a rigid de Sitter background, which we take to be four-dimensional for simplicity. We take the superalgebra to be a graded extension of the de Sitter algebra $\mf{so}(1,4)$ of the associated de Sitter group $SO(1,4)$. The supercharges $Q^i_\alpha$ are taken to be complex operators transforming in the fundamental spinorial representation of the $Spin(1,4) \cong {Sp}(1,1)$ double cover  of $SO(1,4)$.  They are labeled by an extended supersymmetry index $i=1,\ldots,\mathcal{N}$, and a spinor index $\alpha=1,\ldots,4$. The basic obstruction to the realization of the superalgebra is that the Lie algebra generators of $SO(1,4)$ are built from the square of the supercharges  $Q^i_\alpha$. The latter can be arranged to be a bounded quantity. On the other hand, at least when realised across the whole de Sitter manifold, none of the de Sitter generators are bounded. Relatedly, as shown in \cite{Lukierski:1984it}, the de Sitter superalgebra obeys the relation $\sum_{\alpha,i} \{ Q_\alpha^i, Q^{i,\dag}_\alpha \}  = 0$, which cannot be realised on a Hilbert space carrying non-trivial representations of $SO(1,4)$ equipped with a  positive definite inner product.\vspace{0.2cm}\newline
One might imagine realising the supersymmetry only in a portion of de Sitter space, such as the static patch, given that the $SO(1,1) \subset SO(1,4)$ generator of static patch time translations obeys positivity properties. However, in doing so one observes that the de Sitter invariant state appears thermal \cite{Figari:1975km,Gibbons:1977mu} to local operators. The thermal property of the de Sitter invariant state clashes with supersymmetry. Performing a Wick rotation  to Euclidean signature and periodically identifying the Euclidean static patch time, one is led to place the theory on a Euclidean four-sphere. Here, smoothness of field configurations requires that the fermions are necessarily anti-periodic around a circle generated by $SO(2) \subset SO(5)$, a boundary condition that in the absence of further structure, such as $R$-symmetry, breaks supersymmetry. Relatedly, at least for $\mathcal{N}=1$ supersymmetry, there are no Majorana Killing spinors on the round $S^4$. Trying to build, instead, a zero temperature vacuum state within the static patch leads to singular behaviour at the de Sitter horizon, much like the Fulling--Rindler vacuum is singular at the Rindler horizon, and in any case, such a state would break the de Sitter isometries which are part of the superalgebra.
\vspace{0.2cm}\newline
Having said all this, there is at least one class of unitary quantum field theories that evade these obstructions with any amount of supersymmetry. These are superconformal quantum field theories \cite{Hristov:2013spa, Anous:2014lia}. Here, one extends the kinematic group to $SO(2,4) \supset SO(1,4)$. There are now additional conformal generators which are indeed  bounded. Relatedly, there is no obstruction to placing a superconformal field theory on a Euclidean sphere, which is conformally flat, due to the existence of conformal Killing spinors, which can be Majorana \cite{Anous:2014lia}.

\subsection{\texorpdfstring{dS$_4$}{dS4} supergravity}

As mentioned, a general obstruction \cite{Lukierski:1984it} to placing a supersymmetric quantum field theory on de Sitter is the absence of a bounded de Sitter generator. One might imagine that gauging the supersymmetry might aid with this, since all physical states would have to be annihilated by the supercharges, thus relaxing the boundedness obstruction. This is the supergravity analog of the Hamiltonian constraint in general relativity that enforces physical states to be annihilated by the generator of time translations. Concretely, this is the problem of building a supergravity theory out of the gauged de Sitter superalgebra. One would moreover require the theory to admit stable de Sitter solutions, potentially equipped with a positive cosmological constant. 
\vspace{0.2cm}\newline
 Supergravity Lagrangians with a negative cosmological constant were constructed in \cite{Freedman:1976aw,Townsend:1977qa}. An early attempt to construct a four-dimensional supergravity  with a positive cosmological term along with $\mathcal{N}=1$ or $\mathcal{N}=2$ supersymmetry was addressed in \cite{Pilch:1984aw}. It was noted that for $\mathcal{N}=1$ this would require the appearance of a gravitino with imaginary `mass',\footnote{Somewhat remarkably, recent considerations of unitary irreducible representations of the four-dimensional de Sitter group $SO(1,4)$ \cite{Deser:2001xr, Letsios:2023qzq} suggest that a field theoretic realisation of the spin-3/2 discrete series unitary representation invokes a de Sitter Rarita--Schwinger field with imaginary mass.} while for $\mathcal{N}=2$ one would have to confront a wrong sign kinetic term for the graviphoton. Given that the graviphoton and gravitino are physical propagating local degrees of freedom, such unusual reality properties make it hard to make sense of four-dimensional supergravity theories. One might, as for quantum field theory, attempt to build a four-dimensional conformal supergravity theory to overcome some of these obstructions. However, such theories are of a higher-derivative type and as such, present additional challenges due to ghost-like excitations. Moreover, there would be no clear gauge-invariant sense in which one can distinguish a de Sitter solution from a conformally equivalent one in a conformal supergravity setting.
\vspace{0.2cm}\newline
Despite  the above issues, it is worth noting that perturbatively stable de Sitter vacua have appeared in a variety of contexts in supergravity. It is not our aim to review these models. Generally, the de Sitter solution does not obey the BPS condition. In a particularly  simple instance \cite{Bergshoeff:2015tra}, which has $\mathcal{N}=1$ supersymmetry in four-dimensions,  one requires the presence of a Volkov--Akulov nilpotent supermultiplet leading to a model with non-linearly realised supersymmetry. In other models such as \cite{Fre:2003hsb} with extended supersymmetry, one has to also grapple with gauged supergravity models endowed with a non-compact and non-Abelian gauging. The embedding of these models in superstring theory, which is different to the question of the non-perturbative existence of de Sitter vacua in superstring theory explored, for example in \cite{Kachru:2003aw,DeLuca:2021pej}, remains an important open question. The main obstruction here is the no-go theorems of Gibbons \cite{Gibbons:2003gb} and Maldacena--N\'u\~nez \cite{Maldacena:2000mw} forbidding four-dimensional de Sitter vacua   from smooth, classical compactifications of eleven-dimensional supergravity. 

\subsection{\texorpdfstring{dS$_2$}{dS2} supergravity}

The obstructions that emerge in building a four-dimensional supergravity with positive cosmological constant and linearly realised supersymmetry, as well as their embedding in a more ultraviolet complete setting, are alleviated in two spacetime dimensions. Group theoretically, given that the isometry algebras of Lorentzian dS$_2$ and AdS$_2$ are both isomorphic to the $\mathfrak{so}(1,2)$ algebra, the generators of the algebra are liberated from the boundedness constraint. It was already noted in \cite{Lukierski:1984it} that a supersymmetric extension of the $\mathfrak{so}(1,2)$ is not necessarily accompanied by the appearance of negative norm states, at least on group theoretic grounds. Although this might suggest that one can build a supersymmetric quantum field theory in two dimensions, to the best of our knowledge, unitary examples remain unknown.\footnote{In part, this is because the operator that is required to be positive is the generator of the $SO(2)$ subgroup of $SO(1,2)$ which appears as a spatial momentum across the global circle of dS$_2$ (as opposed to time translations in global AdS$_2$). Consequently, a dS$_2$ state space that has boundedness properties for the $SO(2)$, would  have to permit some type of chirality properties. (See also related comments after \eqref{Zfluc}.) Moreover, the Euclidean groups of dS$_2$ and AdS$_2$, namely $SO(3)$ and $SO(1,2)$ respectively, differ. As is the case for the round $S^4$, the round $S^2$ does not admit Majorana Killing spinors.} On the other hand, as for the four-dimensional case, superconformal field theories can certainly be placed on dS$_2$ and its Euclidean counterpart.
\vspace{0.2cm}\newline
At the level of general relativity, the topological nature of two-dimensional gravity implies that the metric field is not a locally propagating degree of freedom. As noted in section \ref{N1sugra}, the same is true of the entire gravity multiplet in two-dimensional supergravity. Indeed, the supergravity kinetic term \eqref{eq:SuperEulerNumber} is just a topological quantity, and the cosmological constant obtained after the integration of the auxiliary field in \eqref{N1Liouville} appears due to the presence of the integral of the Berezinian \eqref{eq:IntegralBerezinian}, which is supersymmetric on its own. There is no conflict between a positive cosmological constant and the linear realisation of two-dimensional supergravity transformations. In order to preserve supersymmetry upon coupling matter fields to the supergravity theory, one can further select a two-dimensional {superconformal field theory}. Notice that the resulting supergravity theory is not itself a conformal supergravity. The presence of  \eqref{eq:IntegralBerezinian} breaks the local super-Weyl transformations. 
\vspace{0.2cm}\newline
We thus see how the somewhat ``unworldly'' \cite{Schwinger:1962tp} nature of two dimensions may  permit the presence of a positive cosmological constant in supergravity, with linearly realised supergravity transformations. What is left to assess is whether the theory permits a dS$_2$ solution. To this end, as was observed in (\ref{eq:N1_ClassicalSolution}),  when the superconformal field theory has large and positive central charge, $c_m \gg 1$, this is indeed possible. Interestingly, as noted below (\ref{GaussianL}), the  appearance of a  dS$_2$ solution is accompanied by a gravitino fluctuation of imaginary `mass' in Lorentzian signature. This resonates  with the behaviour of the gravitino in four-dimensional $\mathcal{N}=1$ de Sitter supergravity \cite{Pilch:1984aw}. The essential difference is that the two-dimensional gravitino is a topological degree of freedom, rendering the unusual reality condition less severe. In fact, the non-unitary nature of the two-dimensional gravitino is of the same type as that of the Weyl factor (already appearing in the non-supersymmetric case) whose wrong sign kinetic term necessitates a complexification of the field space. Moreover, as we shall see in what follows for the $\mathcal{N}=2$ case, the kinetic term of the graviphoton is absent alltoghether \cite{Brink:1976vg}, alleviating the wrong sign issue of \cite{Pilch:1984aw}. It seems the fate of unitarity of dS$_2$ supergravity is in the same hands as that of non-supersymmetric gravity with $\Lambda>0$. 
\vspace{0.2cm}\newline
In this way, we are equipped with a dS$_2$ supergravity theory coupled to superconformal matter fields. At the quantum level, such theories are better behaved in the ultraviolet than their four-dimensional counterparts and one has the good fortune of a concrete proposal \cite{Distler:1989nt} for the quantum measure in the super-Weyl gauge. None of this implies that the theories exist non-perturbatively, but it provides us with a reasonable starting point.

\section{\texorpdfstring{$\mathcal{N}=2$ dS$_2$}{N2 dS2} supergravity}
\label{sec:N2_SUGRA}

Having discussed $\mathcal{N}=1$ dS$_2$ supergravity, we now proceed to consider $\mathcal{N}=2$ dS$_2$ supergravity. The model provides us with yet another example of a dS$_2$ supergravity, now endowed with additional gauged bosonic internal symmetries. Furthermore, the super-Weyl gauge-fixed form is captured by an $\mathcal{N}=2$ super-Liouville theory, preserving eight real Poincaré supercharges, which is potentially amenable to analysis with additional tools  such as supersymmetric localization methods. Our discussion follows steps that parallel those in section \ref{subsec:N1_SUGRA}. Consequently, our treatment will be brief.

\subsection{\texorpdfstring{$\mathcal{N}=2$}{N2} supergravity multiplet \& action}

We begin by recalling that there are two minimal versions of $\cN=2$ supergravity in two dimensions, depending on whether one gauges the vectorial, $U(1)_V$, or axial, $U(1)_A$, $R$-symmetry \cite{Howe:1987ba, Gates:1988tn}. We shall focus on the $U(1)_A$ preserving supergravity -- the one obtained performing dimensional reduction of four-dimensional $\mc{N}=1$ supergravity. 
\vspace{0.2cm}\newline
In Lorentzian signature, the two-dimensional off-shell ${\mathcal{N}=(2,2)}$ gravity multiplet is composed of the zweibein $\rme^a_\mu$, a Dirac gravitino $\chi_\mu$, a $U(1)$ gauge field $A_\mu$ and a complex scalar field $B$. The larger supermultiplet is accompanied by two additional real-valued Grassmann coordinates in superspace, grouped in a Majorana spinor $\tilde{\theta}$. 
The supersymmetry transformations, along with their ensuing constraints, can be used to locally remove all degrees of freedom, rendering the pure $\mathcal{N}=2$ supergravity  topological. In Euclidean, as in Lorentzian signature, the canonical gravitational action is the $\mathcal{N}=2$  supersymmetric extension of the Einstein--Hilbert term. As in the $\mc{N}=1$ case, discussed around \eqref{eq:SuperEulerNumber}, this is just a topological invariant of the bosonic background. The difference with the $\cN=1$ case, though, is that for $\cN=2$ there is also a contribution from the gauge bundle since the curvature superfield includes the $U(1)$ curvature paired to the Ricci scalar \cite{Antoniadis:1990mx} 
\begin{equation}
\label{eq:SuperEulerNumber2}
	\frac{1}{\pi}\int_{\Sigma_h} \rd^2 x \, \rme \left(  {\vartheta} \frac{R}{4} + \ii \frac{\upsilon}{4} \epsilon^{\mu\nu} F_{\mu\nu} \right) =  \vartheta  \chi_h  + \ii  \upsilon n_h ~,
\end{equation}
where $n_h \in \Z$ is the first Chern number of the $U(1)$ bundle, and the associated coupling $\upsilon\in (0,2\pi]$ is taken to be real and compact to ensure boundedness of the gravitational path integral. 
\vspace{0.2cm}\newline
In addition to the topological contribution, we can still insert a purely gravitational term considering the integral of the superdeterminant $\text{E}$ of the superspace zweibein. 
On the other hand, a kinetic term for the $U(1)$ gauge field $A_\mu$ is interestingly forbidden by $\cN=2$ supersymmetry \cite{Brink:1976vg}.
\vspace{0.2cm}\newline
\textbf{$\mathcal{N}=2$ matter content.} In order to couple matter fields to our $\mathcal{N}=2$ supergravity theory, while preserving the same amount of supersymmetry, we must consider an $\mathcal{N}=2$ supersymmetric quantum field theory in the presence of  a background $\mathcal{N}=2$ gravity multiplet. For the sake of concreteness and computational control, we will further take the matter theory to be an  $\cN=2$ superconformal field theory (SCFT) with central charge $c_m$, whose coupling to the $\mathcal{N}=2$ gravity multiplet was first considered in \cite{Brink:1976vg} for a free theory. Apart from the value of $c_m$, the details of the $\mathcal{N}=2$ matter theory will not play an important role in what follows, as we will be focusing mostly on an $S^2$ topology, and vanishing $U(1)$ instanton number, and study properties of the undecorated sphere path integral. We note that although aspects of superconformal minimal models with $\mathcal{N}=2$ supersymmetry have been constructed in the literature \cite{Zamolodchikov:1986gh,Boucher:1986bh,Mussardo:1988av}, there is no  matrix model completion of $\mathcal{N}=2$ supergravity coupled to an $\mathcal{N}=2$ minimal model that we are aware of.\footnote{See however \cite{Turiaci:2023jfa} for recent developments in the context $\mathcal{N}=2$ JT supergravity. Here an important role is played by first fixing the charge under $U(1)_R$.}
\newline\newline
All in all, the  undecorated gravitational path integral over the $\mathcal{N}=2$ supergravity action \cite{Fradkin:1981dd}, reads
\begin{equation}\label{N2sugraPI}
\mathcal{Z}_{\text{grav}}^{\mathcal{N}=2} = \sum_{h=0}^\infty \sum_{n_h \in \mathbb{Z}} e^{\vartheta \chi_h + \ii \upsilon n_h} \int [\mathcal{D}\text{e}_\mu^a][\mathcal{D}\chi_\mu] [\mathcal{D}A_\mu][\mathcal{D}B] \, e^{- \ii \boldsymbol{\mu} \int_{\Sigma_h} \rd^2x\,  \rd^2\theta \rd^2\tilde{\theta} \left( {\rm E}  + \text{h.c.} \right)}\times Z_{\text{SCFT}}^{(h,n_h)}[\rme_\mu^a,\chi_\mu, A_\mu,B]~.
\end{equation}
The topological sum is over the Euler character $\chi_h$ \cite{Antoniadis:1990mx} as well as the first Chern number of the $U(1)$ bundle. In what follows, for the sake of simplicity, we will restrict ourselves to the trivial topological sector with $h=0$ and $n_{h=0}=0$.

\subsection{Higher genus topologies}

It is natural to also consider surfaces with higher genus, so we summarize some considerations on this.
\vspace{0.2cm}\newline
Every closed two-dimensional manifold is locally conformally flat, and the same is generalized to the $\cN=1$ and $\cN=2$ superspaces \cite{Howe:1978ia, Howe:1987ba}. Therefore, one can locally impose the Weyl gauge on a surface of arbitrary genus and study the superconformal field theory there. As mentioned, though, to compute the partition function we introduce a regulator, breaking conformal invariance, and the question then becomes whether it is possible to preserve supersymmetry on any Riemann surface, in particular whether it is possible to preserve the massive supersymmetry $\mf{su}(2|1)$ algebra preserving the vector-like $R$-symmetry. This is achieved by coupling the $\mc{R}$ multiplet of the field theory to the off-shell linearized supergravity, and finding a supersymmetric classical background \cite{Festuccia:2011ws}. A special case of this leads to global supercharges that are constant sections of the trivial bundle obtained via the topological twist \cite{Witten:1988xj}. This possibility exists on every Riemann surface. Another degenerate case is the flat torus.
\vspace{0.2cm}\newline
It turns out that a Riemann surface of genus higher than one does not admit a global supercharge defined in a way different from the topological twist, and on the sphere there are only three possible supersymmetric backgrounds \cite{Closset:2014pda}. One background, which is the one considered here and in  \cite{Benini:2012ui, Doroud:2012xw}, is different from the topological twist, and indeed the $R$-symmetry background gauge field has vanishing flux (see also the discussion below \eqref{N2sugraPI}). In the other two backgrounds, the $R$-symmetry gauge field has flux $\pm 1$, to cancel the spin connection. Therefore, these are the topological twists and $U(1)$ deformations thereof, such as $\Omega$ backgrounds.
\vspace{0.2cm}\newline
In this sense, we have considered the only non-topological supersymmetry-preserving spherical background, and thus the assumptions in \eqref{N2sugraPI} lead to a somewhat distinguished configuration. On the other hand, we should remember that from the gravitational path integral point of view, there is no a priori reason to restrict to supersymmetry preserving configurations. 

\subsection{\texorpdfstring{$\mathcal{N}=2$}{N2} Super-Weyl gauge}
The analogue of the $\mathcal{N}=1$ super-Weyl gauge (\ref{eq:SuperWeylGauge}) for the field content of $\mathcal{N}=2$ supergravity is given by 
\begin{equation}\label{N2sW}
\rme_\mu^a = e^{\tfrac{1}{2}{b(\varphi+\tvarphi)}}\tilde{\rme}_\mu^a~, \quad \, A^\mu = - {\frac{\ii}{2}} \epsilon^{\mu\nu} \partial_\nu \left( \varphi-\tvarphi \right)~,  
\end{equation}
as well as
\begin{equation}\label{N2sW2}
\chi_\mu = e^{\tfrac{1}{2} b \varphi} \gamma_\mu \psi~,\quad \chi^*_\mu = e^{\tfrac{1}{2} b \tvarphi} \gamma_\mu \tpsi~, \quad  B= e^{-b \varphi} F~, \quad  B^* = e^{-b \tvarphi} \widetilde{F}~.
\end{equation}
Again $\tilde{\rme}$ is the zweibein for the round metric on the sphere of radius $r$, while $\varphi$, $F$ and $\psi$ are two complex scalars, and a spin one-half Dirac spinor respectively. The $U(1)$ gauge field is in the Lorenz gauge. The tilded counterparts are equally two complex scalars and a spin one-half Dirac spinor. Whether or not the tilded fields are complex conjugates of the untilded fields is related to our choice of path integration contour. For the time being, we take them to be independent of each other. 
\vspace{0.2cm}\newline
As for the non-supersymmetric case, and the case with $\mathcal{N}=1$ supersymmetry, at least on the two-sphere with vanishing first Chern number for the $U(1)$ bundle, the concise form of the $\mathcal{N}=2$ SCFT path integral is known. It is governed by the superconformal anomaly
\begin{equation}
Z_{\text{SCFT}}^{(0)}[\Phi,\widetilde{\Phi}] = \mathcal{C}\, e^{-S_{\text{s-anomaly}}[\Phi,\widetilde{\Phi}]}~,
\end{equation}
where $\Phi$ and $\widetilde{\Phi}$ combine the chiral and anti-chiral multiplet $(\varphi,\psi,F)$ and $(\tvarphi,\tpsi,\tF)$ respectively and $\mathcal{C}$ is a normalization constant independent of $c_m$.
\vspace{0.2cm}\newline
Incorporating the  contribution from the gravitational path integration measure  leads to the action of $\cN=2$ super-Liouville theory is
\begin{multline}
\label{eq:N22_SpacelikeLiouvilleSphere}
S_{\text{L}}^{\mathcal{N}=2} = \frac{1}{4\pi} \int_{S^2} \dd^2 x \, \tilde{\rme} \left(  \partial_\mu \tvarphi \partial^\mu \varphi - \ii \btpsi \slashed{D}\psi - F \tF + \frac{Q\widetilde{R}}{2} \left( \varphi + \tvarphi \right) \right. \\ \left.
    + \mu  e^{b\varphi} F - \frac{\ii}{2}\mu b e^{b\varphi} \overline{\psi}\psi +  \mu^*  e^{b\tvarphi}\tF - \frac{\ii}{2}  \mu^* b e^{b\tvarphi} \btpsi\tpsi \right)~,
\end{multline}
where $Q=1/b$. The above action describes the interaction of a chiral multiplet, composed of a complex scalar field $\varphi$, a Dirac spinor $\psi$ and an auxiliary complex scalar $F$, and a corresponding anti-chiral multiplet $(\tvarphi, \tpsi, \tF)$, with an exponential superpotential $W = \mu e^{b\varphi}/b$. {A priori}, in Euclidean signature there is no relation between the field content of the chiral and anti-chiral multiplets, whereas in Lorentzian signature they are complex conjugate (or charge conjugate in the case of the spinor). At least for the case of real-valued $Q$, viewed as a quantum field theory, the $\mathcal{N}=2$ super-Liouville theory describes a unitary two-dimensional superconformal field theory. The theory has a $U(1)_V\times U(1)_A$ internal global symmetry acting on the fields as follows  
\begin{equation}
\label{eq:N2_RSymmetry}
\begin{aligned}
&U(1)_V &\, &: &\, \varphi &\to \varphi + 2  \ii Q \rho~,  &\ \tvarphi &\to \tvarphi - 2  \ii Q \rho~, &\  \psi &\to e^{ -\ii\rho } \psi~, &\  \tpsi &\to e^{\ii\rho} \tpsi ~,  \\ 
& & & &  F &\to e^{ -2\ii\rho } F~, &\  \tF &\to e^{2\ii\rho} \tF ~, \\[5pt]
&U(1)_A &\, &: &\, \psi &\to e^{\ii \alpha \gamma_*} \psi~,  &\  \tpsi &\to e^{-\ii \alpha \gamma_*} \tpsi ~.
\end{aligned}
\end{equation}
The corresponding central charge of the theory is $c_{\text{L}}^{\mathcal{N}=2}= 3+6Q^2$, as we will verify via a one-loop analysis in a later section. 
\vspace{0.2cm}\newline
Going back to the gravitational origin, we must cancel $c_{\text{L}}^{\mathcal{N}=2}$ against the matter central charge $c_m$ and the central charge of the $\beta\gamma$, $\mathfrak{b}\mathfrak{c}$, and $U(1)$ ghost theories stemming from the gauge fixing procedure. This leads to the condition
\begin{equation}\label{cN2}
c_{\text{L}}^{\mathcal{N}=2} + c_m +c_{\mathfrak{b}\mathfrak{c}} + 2\times  c_{\beta\gamma}+ c_{U(1)_V} = 3+6Q^2 + c_m -26 + 2\times 11 - 2 =0~.
\end{equation}
In particular we obtain the relation
\begin{equation}\label{QbN2}
Q= \frac{1}{b}=\sqrt{\frac{3-c_m}{6}}~.
\end{equation}
Similarly, given that all symmetries are gauged in the underlying gravitational theory, the axial anomaly from $U(1)_A$ must also be cancelled against the matter and ghost axial anomalies \cite{Distler:1989nt}. Interestingly, $Q$ exhibits a simpler structure as compared to its $\mathcal{N}=1$ counterpart \eqref{charge_conservation}. There is no intermediate region between spacelike and timelike $\mathcal{N}=2$ Liouville theory. For $c_m < 3$ $Q$ and $b$ are real-valued and we are in the regime of spacelike $\mathcal{N}=2$ Liouville theory. Once $c_m>3$ both $Q$ and $b$ become imaginary. In what follows, our main focus will be on timelike $\mathcal{N}=2$ super-Liouville theory whose central charge $c_{\text{tL}}^{\mathcal{N}=2} \leq 3$ approaches negative infinity in the semiclassical $\beta \equiv \ii b \rightarrow 0^+$ limit.
\begin{figure}[H]
\begin{center}
\begin{tikzpicture}
\draw[line width=0.3mm] (-5,0) -- (6,0);
\draw[line width=0.5mm, red,->] (0,0) -- (8,0);
\draw[line width=0.5mm,blue,<-] (-8,0) -- (0,0);
\draw[line width=0.5mm,purple] (0,.2) -- (0,-.2);
\node[scale=1,purple] at (0,.5)   {$c_m=3$ ~ };
\node[scale=1,blue] at (-5.5,-.5)   {spacelike $\mathcal{N}=2$ Liouville ~ };
\node[scale=1,red] at (5.5,-.5)   {timelike $\mathcal{N}=2$ Liouville ~ };
\end{tikzpicture}
\end{center}
\caption{The line shows the central charge of the superconformal field theory coupled to $\mathcal{N}=2$ supergravity. For $c_m < 3$ the Liouville parameters $b$ and $Q$ are real-valued and one has  the spacelike Liouville regime. For $c_m> 3$ both $b$ and $Q$ become imaginary. In this regime we can continue the parameters and fields to obtain timelike $\mathcal{N}=2$ Liouville theory. We stress that since for $\mathcal{N}=2$ Liouville theory the parameter $Q$ in Liouville theory does not get a quantum correction, but $Q=1/b$, there is no intermediate regime between spacelike and timelike Liouville theory.}
\label{fig:centralchargesN2}
\end{figure}
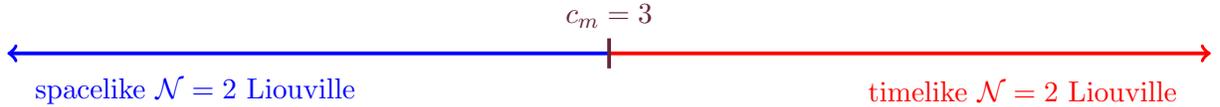
\vspace{0.2cm}
\noindent
\textbf{$U(1)$ gauge field transformations.} Upon inspecting (\ref{eq:N22_SpacelikeLiouvilleSphere}) along with (\ref{N2sW}) we note that the interaction of the gauge-fixed form of the gauge field $A_\mu$ with the gauge-fixed gravitino  is non-local once expressed in terms of the   fields prior to gauge fixing. Moreover, the kinetic term of $\psi$ is independent of $A_\mu$ in its gauge fixed form. One can restore the local form of the interaction by performing the following field dependent transformation on the Dirac fermions $\psi$, $\tpsi$ and auxiliary scalars $F$, $\widetilde{F}$ 
\begin{equation}\label{fujikawa}
\psi \to e^{ -\tfrac{\ii}{2} b \varphi_2 } \psi~, \quad  \tpsi \to e^{ \tfrac{\ii}{2} b \varphi_2 } \tpsi~, \quad F \to e^{- \ii b \varphi_2 }  F~, \quad  \tF \to e^{ \ii b \varphi_2 }  \tF~,
\end{equation}
where $\varphi_2  \equiv - \ii (\varphi - \tvarphi)/2$. The above transformation is of the  Fujikawa type which can be used  to realise the chiral anomaly from a path integral perspective \cite{Fujikawa:1979ay} . Here the Fujikawa method is applied to a two-dimensional setup, as was done for the Schwinger model in \cite{Roskies:1980jh}. Upon performing the transformation (\ref{fujikawa}), the dependence on $\varphi_2$ is removed from the superpotential part of the Lagrangian and instead renders the  fermionic kinetic term gauge covariant for a $U(1)$ axial gauge symmetry. The kinetic term for $\varphi_2$ itself, which is not gauge-invariant, encodes the $U(1)$ gauge anomaly. As already mentioned, the gauge anomaly must cancel against that of the ghost and matter SCFT to render the underlying $\mathcal{N}=2$ consistent at the quantum level. Technically, the  advantage in analysing the theory in the form (\ref{eq:N22_SpacelikeLiouvilleSphere}) is that $\varphi_2$ appears in a structurally simple way, essentially complexifying the Weyl factor.

\subsection{Residual gauge supersymmetries}

As is the case for the $\mathcal{N}=1$ theory,  the super-Weyl gauge choice in (\ref{N2sW}) and (\ref{N2sW2}) does not fully fix the redundancies of the underlying $\mathcal{N}=2$ supergravity. The remaining redundancy is given by the global subgroup of two-dimensional $\mathcal{N}=2$ superconformal transformations, namely $OSp(2|2,\C)$ $\mathcal{N}=2$ super-Moebius transformations. The bosonic subgroup, up to discrete factors, is $PSL(2,\C) \times U(1)_V \times U(1)_A$, which is generated by $J_m$ and $K_m$ in \eqref{eq:osp12C_v2}, as well as the two $R$-symmetry generators $\mc{V}$ and $\mc{A}$ that act on the supercharges in vector-like and axial-like way. As compared to the $\mf{osp}(1|2,\C)$ Lie super-algebra, the fermionic generators are doubled -- in addition to $Q^\alpha$ and $S^\alpha$, one also has $\widetilde{Q}^\alpha$ and $\widetilde{S}^\alpha$. The full set of commutation relations can be found in \cite{Doroud:2012xw}.\footnote{It may be more familiar to the reader to write the $\mf{osp}(2|2,\C)$ algebra in terms of the generators adapted to flat space, namely the bosonic set $\{ J_0, L_0, L_\pm; \widetilde{J}_0, \widetilde{L}_0, \widetilde{L}_\pm\}$ and the fermionic $\{ G^\alpha_s; \widetilde{G}^\alpha_s \}$ with $s=\pm$. The map between the two bases can be found in \cite{Hori:2013ika, Doroud:2013pka}.} 
\vspace{0.2cm}\newline
{\textbf{$\mathcal{N}=2$ super-Moebius group.}} For the sake of completeness, we can also express $OSp(2|2,\C)$ in the superspace picture. The {$OSp(2|2,\mathbb{C})$ supergroup}  extends the $OSp(1|2,\mathbb{C})$  group of $\mathcal{N}=1$ super-Moebius transformations (\ref{SM}). It can be defined  \cite{Shu_2007} by the set of  supermatrices
\begin{equation}\label{SM2}
M = 
\begin{pmatrix}
a & b  & \alpha & \beta  \\
c & d   & \gamma  & \delta  \\ 
\zeta & \eta    & e & f  \\
\rho & \sigma    & g & h
\end{pmatrix} \equiv \begin{pmatrix}
A   & B \\ 
C      & D
\end{pmatrix} ~,
\end{equation}
where the $a,b,c,d, e,f,g,h \in \mathbb{C}$,  and  $\alpha$, $\beta$, $\gamma$, $\delta$, $\zeta$, $\eta$, $\rho$, $\sigma$ are complex-valued Grassmann numbers, subject to the condition
\begin{equation}
\begin{pmatrix}
A^T   & C^T \\ 
-B^T      & D^T 
\end{pmatrix} \cdot
\begin{pmatrix}
\Omega_2   & 0 \\ 
0     & \mathbb{S}_2 
\end{pmatrix} \cdot 
\begin{pmatrix}
A   & B \\ 
C     & D 
\end{pmatrix}  = 
\begin{pmatrix}
\Omega_2   & 0 \\ 
0     & \mathbb{S}_2 
\end{pmatrix}~,
\end{equation}
with $\Omega_2$ the $2\times 2$ symplectic matrix and $\mathbb{S}_2$ the $2\times 2$ symmetric matrix with vanishing diagonals and unit off-diagonals. The group elements of the $SO(2) \cong U(1)$ bosonic subgroup are given by the $D$ matrices. The group elements of the $SL(2,\mathbb{C})$ subgroup are given by the $A$ matrices. The Berezinian is again $\text{ber}(M)=1$.
\vspace{0.2cm}\newline
At the level of superspace coordinates one must add an additional  Grassmann coordinate $\tilde{\theta}$ to the $\mathcal{N}=1$ superspace coordinate introduced above (\ref{st}). This reflects the additional supercharges in the $\mathcal{N}=2$ theory. The resulting superspace coordinate is now given by $Z = (z|\theta,\tilde{\theta})$, along with its anti-holomorphic counterpart. The Grassmann coordinates carry charge under the $SO(2)$ subgroup, reflecting the fact that the fermions $\psi$ and $\tpsi$, as well as the auxiliary scalars $F$ and $\widetilde{F}$, are themselves charged under the $SO(2)$. The corresponding chiral and anti-chiral superfields are given by $\Phi = \varphi + \bar{\theta} \psi + \tfrac{1}{2} \bar{\theta}\theta F$ and $\widetilde{\Phi} = \widetilde{\varphi} + \tilde{\theta}\widetilde{\psi} + \tfrac{1}{2} \bar{\tilde{\theta}}\tilde{\theta} \widetilde{F}$, and transform by a superspace shift of the type (\ref{SMPhi}) under the $\mathcal{N}=2$ super-Moebius transformations.\footnote{The explicit expression for the $\mathcal{N}=2$ super-Moebius transformation of the superspace coordinate, along with the corresponding transformation of the superfields, can be found in eqns. (24) and (25) of \cite{Gieres:1997ia}.}

\subsection{Precise \texorpdfstring{$\mathcal{N}=2$}{N2} supergravity path integral}

Once the dust settles, and upon path integration of the ghost and matter fields, the genus zero contribution to the $\mathcal{N}=2$ supergravity path integral (\ref{N2sugraPI}) with vanishing first Chern number for the $U(1)$ bundle, takes the following form
\begin{equation}
\label{eq:Z_grav_N2_0}
	\mc{Z}^{\cN=2}_{{\rm grav}, (0)} = e^{2\vartheta} \times {\mc{A}} \times \left( \frac{r}{\ell_{\text{uv}}}\right)^{(c_m+c_{\mf{bc}} + 2c_{\beta\gamma}+ c_{U(1)})/3} \times \int \frac{[\mc{D}\Phi][\mc{D}\widetilde{\Phi}]}{{\rm vol}_{OSp(2|2;\C)}} e^{- \mc{S}_{\text{L}}^{\cN=2}} ~,
\end{equation}
where $\mc{A}$ is a $\mu$ independent prefactor stemming from the partition functions of the SCFT and the ghost systems on the round sphere with unit radius and any additional normalization choices from the path integral measure. We can absorb $\mc{A}$ into the definition of the physical coupling $\vartheta$. The length scale $\ell_{\text{uv}}$ is a reference scale, such as the ultraviolet cutoff scale, necessary to render the expression dimensionless.  In light of (\ref{cN2}) we note that $\mc{Z}^{\cN=2}_{{\rm grav}, (0)}$ is independent of $r$, as well as $\ell_{\text{uv}}$, as it should be. Given that the leading ultraviolet divergences can be arranged to cancel in supersymmetric path integrals due to the boson/fermion degeneracy, the supergravity theory at hand stands a chance to be ultraviolet finite altogether, at least when placed on an $S^2$ topology.
\vspace{0.2cm}\newline
At this point, we are invited to direct our attention to the super-Liouville path integral
\begin{equation}
\label{eq:Z_L_N2}
	\mathcal{Z}_{\text{L}}^{\mathcal{N}=2}[\mu] \equiv \int \frac{[\mc{D}\varphi][\mc{D}\psi] [\mc{D}F]}{{\rm vol}_{OSp(2|2;\C)}} e^{- \mc{S}_{\text{L}}^{\cN=2}} ~,
\end{equation}
and related quantities, always bearing in mind the relationship to the underlying $\mathcal{N}=2$ supergravity theory and ensuring that the volume ${\rm vol}_{OSp(2|2;\C)}$ in the measure is properly accounted for.

\section{Timelike \texorpdfstring{$\mathcal{N}=2$}{N2} super-Liouville theory}\label{tSLN2}

Motivated by our considerations of  $\mathcal{N}=2$ supergravity in  two dimensions coupled to $\mathcal{N}=2$ superconformal matter, in this section we consider $\cN=2$ super-Liouville theory. For the sake of concreteness, and especially due to the physical motivation of having real-valued dS$_2$ solutions, we will take our SCFT to have large and positive central charge $c_m$. As we will explore in detail, the limit of large $c_m$ governs a semi-classical limit governing small geometric fluctuations around a classical saddle point solution.

\subsection{Action and symmetries}
\label{subsec:N2Action}

It follows from (\ref{QbN2}) that in the regime $c_m >3$ the Liouville parameter $Q = 1/b$ in (\ref{eq:N22_SpacelikeLiouvilleSphere}) becomes pure imaginary (\ref{QbN2}). To restore the reality of the purely bosonic part of the action (upon integrating out $F$ and $\widetilde{F}$) we  can analytically continue the chiral and anti-chiral superfields as $\Phi \to \ii \Phi$ and $\widetilde{\Phi} \to \ii \widetilde{\Phi}$. This results in the following action
\begin{align}
\mc{S}^{\cN=2}_{\text{tL}} = \frac{1}{4\pi} \int_{S^2} \rd^2x \, \tilde{\rme}\, \bigg( & - \partial_\mu \tvarphi \partial^\mu \varphi + \ii \btpsi \slashed{D}\psi + F \tF - \frac{q\widetilde{R}}{2} \left( \varphi + \tvarphi \right) \nonumber \\
\label{eq:N22_TimelikeLiouvilleSphere}
& + \ii \mu  e^{\beta\varphi} F  + \frac{1}{2}\mu \beta e^{\beta\varphi} \overline{\psi}\psi + \ii \mu^*  e^{\beta \tvarphi} \tF + \frac{1}{2}\mu^* \beta e^{\beta\tvarphi} \btpsi\tpsi \bigg) ~,
\end{align}
where we have further defined $\beta \equiv \ii b$. We refer to (\ref{eq:N22_TimelikeLiouvilleSphere}) as   $\mathcal{N}=2$ timelike super-Liouville theory. In terms of a superpotential $W(\varphi) \equiv \mu e^{\beta\varphi} /\beta$, and its anti-chiral counterpart $\widetilde{W}(\tvarphi) \equiv \mu^* e^{\beta\tvarphi} /\beta$, the action reads
\begin{align}
\mc{S}^{\cN=2}_{\text{tL}} = \frac{1}{4\pi} \int_{S^2} \rd^2x \, \tilde{\rme}\, \bigg[ & - \partial_\mu \tvarphi \partial^\mu \varphi + \ii \btpsi \slashed{D}\psi + F \tF - \frac{q\widetilde{R}}{2} \left( \varphi + \tvarphi \right) \cr  
&+\ii\left(F W'(\varphi) - \frac{\ii}{2}W''(\varphi) \bpsi \psi\right)+ \ii\left(\widetilde{F} \widetilde{W}'(\tvarphi) - \frac{\ii}{2}\widetilde{W}''(\tvarphi) \btpsi \tpsi\,\right)  \bigg]~.
\end{align}
When viewed as a quantum field theory, it is natural to postulate that timelike $\mathcal{N}=2$ super-Liouville theory is a non-unitary superconformal field theory. The corresponding central charge is given by
\begin{equation}\label{eq:N2 timelike central charge}
c_{\text{tL}}^{\mathcal{N}=2} = 3-\frac{6}{\beta^2}~, 
\end{equation}
as we will verify via a one-loop analysis. In contrast to its $\mathcal{N}=1$ counterpart (\ref{eq:N1_TimelikeLiouville}), the relation $q = 1/\beta$ is exact. If we restrict to $\tvarphi=\varphi$, $\tpsi = \psi$, $\tF=F$, the chiral multiplet reduces to the $\cN=1$ multiplet and the action \eqref{eq:N22_TimelikeLiouvilleSphere} to twice \eqref{eq:N1_TimelikeLiouville}, at least formally since the relation between $q$ and $\beta$ is different for the two theories.
\vspace{0.2cm}\newline
The theory (\ref{eq:N22_TimelikeLiouvilleSphere}) is invariant under the supersymmetry transformations
\begin{equation}
\label{eq:SuperConformalTransformations_Timelike}
\begin{aligned}
\delta \varphi &= \btepsilon \psi~, & \quad \delta \tvarphi &=  \overline{\epsilon} \tpsi~, \\
\delta \psi &=  \ii \slashed{\partial}\varphi \, \epsilon + \ii q \slashed{D}\epsilon  -\ii F \, \tepsilon  ~, &\quad  \delta \tpsi &= \ii \slashed{\partial}\tvarphi \,  \tepsilon + \ii q \slashed{D}\tepsilon  - \ii \tF \, \epsilon~, \\
\delta F &= - \overline{\epsilon} \slashed{D}\psi ~, &\quad \delta \tF &= \btepsilon \slashed{D} \tpsi ~.
\end{aligned}
\end{equation}
We remark, again, that in Euclidean signature there is no {a priori} relation between the fields of the chiral and anti-chiral multiplet. Correspondingly, the Euclidean supersymmetry parameters $\epsilon$ and $\tepsilon$ are independent. For the purposes of our analysis, we will impose the following reality condition on the fields
\begin{equation}
\label{eq:ChiralRealContour}
	\tvarphi = \varphi^* ~, \qquad \tpsi = \gamma_* \psi^C ~, \qquad \tF = F^* ~.
\end{equation}
Focusing on the chiral multiplet for concreteness, we find the following superalgebra\footnote{As for the $\mc{N}=1$, there is a term proportional to $\overline{\epsilon}_{[1}D^2 \epsilon_{2]}$ in $[\delta_{\epsilon_1}, \delta_{\epsilon_2}]F$ which vanishes upon choosing the supersymmetry parameters to be Killing spinors on the sphere \cite{Benini:2012ui}.}
\begin{equation}
\begin{split}
[\delta_\epsilon, \delta_{\tilde{\epsilon}}] \varphi &= \cL_\xi \varphi + q \Delta + 2\ii q \rho ~, \\
[\delta_\epsilon, \delta_{\tilde{\epsilon}}] \psi &= \cL_\xi \psi + \frac{1}{2} \Delta \, \psi -\ii \rho \, \psi + \ii \alpha \, \gamma_* \psi ~, \\
[\delta_\epsilon, \delta_{\tilde{\epsilon}}] F &= \cL_\xi F + \Delta \, F - 2\ii \rho \, F ~,
\end{split}
\end{equation}
where we have defined the following parameters
\begin{equation}
\label{eq:SCA_LiouvilleChiral}
\begin{aligned}
\xi^\mu &= \ii \, \btepsilon \gamma^\mu \epsilon ~, &\quad \Delta &= \frac{1}{2} \nabla_\mu \xi^\mu ~, \\
\rho &= - \frac{1}{4}\left( D_\mu \btepsilon \gamma^\mu \epsilon - \btepsilon \gamma^\mu D_\mu \epsilon \right) ~, &\quad \alpha &= \frac{1}{4} \left( D_\mu \btepsilon \gamma_*\gamma^\mu \epsilon - \btepsilon \gamma_* \gamma^\mu D_\mu \epsilon \right) ~.
\end{aligned}
\end{equation}
Here $\xi^\mu$ is a parameter for translations, $\Delta$ for dilation, $\rho$ for vector-like $R$-symmetry, and $\alpha$ for the axial-like $U(1)$. 
\vspace{0.2cm}\newline
A few observations are in order, extending those made around the $\mc{N}=1$ commutators in \eqref{eq:N1SCA_Fields}. First, we remark that these anti-commutation relations correspond to the anti-commutation relations of the fermionic generators of $OSp(2|2,\C)$, where on the right-hand side we find $J_m$ and $K_m$ generating the isometries and conformal isometries of the sphere, respectively, and the generators $\mc{V}$ and $\mc{A}$ of the $R$-symmetries. Moreover, we notice that choosing $\epsilon$ and $\tepsilon$ to be both positive (or negative) Killing spinors means that $\xi$ generates an isometry of $S^2$, so $\Delta=0$, but also $\rho\neq 0$ and $\alpha = 0$. This choice (once extended to the bosonic generators) corresponds to a choice of $\mf{su}(2|1)_V$ subalgebra of the superconformal algebra, and it will be relevant later for the localization computation. In fact, there is another allowed choice of $\mf{su}(2|1)$ subalgebra, where the axial $R$-symmetry is preserved (see \cite{Gerchkovitz:2014gta} for a discussion). This structure mirrors the existence of two minimal supergravities as discussed at the beginning of section \ref{sec:N2_SUGRA}.
\vspace{0.2cm}\newline
The structure of \eqref{eq:SCA_LiouvilleChiral} is analogous to that of the chiral multiplet found in \cite{Benini:2012ui, Doroud:2012xw}, except for the fact that dilation and the vector $R$-symmetry do not act on $\varphi$ linearly, as already observed in \eqref{eq:N2_RSymmetry}. This is well-known to happen for the dilation action on the Liouville field, and we already observed it for the $\mc{N}=1$ case. Compared to the $\mc{N}=1$ case, though, here the fact that $q=\beta^{-1}$ implies that the finite scale transformation
\begin{equation}
\begin{aligned}
	\varphi &\to \varphi + q \Delta ~, &\quad \psi &\to e^{\frac{\Delta}{2}}\psi ~, &\quad F &\to e^{\Delta}F ~, \\
	\tvarphi &\to \tvarphi - q \Delta ~, &\quad \tpsi &\to e^{-\frac{\Delta}{2}} \tpsi ~, &\quad \tF &\to e^{-\Delta}\tF~,	
\end{aligned}
\end{equation}
is an actual symmetry of the action \eqref{eq:N22_TimelikeLiouvilleSphere} without the need of quantum corrections. Because of the peculiar representation of the scale transformation on the scalar field of the supermultiplet $(\varphi, \psi, F)$, the latter does not behave as a canonical chiral supermultiplet. If we need to highlight this, we shall refer to it as Liouville chiral multiplet.

\subsection{\texorpdfstring{dS$_2$}{dS2} saddle point \texorpdfstring{\&}{and} Gaussian fields}

In the limit $\beta \to 0^+$, the $\mathcal{N}=2$ super-Liouville theory (\ref{eq:N22_TimelikeLiouvilleSphere}) lends itself to a saddle point approximation. The corresponding saddle point equations of motion are given by
\begin{equation}
\label{SUSY_transformations_sL}
\begin{aligned}[c]
0&= \nabla^2\varphi - \frac{1}{\beta r^2} +\ii \mu \beta e^{\beta \varphi}F +\frac{1}{2}\mu \beta^2\overline{\psi} \psi~, \cr
0&= - \ii \slashed{D}\btpsi + \frac{1}{2}\mu\beta e^{\beta\varphi}\bpsi~,\cr
0&=\widetilde{F}+  \ii \mu e^{\beta\varphi}~,
\end{aligned}
\qquad
\begin{aligned}[c]
0&=\nabla^2\tvarphi - \frac{1}{\beta r^2} +\ii \mu^* \beta e^{\beta \tvarphi}\widetilde{F} +  \frac{1}{2}\mu^* \beta^2\btpsi \tpsi~,\cr
0&= \ii \slashed{D}\psi +\frac{1}{2}\mu^* \beta e^{\beta\tvarphi}\tpsi~\cr
0&=F+\ii \mu^*  e^{\beta \tvarphi}~.
\end{aligned}
\end{equation}
Taking $\psi_*=0$ and $\tpsi_*=0$, and placing $F$ and $\widetilde{F}$ on-shell, one finds the following saddle point solution
\begin{equation}\label{saddleN2varphi}
(\varphi + \tvarphi)_* = \frac{1}{\beta}\log \frac{1}{|\mu|^2 \beta^2 r^2}~, \quad\quad (\varphi - \tvarphi)_* = c~,
\end{equation}
where $c$ is an unfixed constant. Given the invariance of the saddle point equations under $OSp(2|2,\mathbb{C})$ transformations, there will be a $OSp(2|2,\mathbb{C})$ family of solutions, which are all gauge-equivalent from the perspective of the underlying $\mathcal{N}=2$ supergravity. As such, the unfixed constant $c$ can be set to zero. Going back to the underlying $\mathcal{N}=2$ supergravity picture, one can identify $(\varphi + \tvarphi)$ with the Weyl mode (\ref{N2sW}) of the physical metric such that the above saddle corresponds to the round two-sphere, or Euclidean dS$_2$, whose volume scales as  $\sim 1/|\mu|^2\beta^2$.
\vspace{0.2cm}\newline
Upon integrating out the auxiliary fields $F$, $\tF$ we obtain
\begin{multline}
	\mc{S}^{\cN=2}_{\text{tL}} = \frac{1}{4\pi} \int_{S^2} \rd^2 x \sqrt{\tilde{g}} \bigg(  - \partial_\mu \tvarphi\partial^\mu \varphi + \ii \btpsi \slashed{D}\psi + \lvert \mu \rvert^2 e^{\beta(\varphi + \tvarphi)}  \\ 
	- \frac{1}{\beta r^2} \left( \varphi + \tvarphi \right)  +  \frac{\beta}{2} \left( \mu e^{\beta\varphi} \overline{\psi}\psi + \mu^* e^{\beta\tvarphi} \btpsi\tpsi \right) \bigg) ~.
\end{multline}
Similar considerations to those made in section \ref{subsec:SUSYdS2} apply in this case. Again we notice the negative sign of the kinetic term for the scalars. Identifying the cosmological constant $\Lambda = |\mu|^2 /4\pi>0$,  this is a supersymmetric theory with positive cosmological constant. 
\vspace{0.2cm}\newline
Evaluating the on-shell action using (\ref{saddleN2varphi}) we can obtain the leading saddle point approximation to the $\mathcal{N}=2$ super-Liouville sphere path integral
\begin{equation}
\mathcal{Z}_{\text{tL}}^{\mathcal{N}=2}[\mu] \approx \mathcal{Z}_{\text{saddle}}^{\mathcal{N}=2} = \left(\frac{1}{e |\mu|^2 \beta^2 r^2}\right)^{\frac{1}{\beta^2}}~.
\end{equation}
Due to the simple relation $q=1/\beta$, the structure of $\mathcal{Z}_{\text{saddle}}^{\mathcal{N}=2}$ as compared to $\mathcal{Z}_{\text{saddle}}^{\mathcal{N}=1}$ in (\ref{ZtLs}) is significantly simpler. 
\vspace{0.2cm}\newline
\textbf{Gaussian theory.} We now consider the leading effect of fluctuations about the saddle point solution  (\ref{saddleN2varphi}). To study the fluctuations, it is convenient to introduce to following field variables 
\begin{equation}\label{eq:def deltaphi12}
\varphi_1 \equiv {\frac{1}{2}}(\varphi + \tvarphi) ~,\quad \varphi_2 \equiv - {\frac{\ii}{2}}(\varphi - \tvarphi)~,
\end{equation}
such that the fluctuating fields are given by
\begin{equation}
\varphi_1 = \varphi_{1,*}+ \delta\varphi_1~,\quad \varphi_2= \varphi_{2,*}+ \delta\varphi_2~,\quad \psi= \psi_*+ \delta \psi~,\quad \tpsi= \tpsi_*+ \delta \tpsi~.
\end{equation}
To quadratic order, one finds the following Gaussian action
\begin{multline}\label{S_gauss}
S_{\text{pert}}^{(2)}=\frac{1}{4\pi}\int_{S^2} \dd^2 x \, \tilde{\text{e}} \bigg[-\delta\varphi_1\left(-\nabla^2-\frac{2}{r^2}\right)\delta\varphi_1 -\delta\varphi_2(- \nabla^2)\delta\varphi_2  \\  + \ii \delta\btpsi \slashed{D}\delta\psi  + \frac{1}{2r}\frac{\mu}{|\mu|} \delta\bpsi\delta \psi + \frac{1}{2r}\frac{\mu^*}{|\mu|}\delta\btpsi \delta\tpsi \,  \bigg]~.
\end{multline}
As compared to the $\mathcal{N}=1$ theory (\ref{GaussianL}) governing the quadratic fluctuations, we will see  that the fluctuations exhibit exact zero modes rather than almost zero modes. This is due to the $\mathcal{N}=2$ relation $q=1/\beta$. 

\subsection{One-loop contribution}

As for the $\mathcal{N}=1$ theory, in order to compute the one-loop contribution we can decompose the various fields into spherical harmonics. For the bosonic fluctuations we use the decomposition (\ref{phiexp}), while for the fermions we use the decomposition (\ref{psiexp}) with slightly modified reality conditions. 
\vspace{0.2cm}\newline
{\textbf{Bosonic one-loop determinant}.} With the bosonic measure (\ref{fermionMeasure}) for both $\varphi_1$ and $\varphi_2$, which guarantees that
\begin{equation}
1= \int [\mathcal{D}\varphi_1][\mathcal{D}\varphi_2]\, e^{-\Lambda_{\text{uv}}\int_{S^2} \dd^2 x\,\tilde{\text{e}}\,\varphi(x)\tvarphi(x)}~,
\end{equation}
we obtain for the bosonic Gaussian part 
\begin{align}\label{N2bprod}
\hspace{-7mm}
\mathcal{Z}_{\text{tL,bos,1-loop}}^{\mathcal{N}=2}
=& \pm \ii \left({2\pi r^2\Lambda_{\mathrm{uv}}}\right)^{\frac{1}{2}}  \left({\frac{r^2\Lambda_{\mathrm{uv}}}{\pi}}\right)^{2} \times \prod_{l = 2}^\infty\left(\frac{4\pi r^2\Lambda_{\mathrm{uv}}}{l(l+1)-2}\right)^{l+\frac{1}{2}}\prod_{l = 1}^\infty \left(\frac{4\pi r^2\Lambda_{\mathrm{uv}}}{l(l+1)}\right)^{l+\frac{1}{2}}\,,
\end{align}
where the overall $\pm \ii$ accounts for the Wick rotation of the $l=0$ mode of $\delta\varphi_1$ analogously to the $\mathcal{N}=1$ case. From (\ref{S_gauss}) we infer that for $\mathcal{N}=2$ we have three zero modes associated to the $l=1$ mode of $\delta\varphi_1$ and one zero mode because of the vanishing eigenvalue of the spherical Laplacian of the $l=0$ mode of $\delta \varphi_2$. Since $q=1/\beta$ does not have a term linear in $\beta$, in the $\mathcal{N}=2$ case these are exact zero modes, contrarily to the Liouville theories with less supersymmetry where the zero modes are lifted at order $\mathcal{O}(\beta^2)$ (see discussion above (\ref{Zbos_loop_N1}) and \cite{Anninos:2021ene}).
Following the reasoning for $\mathcal{N}=1$ discussed around (\ref{heatkernel}), one finds
\begin{equation}
-\frac{1}{2}\sum_{l= 2}^\infty(2l+1)\log \frac{l(l+1)-2}{4\pi r^2 \Lambda_{\text{uv}}} =  \int_{0}^\infty\frac{\dd t}{2t}\left[\frac{1+e^{-t}}{(1-e^{-t})}\left(\frac{2e^{-2t}}{(1-e^{-t})}+5e^{-t}-5e^{-2t}+3e^{-3t}\right)\right]~,
\end{equation}
for the logarithm of the first product in (\ref{N2bprod}), and
\begin{equation}
-\frac{1}{2}\sum_{l= 1}^\infty(2l+1)\log \frac{l(l+1)}{4\pi r^2 \Lambda_{\text{uv}}} = \int_{0}^\infty\frac{\dd t}{2t}\left[\frac{1+e^{-t}}{(1-e^{-t})}\left(\frac{2e^{-t}}{(1-e^{-t})}+e^{-t}\right)  \right]~,
\end{equation} 
for the logarithm of the second product in (\ref{N2bprod}). From the above expressions, we can identify
\begin{equation}
\chi_{\Delta=2}(t) = \frac{e^{-{2t}}}{(1-e^{-t})}\quad  \text{and} \quad \chi_{\Delta=1}(t) = \frac{e^{-{t}}}{(1-e^{-t})}~,
\end{equation}
as the Harish--Chandra character of the discrete series representation of $SO(1,2)  \cong PSL(2,\mathbb{R})$ with $\Delta=2$ and $\Delta=1$. 
We thus obtain 
\begin{equation}\label{ZgausbosN2}
\log \mathcal{Z}_{\text{tL,bos,1-loop}}^{\mathcal{N}=2}= 2\times  \frac{2}{\varepsilon^2}  -\frac{8}{3}\log \varepsilon + \text{constants}~,
\end{equation}
where $\varepsilon = {e^{-\gamma_E}}/{\sqrt{2\pi r^2 \Lambda_{\mathrm{uv}}}}$ as follows from analogous steps to those in (\ref{bessel}).
\vspace{0.2cm}\newline
{\textbf{Fermionic one-loop determinant}.} 
To obtain the one-loop contribution and eventually the fermionic propagators we decompose the Dirac fields $\psi$ and $\tpsi$ in a basis of eigenspinors of the Dirac operator on the two-sphere:
\begin{equation}
\label{eq:psis_N2}
\begin{split}
\psi(\Omega)&= \frac{1}{\sqrt{r}} \sum_{l = 0}^\infty \sum_{m= 0}^l \sum_{\pm}\left(\alpha_{\pm\,lm} \psi^{(+)}_{\pm \, lm}(\Omega) + \beta_{\pm \,lm}\psi^{(-)}_{\pm \, lm}(\Omega)\right)~,\cr
\tpsi(\Omega) &= - \frac{\ii}{\sqrt{r}} \sum_{l = 0}^\infty \sum_{m= 0}^l \sum_{\pm}\left(\beta^*_{\mp \,lm} \psi^{(+)}_{\pm \,lm}(\Omega) +  {\alpha}_{\mp \, lm}^*\psi^{(-)}_{\pm \, lm}(\Omega)\right)~.
\end{split}
\end{equation}
In the above we implemented the reality condition \eqref{eq:ChiralRealContour} (and thus \eqref{eq:RelationsGrassmann_1}).
The coefficients $\alpha_{\pm\,lm}$ and $\talpha_{\pm\,lm}$ are complex-valued Grassmann variables. In terms of these Grassmann variables, the path integration measure is given by
\begin{equation}\label{Measure2}
 [\mathcal{D}\psi] =  \prod_{{\substack{l\geq 0\\ 0\leq m\leq l}}} \left(\frac{4\pi}{\Lambda_{\text{uv}}r^2}\right)\prod_{\pm}\dd{\alpha}_{\pm\,lm}\dd \beta_{\pm\,lm}~,\quad 
  [\mathcal{D}\tpsi] =   \prod_{{\substack{l\geq 0\\ 0\leq m\leq l}}} \prod_{\pm}\left(\frac{4\pi}{\Lambda_{\text{uv}}r^2}\right)\dd{\alpha}^*_{\pm\,lm}\dd \beta_{\pm\,lm}^*~.
\end{equation}
In order to diagonalise the Gaussian action for the fermionic fluctiations, for each $l\geq 0$ and $0 \leq m \leq l$, it is convenient to define $\boldsymbol{x}_{lm}$ as 
\begin{equation}
\boldsymbol{x}_{lm}^T \equiv  \begin{pmatrix} \alpha_{+\,lm}& \alpha_{+\,lm}^* & \alpha_{-\,lm}  & \alpha_{-\,lm}^* & \beta_{+\,lm}  & \beta_{+\,lm}^* & \beta_{-\,lm} & \beta_{-\,lm}^*  \end{pmatrix}~.
\end{equation}
The Gaussian action for fermionic fluctuations, is then given by
\begin{equation}\label{N2fG}
S_{\mathrm{pert}}^{(2)}[\delta\psi,\delta \tpsi]\equiv\frac{1}{4\pi}\int_{S^2} \dd^2 x\, \tilde{\rme}\,\bigg( \ii \delta\btpsi \slashed{D}\delta\psi  + \frac{1}{r}\frac{\mu}{2|\mu|} \delta\bpsi\delta \psi + \frac{1}{r}\frac{\mu^*}{2|\mu|}\delta\btpsi \delta\tpsi  \bigg) = \frac{1}{4\pi} \sum_{l = 0}^\infty\sum_{m=0}^l \boldsymbol{x}_{lm}^TB_{lm}\boldsymbol{x}_{lm}~,
\end{equation}
where the matrix $B_{lm}$ reads  as follows
\begin{equation}\label{eq:Bmatrix}
B_{lm}\equiv \begin{pmatrix}  
0 & 0 & 0 & \frac{1}{2}(l+1) & -\ii \frac{\mu}{2|\mu|} & 0 & 0 & 0 \\ 
0 & 0 & \frac{1}{2}(l+1) & 0 & 0 & -\ii \frac{\mu^*}{2|\mu|}  & 0 & 0 \\ 
0 & -\frac{1}{2}(l+1) & 0 & 0 & 0 & 0 & -\ii \frac{\mu}{2|\mu|}  & 0 \\
-\frac{1}{2}(l+1) & 0 & 0 & 0 & 0 & 0 & 0 & -\ii \frac{\mu^*}{2|\mu|}\\
\ii \frac{\mu}{2|\mu|}   & 0 & 0 & 0 &0& 0 & 0 & -\frac{1}{2}(l+1) \\ 
0  & \ii \frac{\mu^*}{2|\mu|} & 0 & 0 &0& 0 & -\frac{1}{2}(l+1) & 0 \\ 
0  & 0 & \ii \frac{\mu}{2|\mu|}  & 0 &0& \frac{1}{2}(l+1)  & 0 & 0 \\ 
0  & 0 & 0 & \ii \frac{\mu^*}{2|\mu|} & \frac{1}{2}(l+1) & 0  & 0 & 0 \\ 
\end{pmatrix}~.
\end{equation}
Assembling the pieces together, we  arrive at the one-loop fermionic contribution
\begin{equation}\label{fermionGaussianN2}
\mathcal{Z}_{\text{fer,1-loop}}^{\mathcal{N}=2} =\left(\frac{4\pi}{\Lambda_{\text{uv}}r^2}\right)^2 \prod_{l= 1}^\infty\left(\frac{l(l+2)}{4\pi \Lambda_{\text{uv}}r^2}\right)^{2(l+1)}~.
\end{equation}
In contrast to the $\mathcal{N}=1$ case (\ref{fermionGaussianN1}), the $l=0$ zero modes for the $\mathcal{N}=2$ fermionic fluctuations  are not lifted at order $\mathcal{O}(\beta^2)$. 
We can borrow the result for $\mathcal{N}=1$ and obtain
\begin{equation}\label{ZgausferN2}
\sum_{l= 1}^\infty 2(l+1) \log \frac{l(l+2)}{4 \pi\Lambda_{\text{uv}}r^2}   =-\int_{0}^\infty\frac{\dd t}{2t}\left[\frac{2e^{-\frac{t}{2}}}{(1-e^{-t})}\left(4 \chi_{\Delta=3/2}(t)+ 4e^{-\frac{t}{2}}- 2e^{-\frac{3t}{2}}+2 e^{-\frac{5 t}{2}}\right)\right]~,
\end{equation}
with the character $\chi_{\Delta=3/2}(t)$ given in (\ref{chi32}).
Expanding the above expression for small $t$ and combining it with (\ref{fermionGaussianN2}) we obtain
\begin{equation}
\log \mathcal{Z}_{\text{fer,1-loop}}^{\mathcal{N}=2} = -\frac{4}{\varepsilon^2} +\frac{5}{3}\log \varepsilon + \text{constants}~.
\end{equation}
Further combining the bosonic (\ref{ZgausbosN2}) and fermionic (\ref{ZgausferN2}) one-loop contribution we observe the supersymmetric cancellation of the leading ultraviolet divergence.
\newline\newline
Piecing the bosonic and fermionic contributions together, we find the following expression for $\mathcal{Z}_{\text{tL}}^{\mathcal{N}=2}[\mu]$ up  to one-loop order
\begin{equation}\label{ZtLN2}
\mathcal{Z}_{\text{tL}}^{\mathcal{N}=2}[\mu] \approx \pm \ii \times \text{const}\times \left(\frac{\mu\mu^*}{\beta^2}\right)^{-\frac{1}{\beta^2}} \, \,  \Lambda_{\text{uv}}^{\frac{1}{2}}\, \, r^{{c_{\text{tL}}^{\mathcal{N}=2}}/{3}}\times e^{-\frac{1}{\beta^2} - \frac{2}{\beta^2}\log \beta^2} \times  \left( 1 +\mathcal{O}(\beta^2) \right)~.
\end{equation}
where `const' denotes a real constant independent of $\beta$, $\mu$,  $\mu^*$, and $r$. The  $\mathcal{N}=2$ timelike Liouville central charge is $c_{\text{tL}}^{\mathcal{N}=2}= 3-6/\beta^2$. One no longer needs to divide by the volume of $OSp(2|2,\mathbb{C})$ in (\ref{ZtLN2}), as it has been used to fix the three  bosonic and the four fermionic zero modes. A proper treatment of this would require computing a Faddeev--Popov superdeterminant for the gauge fixing procedure, which again we leave to the future. Inspection of (\ref{SMPhi}) suggests that to leading order in the small $\beta$ expansion, the Faddeev--Popov superdeterminant contributes a factor of $1/\beta$ for each bosonic zero mode and a factor of $\beta$ for each fermionic zero mode yielding the prefactor of $\beta^{4-4} = 1$ in $\mathcal{Z}_{\text{tL}}^{\mathcal{N}=2}$. 

\subsection{Non-Gaussian interactions}

We now consider the leading effect due to the interactions, which is a two-loop contribution. Expanding the action up to $\mathcal{O}(\beta^2)$, we encounter interaction terms between the fermionic and bosonic fields. The interacting part of the action governing the fluctuations is given by
\begin{multline}\label{S_2loop}
 S_{\text{L,int}}^{\mathcal{N}=2}= \frac{1}{\color{black}4\pi}\int_{S^2} \dd^2 x \, \tilde{\rme}\,\bigg[
 - \frac{4\ii}{3r^2}\beta \delta\varphi_1^3 +\frac{2}{3r^2}\beta^2\delta\varphi_1^4 + \frac{\beta}{2r}\frac{\mu}{|\mu|} \delta\bpsi \delta\psi(\ii \delta\varphi_1 -\delta\varphi_2) \cr + \frac{\beta}{2r}\frac{\mu^*}{|\mu|}\delta \btpsi \delta\tpsi (\ii \delta\varphi_1+ \delta\varphi_2)  - \frac{1}{4r}\beta^2 \frac{\mu}{|\mu|}\delta\bpsi \delta\psi( \delta\varphi_1^2 -\delta\varphi_2^2) -  \frac{1}{4r}\beta^2 \frac{\mu^*}{|\mu|}\delta\btpsi \delta\tpsi ( \delta\varphi_1^2- \delta\varphi_2^2)\bigg]~.
\end{multline}
In the above we already performed the analytic continuation $\delta \varphi_1 \rightarrow \ii \delta \varphi_1$, $\delta \varphi_2 \rightarrow \ii \delta \varphi_2$. For convenience we choose the counterclockwise rotation, but none of the following results depend on this choice.
To proceed with the two-loop expressions, we obtain the relevant propagators stemming from the Gaussian action (\ref{S_gauss}).
\vspace{0.2cm}\newline
{\textbf{Bosonic propagator.}}
First we obtain the bosonic propagators for the bosonic fluctuations $\delta\varphi_1$ and $\delta\varphi_2$. Given two-points $\Omega$, and $\Omega'$ on the round two-sphere, we have
\begin{equation}\label{propagator}
G(\Omega;\Omega')  = 2\pi \sum_{l \neq 1}  \sum_{-l\leq m\leq l} \frac{{Y}_{l m}(\Omega) {Y}_{l m}(\Omega')}{l(l+1)-2}~.
\end{equation}
Even though we removed the three-fold degenerate $l=1$ mode the two-point function $G(\Omega;\Omega')$ remains $SO(3)$ invariant. At the coincident point $\Omega=\Omega'$ one finds
\begin{equation}\label{divloop}
G(\Omega_0,\Omega_0) = \frac{1}{4\pi}\int_{S^2}\dd\Omega \, G(\Omega;\Omega) = \frac{1}{2}\sum_{l \neq 1 }  \frac{2l+1}{l(l+1)-2}~. 
\end{equation}
The above sum diverges logarithmically, as expected for coincident fields in two dimensions. For $\delta\varphi_2$ we have 
\begin{equation}\label{propagator_deltaphi2}
\widetilde{G}(\Omega;\Omega') = 2\pi \sum_{l = 1}^\infty   \sum_{-l\leq m\leq l} \frac{{Y}_{l m}(\Omega) {Y}_{l m}(\Omega')}{l(l+1)} ~,
\end{equation}
which is again $SO(3)$ invariant and logarithmically divergent at coincident points.
\vspace{0.2cm}\newline
{\textbf{Fermionic propagator.}} For the fermionic propagator it is convenient to work in momentum space where the relevant quadratic action is given by (\ref{N2fG}). We obtain for the propagators
\begin{multline}
H(\Omega;\Omega') \equiv  \frac{1}{\mathcal{Z}_{\text{fer},1\text{-loop}}^{\mathcal{N}=2}}  \left(\frac{4\pi}{\Lambda_{\text{uv}}r^2}\right)^2 \int [\mathcal{D}'\delta\psi] [\mathcal{D}'\delta\tpsi] \, e^{-S_{\mathrm{pert}}^{(2)}[\delta\psi,\delta \tpsi]} \delta\bpsi(\Omega) \delta\psi(\Omega') ~\cr
= \frac{4\pi\ii}{r}\,\frac{\mu^*}{|\mu|}\sum_{l\geq 1}\sum_{ 0\leq m \leq l} \frac{1}{l(l+2)} \Big(\bpsi^{(+)}_{+\,lm}(\Omega)\psi^{(-)}_{+\,lm}(\Omega') + \bpsi^{(+)}_{-\,lm}(\Omega)\psi^{(-)}_{-\,lm}(\Omega') \cr
- \bpsi^{(-)}_{+\,lm}(\Omega) \psi^{(+)}_{+\,lm}(\Omega') - \bpsi^{(-)}_{-\,lm}(\Omega)\psi^{(+)}_{-\,lm}(\Omega')\Big)~.
\end{multline}
Similarly, we have
\begin{multline}
\widetilde{H}(\Omega;\Omega') \equiv  \frac{1}{\mathcal{Z}_{\text{fer},1\text{-loop}}^{\mathcal{N}=2}}  \left(\frac{4\pi}{\Lambda_{\text{uv}}r^2}\right)^2 \int [\mathcal{D}'\delta\psi] [\mathcal{D}'\delta\tpsi] \,  e^{-S_{\mathrm{pert}}^{(2)}[\delta\psi,\delta \tpsi]} \delta\btpsi(\Omega) \delta\tpsi(\Omega') ~\cr
=\frac{4\pi\ii}{r}\, \frac{\mu}{|\mu|}\,\sum_{l\geq 1}\sum_{ 0\leq m \leq l} \frac{1}{l(l+2)} \Big(\bpsi^{(+)}_{+\,lm}(\Omega)\psi^{(-)}_{+\,lm}(\Omega') + \bpsi^{(+)}_{-\,lm}(\Omega)\psi^{(-)}_{-\,lm}(\Omega')\cr
-\bpsi^{(-)}_{+\,lm}(\Omega) \psi^{(+)}_{+\,lm}(\Omega') - \bpsi^{(-)}_{-\,lm}(\Omega)\psi^{(+)}_{-\,lm}(\Omega') \Big)~,
\end{multline}
where the fermionic Gaussian action $S_{\mathrm{pert}}^{(2)}[\delta\psi,\delta \tpsi]$ for the $\mathcal{N}=2$ action is given in (\ref{N2fG}).
The above in particular imply the following logarithmic divergent coincident point limits 
\begin{equation}
\label{eq:Fermion equal points}
\begin{split}
H(\Omega_0,\Omega_0) &= \frac{1}{4\pi}\int_{S^2} \dd \Omega \,  H(\Omega,\Omega) =  \frac{4}{r}\frac{ \mu^*}{|\mu|} \sum_{l =  1}^\infty \frac{l+1}{l(l+2)}~,\cr
\widetilde{H}(\Omega_0,\Omega_0) &= \frac{1}{4\pi}\int_{S^2} \dd \Omega \,  \widetilde{H}(\Omega,\Omega)   =    \frac{4}{r}  \frac{ \mu}{|\mu|} \sum_{l = 1}^\infty \frac{l+1}{l(l+2)}~.
\end{split}
\end{equation}
On the other hand, the propagator $\langle \btpsi(\Omega) \psi(\Omega')\rangle$ vanishes in the coincident point limit.
\vspace{0.2cm}\newline
{\textbf{Two-loop contributions.}} We can now compute the various contributions. The diagrams are split into four categories. The first kind of diagrams are purely bosonic and are the diagrams that appear in the non-supersymmetric case. The difference in the $\mathcal{N}=2$ case is that the relation $q=1/\beta$  renders the $l=1$ mode for $\delta\varphi_1$ an exact zero mode. We have
\begin{equation}
-\frac{1}{3\pi^2} \beta^2\int_{S^2} \dd\Omega\dd\Omega' \, G(\Omega,\Omega')^3-\frac{1}{2\pi^2} \beta^2\int_{S^2}\dd\Omega\dd\Omega' \, G(\Omega,\Omega)G(\Omega',\Omega')G(\Omega,\Omega')
- \frac{1}{2\pi}\beta^2\int_{S^2} \dd{\Omega} \, G(\Omega,\Omega)^2~.
\end{equation}
These diagrams already appeared in the $\mathcal{N}=1$ case (\ref{loops_b_N1}) as well as the non-supersymmetric case \cite{Anninos:2021ene}. They represent melonic, double tadpole and cactus diagrams. The cactus and double tadpoles mutually cancel each other while the melonic diagrams lead to an ultraviolet finite contribution (\ref{melons_N1}).
The second category consists of the cactus diagrams
\begin{align}
&-\frac{r}{16\pi}\frac{\mu}{|\mu |}\beta^2\int_{S^2} \dd\Omega \, H(\Omega,\Omega){G}(\Omega,\Omega)- \frac{r}{16\pi}\frac{\mu^*}{|\mu |}\beta^2\int_{S^2} \dd\Omega \, \widetilde{H}(\Omega,\Omega){G}(\Omega,\Omega)\cr
&+\frac{r}{16\pi}\frac{\mu}{|\mu |}\beta^2 \int_{S^2}\dd\Omega \, {H}(\Omega,\Omega)\widetilde{G}(\Omega,\Omega)+ \frac{r}{16\pi}\frac{\mu^*}{|\mu |}\beta^2 \int_{S^2}\dd\Omega \, \widetilde{H}(\Omega,\Omega)\widetilde{G}(\Omega,\Omega)~.
\end{align}
These diagrams have one fermionic and one bosonic loop. The leading $\log^2$ divergence cancels between the four diagrams as can be easily seen by plugging in the coincident point limits of the fermion (\ref{eq:Fermion equal points})  and boson propagators (\ref{divloop}).
Next we have the double-tadpole diagrams
\begin{align}
&+\frac{r}{8\pi^2}\frac{\mu}{|\mu|}\beta^2\int_{S^2} \dd{\Omega} \dd{\Omega'} \, H(\Omega,\Omega) G(\Omega,\Omega') G(\Omega',\Omega')+\frac{r}{8\pi^2}\frac{\mu^*}{|\mu|}\beta^2\int_{S^2} \dd{\Omega} \dd{\Omega'} \, \widetilde{H}(\Omega,\Omega) G(\Omega,\Omega') G(\Omega',\Omega')\cr
 & {\color{black}-\frac{r^2}{128\pi^2}\frac{\mu}{\mu^*}\beta^2\int_{S^2} \dd{\Omega} \dd{\Omega'} \, H(\Omega,\Omega)G(\Omega,\Omega')F(\Omega',\Omega')-\frac{r^2}{128\pi^2}\frac{\mu^*}{\mu}\beta^2\int_{S^2} \dd{\Omega} \dd{\Omega'} \, \widetilde{H}(\Omega,\Omega)G(\Omega,\Omega')\widetilde{H}(\Omega',\Omega')}~\cr
 &-\frac{r^2}{64\pi^2}\beta^2\int_{S^2} \dd{\Omega} \dd{\Omega'} \, H(\Omega,\Omega') \widetilde{H}(\Omega',\Omega) G(\Omega,\Omega')~.
\end{align}
We will not give the explicit expressions for the individual diagrams but just discuss their leading ultraviolet behaviour. The two double tadpoles in the first line diverge as $-4  \log^2 (\Lambda_{\text{uv}}r^2 )$. The three diagrams in the second and third line have a leading divergence that goes as $+2 \log^2 (\Lambda_{\text{uv}}r^2 )$.
Finally we have the melonic type diagrams
\begin{align}\label{diagrams_melons}
 &-\frac{r^2}{128\pi^2}\frac{\mu}{\mu^*}\beta^2\int_{S^2} \dd{\Omega} \dd{\Omega'} \, H(\Omega,\Omega')^2G(\Omega,\Omega') -\frac{r^2}{128\pi^2}\frac{\mu^*}{\mu}\beta^2\int_{S^2} \dd{\Omega} \dd{\Omega'} \,  \widetilde{H}(\Omega,\Omega')^2G(\Omega,\Omega')\cr
&{\color{black}+\frac{r^2}{128\pi^2}\frac{\mu}{\mu^*}\beta^2\int_{S^2} \dd{\Omega} \dd{\Omega'} \, H(\Omega,\Omega')^2\widetilde{G}(\Omega,\Omega')}
{\color{black}+\frac{r^2}{128\pi^2}\frac{\mu^*}{\mu}\beta^2\int_{S^2} \dd{\Omega} \dd{\Omega'} \, \widetilde{H}(\Omega,\Omega')^2\widetilde{G}(\Omega,\Omega')}\cr
&-\frac{r^2}{64\pi^2}\beta^2\int_{S^2} \dd{\Omega} \dd{\Omega'} \, F(\Omega,\Omega') \widetilde{H}(\Omega',\Omega) G(\Omega,\Omega')-\frac{r^2}{64\pi^2}\beta^2\int_{S^2} \dd{\Omega} \dd{\Omega'} \, H(\Omega,\Omega') \widetilde{H}(\Omega',\Omega) \widetilde{G}(\Omega,\Omega')~.\cr
\end{align}
Since the bosonic propagator for $\delta\varphi_1$ and $\delta\varphi_2$ and the fermionic propagators for $\delta\psi$ and $\delta \tpsi$ have the same leading asymptotic behaviour we infer that the leading $\log^2$ divergence of the four melonic type diagrams in the first and second line of (\ref{diagrams_melons}) mutually cancels. The remaining diagrams are of melonic type and mix the chiral and anti-chiral fermion and the bosonic fields $\delta\varphi_1$ and $\delta\varphi_2$ respectively. Numerically, we infer that they diverge as $+2 \log^2 (\Lambda_{\text{uv}}r^2 )$. 
In summary, the leading $\mathcal{O}(\log^2)$ divergence of the two-loop diagrams of $\mathcal{N}=2$ timelike Liouville theory vanishes. We expect this to also apply to the $\mathcal{O}(\log)$ divergence. 
\vspace{0.2cm}\newline
Finally, as for the $\mathcal{N}=1$ case, we note that the spacelike counterpart of the $\mathcal{N}=2$ timelike super-Liouville theory, which we discuss in appendix \ref{spacelikeL}, exhibits a very similar perturbative loop expansion. In particular, the cancellation of two- and higher-loop leading ultraviolet divergences in the small $b$ expansion of one theory imply cancellation in the small $\beta$ expansion of the other.

\subsection{\texorpdfstring{$\mathcal{N}=2$}{N2} summary}
We now summarise the main results obtained from the path integral perspective of $\mathcal{N}=2$ supergravity. This subsection is the $\mathcal{N}=2$ counterpart of subsection \ref{subsec:N1summary}. The two-sphere partition function of $\mathcal{N}=2$ timelike Liouville theory is given by
\begin{equation}\label{eq:ZtLN2}
\mathcal{Z}_{\text{tL}}^{\mathcal{N}=2}[\mu] \approx \pm \ii \times \text{const}\times \left(\frac{\mu\mu^*}{\beta^2}\right)^{-\frac{1}{\beta^2}} \, \,  \Lambda_{\text{uv}}^{\frac{1}{2}}\, \, r^{{c_{\text{tL}}^{\mathcal{N}=2}}/{3}}\times e^{-\frac{1}{\beta^2} - \frac{2}{\beta^2}\log \beta^2} \times  \left( 1 + \text{loops}\,\beta^2 \right)~.
\end{equation}
The overall constant is independent of $\beta$ and $\mu,\mu^*$. We list some remarks about (\ref{eq:ZtLN2}). 
\begin{itemize}
\item As in the $\mathcal{N}=1$ case the path integral is pure imaginary. This is a result of the unbounded $l=0$ mode of $\delta\varphi_1 = 2 (\delta\varphi+ \delta \tvarphi)$ (\ref{eq:def deltaphi12}) which is the Weyl-factor of the $\mathcal{N}=2$ theory (\ref{N2sW}).
\item From the sphere path integral (\ref{eq:ZtLN2}) we can read off the central charge of $\mathcal{N}=2$ timelike Liouville, confirming the aforementioned value (\ref{eq:N2 timelike central charge}).
\item The structure constants for $\mathcal{N}=2$ Liouville theory are not known. Therefore we cannot relate (\ref{eq:ZtLN2}) to the three-point function of Liouville vertex operators as in the $\mathcal{N}=1$ and non-supersymmetric case. 
\item {From a field theory perspective, the sphere partition function of an $\cN=2$ SCFT is a well-defined function of the marginal {couplings}, in contrast to the $\cN=1$ case mentioned in section \ref{subsec:N1summary} \cite{Gerchkovitz:2014gta}.}
\end{itemize}
Combining (\ref{eq:ZtLN2}) with the ghost and matter structure, we obtain for the gravitational path integral of $\mathcal{N}=2$ dS$_2$ supergravity
\begin{equation}
\label{eq:Z_grav_N2_final}
	\log \mc{Z}^{\cN=2}_{{\rm grav}, (0)} = 2\vartheta -\left(\frac{c_m}{6}-\frac{3}{6}\right) \log \frac{|\mu|^2}{\Lambda_{\text{uv}}}+ f_0(c_m)~,
\end{equation}
where the loop expansion is contained in $f_0(c_m)$. The coefficient of the logarithm is exactly $c_m/6 - 3/6$, whose leading term again takes the form \cite{Holzhey:1994we, Calabrese:2004eu, Casini:2011kv} appearing in the entanglement entropy of a two-dimensional conformal field theory across a finite interval. Interestingly, unlike the two other cases, the coefficient of the logarithm truncates in the $\mathcal{N}=2$ case. One can compare the first subleading coefficient $-3/6 = (3_{\text{L}}-26_{\mathfrak{bc}}+2 \times 11_{\beta\gamma}-2_{U(1)})/6$ to the one for the $\mathcal{N}=1$ case which is $-7/4= ((3/2)_{\mathrm{L}}  - 26_{\mathfrak{b}\mathfrak{c}} + 11_{\beta\gamma} )/{6} + 1/2$, as well as the one appearing in the non-supersymmetric case \cite{Anninos:2021ene} which is $-19/6 = (1_{\mathrm{L}} - 26_{\mathfrak{b}\mathfrak{c}} )/{6}+1$. 

\section{Super-Liouville localization}\label{N2Loc}

In this section we discuss the method of supersymmetric localization (see \cite{Pestun:2016zxk} for a review) as an alternative approach to analyse the $\mathcal{N}=2$ super-Liouville theory. Supersymmetric localization methods for a two-dimensional quantum field theory with $\mathcal{N}=2$ supersymmetry on $S^2$ have been developed in \cite{Benini:2012ui, Doroud:2012xw}. An interesting feature of localization is an effective reduction in the degrees of freedom that are path integrated over.

\subsection{General setup}

The theory we are considering \eqref{eq:N22_TimelikeLiouvilleSphere} is the theory of a (Liouville) chiral and anti-chiral multiplet with exponential superpotential $W(\varphi)= \mu e^{\beta \varphi}/\beta$ and $\widetilde{W}(\tvarphi)= \mu^* e^{\beta \tvarphi}/\beta$.
As we saw, it is conformally invariant, so it can be formulated on the sphere preserving the full superconformal algebra $\mf{osp}(2|2,\C)$. However, in computing the path integral we introduce a regulator, and to preserve supersymmetry we must choose the regulator to break the superconformal algebra to a massive supersymmetry algebra. As mentioned in section \ref{subsec:N2Action}, there are two massive subalgebras $\mf{su}(2|1)$ of $\mf{osp}(2|2,\C)$, corresponding to the possible choices of off-shell supergravity one can couple to, in the spirit of \cite{Festuccia:2011ws,Doroud:2012xw, Doroud:2013pka, Gerchkovitz:2014gta}. The bosonic subalgebra of $\mf{su}(2|1)$ is $\mf{su}(2)\times \mf{u}(1)$, generated by the isometries of $S^2$ and one of the two $\mf{u}(1)$ $R$-symmetry factors of $\mf{osp}(2|2,\C)$. We shall consider the subalgebra $\mf{su}(2|1)_V$, preserving the vector-like $R$-symmetry $\mf{u}(1)_V$. This corresponds to choosing in the supersymmetry transformations \eqref{eq:SuperConformalTransformations_Timelike} positive $\epsilon$, $\tepsilon$ satisfying
\begin{equation}
	D_\mu \epsilon = \frac{\ii}{2r}\gamma_\mu \epsilon ~, \qquad D_\mu \tepsilon = \frac{\ii}{2r}\gamma_\mu \tepsilon ~, 
\end{equation}
in which case \eqref{eq:SuperConformalTransformations_Timelike} reduce to 
\begin{equation}
\label{eq:SupersymmetryTransformations_Timelike}
\begin{aligned}
\delta \varphi &= \btepsilon \psi ~, & \quad \delta \tvarphi &= \overline{\epsilon} \tpsi ~, \\
\delta \psi &= \ii \slashed{\partial}\varphi \, \epsilon - \frac{1}{\beta r} \epsilon  -\ii F \, \tepsilon  ~, &\quad  \delta \tpsi &= \ii \slashed{\partial}\tvarphi \,  \tepsilon - \frac{1}{\beta r} \tepsilon  - \ii \tF \, \epsilon  , \\
\delta F &= - \overline{\epsilon} \slashed{D}\psi ~, &\quad \delta \tF &= \btepsilon \slashed{D} \tpsi ~.
\end{aligned}
\end{equation}
with commutators
\begin{equation}
\begin{split}
[\delta_\epsilon, \delta_{\tilde{\epsilon}}] \varphi &= \cL_\xi \varphi + \frac{2\ii}{\beta}\rho ~, \\
[\delta_\epsilon, \delta_{\tilde{\epsilon}}] \psi &= \cL_\xi \psi -\ii \rho \psi ~, \\
[\delta_\epsilon, \delta_{\tilde{\epsilon}}] F &= \cL_\xi F - 2\ii \rho F ~,
\end{split}
\end{equation}
and now indeed $\xi$ is a Killing vector on the two-sphere, $\nabla_\mu \xi_\nu = \frac{1}{r} \btepsilon \gamma_{\mu\nu}\epsilon$, and $\rho = \frac{1}{2r}\btepsilon \epsilon$, as mentioned below \eqref{eq:SCA_LiouvilleChiral}. 

$\,$\vspace{0.2cm}\newline
The path integral of $\cN=2$ supersymmetric field theories on the two-sphere is amenable to supersymmetric localization, as explained in \cite{Benini:2012ui, Doroud:2012xw}, which we follow with minor changes due to the behaviour of the scalar $\varphi$. Supersymmetric localization is an extension of the equivariant localization \cite{BV:1982, Witten:1982im, Atiyah:1984px}, which relies on the following steps \cite{Witten:1991zz}. Suppose there is an action of a group $F$ on the integration space of the path integral $\mc{E}$. If the action is free, it generates a fibration $\mc{E} \to \mc{E}/F$ and, by integrating first over the fibre, we can reduce the integral on $\mc{E}$ to an integral over $\mc{E}/F$. For an $F$-invariant observable $\mc{O}$, this gives
\begin{equation}
	\int_{\mc{E}} [\mc{D}\Phi] \, e^{-S} \mc{O} = {\rm vol}(F) \, \int_{\mc{E}/F} [\widetilde{\mc{D}\Phi}] \, e^{-S} \mc{O} ~.
\end{equation}
In the case at hand, the symmetry group is fermionic generated by the supercharge $\mc{Q}$ with $\{ \mc{Q}, \mc{Q} \} = \mc{B}$ ($\mc{B}$ being a bosonic symmetry), and its volume vanishes. Therefore, the only contributions to the path integral come from the locus $\mc{E}_0$ where the action is not free and the symmetry acts trivially. The integral over the entirety of field space is then reduced to a sum of pullbacks on $\mc{E}_0$ weighted by the Pfaffian of the $F$-equivariant curvature of the normal bundle to $\mc{E}_0$, or, by analogy, a classical term and an integration of the one-loop determinant. To simplify the latter integration we can modify the Lagrangian by a $\mc{Q}$-exact term $+t \mc{Q}V$, provided $\mc{B}V=0$, and the value of the observable is unchanged. For instance, for the partition function
\begin{equation}
	Z(t) = \int_{\mc{E}} [\mc{D}\Phi] \, e^{- S - t \mc{Q} V}
\end{equation}
one finds
\begin{equation}\label{totalD}
	Z'(t) = - \int_{\mc{E}} [ \mc{D}\Phi] \, \mc{Q} \left( V \, e^{-S - t \mc{Q}V} \right) ~.
\end{equation}
One can then invoke an analog of Stokes' theorem in the space of fields, provided the measure is invariant under $\mc{Q}$, and provided there are no boundary contributions in $\mc{E}$. If both conditions hold, then $Z(t)$ is independent of $t$, and if the bosonic part of $\mc{Q}V$ is positive-definite, then in the limit $t\to +\infty$ only the terms with $\mc{Q}V=0$ may contribute, and of these contributions only the supersymmetric ones are non-vanishing. This locus is $\mc{E}_0^{\mc{Q}}$.

\subsection{Various deformations}

Importantly, although the steps of the computations may depend on the choice of $V$, the final result should not (provided the conditions above are satisfied and no saddle point is added), so we shall consider a few deformation terms. 
First, we construct $\cQ$.
We write the supersymmetry variations in \eqref{eq:SupersymmetryTransformations_Timelike} as $\delta_{\epsilon} \equiv \epsilon^\alpha Q_\alpha$ and $\delta_{\tepsilon} \equiv \tepsilon^\alpha \widetilde{Q}_\alpha$. Then, using a \textit{commuting} positive Killing spinor $\epsilon_+$ and its charge conjugate $\epsilon_+^c \equiv \cC^{-1} \epsilon^*_+$, we construct the anti-commuting operators
\begin{equation}
	Q \equiv \epsilon^\alpha_+ Q_\alpha ~, \qquad Q^\dagger \equiv (\epsilon^c_+)^\alpha \widetilde{Q}_\alpha ~.
\end{equation}
For notational purposes, it is also useful to introduce the action of $\dagger$ on the fields
\begin{equation}
	\varphi^\dagger \equiv \tvarphi ~, \qquad \psi^\dagger \equiv - \btpsi ~, \qquad F^\dagger = \tF ~.
\end{equation}
As before, \textit{a priori} in Euclidean signature there is no relation between the fields of the chiral and anti-chiral multiplets, so the operation $\dagger$ is only formal. However, on the real contour \eqref{eq:ChiralRealContour}, it actually corresponds to the Hermitian conjugation of the field.  The action of $Q$ and $Q^\dagger$ on the fields can be straightforwardly recovered 
\begin{equation}
\begin{aligned}
Q\varphi &= 0 ~, &\qquad Q^\dagger \varphi &= - \epsilon^\dagger_+ \psi ~, \\
Q\psi &= \left( \ii \slashed{\partial}\varphi - \frac{q}{r} \right)\epsilon_+ ~, &\qquad Q^\dagger \psi &= -\ii F \epsilon^c_+ ~, \\
QF &= - \overline{\epsilon_+}  \slashed{D}\psi ~, &\qquad Q^\dagger F &= 0 ~, \\[10pt]
Q\varphi^\dagger &= \psi^\dagger \epsilon_+ ~, &\qquad Q^\dagger \varphi^\dagger &= 0 ~, \\
Q\psi^\dagger &= \ii F^\dagger \overline{\epsilon_+} ~, &\qquad Q^\dagger \psi^\dagger &= \epsilon^\dagger_+ \left( - \ii \slashed{\partial}\varphi^\dagger - \frac{q}{r}\right) ~, \\
QF^\dagger &= 0 ~, &\qquad Q^\dagger F^\dagger &= D_\mu \psi^\dagger \gamma^\mu \epsilon_+^c ~.
\end{aligned}
\end{equation}
and therefore on the fields
\begin{equation}
	\{ Q, Q^\dagger \} = J + \frac{1}{2}R ~,
\end{equation}
where $J= - \ii \cL_v$ generates a $U(1)$ inside the $SU(2)$ isometries of the sphere ($v \equiv \epsilon_+^\dagger \gamma^\mu \epsilon$ is a Killing vector and one can construct the azimuthal angle as its adapted coordinate), and $R$ is the generator of the $R$-symmetry transformation with parameter $s \equiv \epsilon^\dagger_+ \epsilon_+$. The supercharge used for the localization is $\mc{Q} \equiv Q + Q^\dagger$, which generates a subalgebra $\mf{su}(1|1)$ of the superalgebra $\mf{su}(2|1)$.
\vspace{0.2cm}\newline
It is tedious but straightforward to show that the kinetic part of the action \eqref{eq:N22_TimelikeLiouvilleSphere} for the Liouville chiral multiplet is $\mc{Q}$-exact  
\begin{align}
	\mc{L}_{\rm kin} &= - \partial_\mu {\varphi}^\dagger \partial^\mu \varphi - \ii \psi^\dagger \slashed{D}\psi + {F}^\dagger F - \frac{q}{r^2} (\varphi + {\varphi}^\dagger) - \frac{1}{\beta^2 r^2} \nonumber \\
	\label{eq:N2_Exactness_KineticLag}
	&= \frac{1}{s} \cQ \left( - \ii \, \epsilon^\dagger_+ \slashed{\partial}\varphi^\dagger \psi - \frac{q}{r}\epsilon^\dagger_+\psi -\ii F \, \psi^\dagger \epsilon_+^c - \frac{1}{r} (\varphi^\dagger - \varphi) \, \epsilon^\dagger_+ \psi \right) ~.
\end{align}
As for the superpotential, one finds, with $W=\mu e^{\beta\varphi}$ and $W^\dagger = \mu^* e^{\beta \varphi^\dagger}$
\begin{equation}
\label{eq:N2_Exactness_Superpotential}
\begin{split}
	\mc{L}_W &\equiv FW'(\varphi) - \frac{\ii}{2} W''(\varphi) \bpsi \psi = \frac{\ii}{s} \mc{Q} \left( W'(\varphi) \overline{\epsilon}_+ \psi \right) ~, \\
	\mc{L}_{W^\dagger} &\equiv F^\dagger W'^\dagger(\varphi^\dagger) - \frac{\ii}{2} W''^\dagger(\varphi^\dagger) \btpsi \tpsi = - \frac{\ii}{s} \mc{Q} \left( W'^\dagger(\varphi^\dagger) \psi^\dagger \epsilon^c_+ \right)	~.
\end{split}
\end{equation}
This was guaranteed by the fact that the superpotential is the top component of a (canonical) chiral supermultiplet with bottom component $\phi_W \equiv W(\varphi)$ and $R$-charge $2$, and the transformation of the top component of such a multiplet is
\begin{equation}
	\delta F_W = - \nabla_\mu \left( \bepsilon \gamma^\mu \psi_W \right) ~.
\end{equation}
In fact, more is true: the top component of a (canonical) chiral supermultiplet with $R$-charge $2$ is a superconformal invariant that represents an exactly marginal operator.\footnote{The superconformal transformation of the auxiliary scalar field in a chiral multiplet with $R$-charge $r$ has the form
\[
\delta F = - \nabla_\mu \left( \bepsilon \gamma^\mu \psi\right) - \frac{r-2}{2} D_\mu \bepsilon \gamma^\mu \psi ~.
\]
} After breaking the superconformal symmetry, it remains supersymmetry-exact only when the massive subalgebra preserved is $\mf{su}(2|1)_V$, as we choose. Since the entire Liouville action \eqref{eq:N22_TimelikeLiouvilleSphere} is exact, and the localization argument implies that the partition function does not depend on $\mc{Q}$-exact deformations, we expect that the partition function would in particular not depend on the couplings in the superpotential, namely $\beta$.
\vspace{0.2cm}\newline
To compute the resulting number, we can consider a deformation $\mc{Q}V$ that is invariant under $J + \frac{1}{2}R$ and has positive-definite bosonic part. The canonical such deformation is
\begin{equation}
	V_{\rm can} = ( \mc{Q}\psi )^\dagger \psi + \btpsi \left( \mc{Q}\tpsi \right)^\dagger ~, 
\end{equation}
and the bosonic terms of $\mc{Q}V_{\rm can}$ are clearly positive-definite from their formal expressions: explicitly, ignoring total derivative
\begin{equation}
\label{eq:QVCanonical_Bosonic}
\begin{split}
\mc{Q} V_{\rm can} \rvert_{\rm bos} &= 2 \left[ \partial_\mu \varphi^\dagger \partial^\mu \varphi + \frac{1}{\beta^2r^2} + \ii \frac{1}{ r} v^\mu \, \varphi^\dagger \partial_\mu \varphi \right] -2 F^\dagger F - 2 \ii \left( \ii F w^\mu \partial_\mu \varphi^\dagger + c.c. \right) ~,
\end{split}
\end{equation}
where $w^\mu \equiv \overline{\epsilon_+}\gamma^\mu \epsilon_+$. As reviewed above, the partition function receives contributions from the field configurations where $V_{\rm can}$ vanishes that are also supersymmetric, which are all, by construction, since we require $\psi=\psi^\dagger=0$, and $\mc{Q}\psi = \mc{Q}\psi^\dagger =0$. In spherical coordinates, choosing the spinor (see \eqref{eq:filippo})
\begin{equation}
    \epsilon_+ = e^{\ii \frac{\phi}{2}} \begin{pmatrix} \cos\frac{\theta}{2} \\ \ii \sin\frac{\vartheta}{2} \end{pmatrix} ~,
\end{equation}
we find that the BPS equations $\mc{Q}\psi = \mc{Q}\psi^\dagger=0$ become
\begin{equation}
\begin{split}
0 &= \sin \frac{\theta}{2} \left( 2 \partial_- \varphi -\ii F e^{- \ii\phi} \right) + \cos\frac{\theta}{2} \frac{1}{\beta r}  ~, \\
0 &= \cos \frac{\theta}{2} \left( 2 \partial_+ \varphi -\ii F e^{- \ii\phi} \right)  - \sin\frac{\theta}{2} \frac{1}{\beta r}  ~, \\
0 &= \cos \frac{\theta}{2} \left( 2 \partial_- \varphi^\dagger + \ii F^\dagger e^{\ii \phi} \right) - \sin \frac{\theta}{2} \frac{1}{\beta r} ~, \\
0 &= \sin \frac{\theta}{2} \left( 2 \partial_+ \varphi^\dagger + \ii F^\dagger e^{\ii \phi} \right) +  \cos \frac{\theta}{2} \frac{1}{\beta r} ~,
\end{split}
\end{equation}
where $\partial_{\pm} \equiv (\partial_1 \pm \ii \partial_2)/2$. However, clearly these cannot be solved at the poles and even away from the poles there is no smooth BPS configuration. Therefore, we conclude that there is no smooth BPS field configuration, and therefore nothing contributes to the partition function, which is then expected to be a {constant independent of any features of the underlying quantum field theory}. 
\vspace{0.2cm}\newline
The same result can be reached using a different deformation: we can take $\mc{L}_{\rm kin}$ (ignoring the constant term, which doesn't contribute). This is also $\mc{Q}$-exact, as showed in \eqref{eq:N2_Exactness_KineticLag}, and is clearly invariant not just under $R$ and $J$, but under the full $SU(2)$ isometry of the sphere, which in more involved theories helps with the computation of the one-loop determinants \cite{Benini:2012ui, Doroud:2012xw}. With this deformation, the BPS locus is found by solving the equations of motion for a Liouville chiral multiplet
\begin{equation}
\label{eq:EOMFreeChiral}
\begin{aligned}
0 &= F = \tilde{F} ~, &\qquad 0 &= \nabla^2\varphi - \frac{1}{\beta r^2} = \nabla^2\tvarphi - \frac{1}{\beta r^2} ~, &\qquad 0 &= \slashed{D}\psi = \slashed{D}\tpsi ~.
\end{aligned}
\end{equation}
However, these do not admit a solution with constant scalar field, so there is no $SU(2)$-invariant solution. This is in contrast with the theory of a `canonical' chiral  multiplet.

\subsection{Boundary terms}

The result from the localization analysis seems to be in sharp contrast to the semiclassical approximation uncovered in (\ref{ZtLN2}). This tension is not however a contradiction, and we now mention two reasons for this.
\vspace{0.2cm}\newline
Firstly, to make sense of the Liouville path integral we had to divide by the volume of $OSp(2|2,\mathbb{C})$, which required separating certain zero modes from the $\mathcal{N}=2$ super-Liouville path integral (\ref{eq:Z_L_N2}). If one does not divide by the volume of $OSp(2|2,\mathbb{C})$, it is conceivable that the $\mathcal{N}=2$ super-Liouville path integral is  no longer well-defined.  This might be amended by considering decorated path integrals that soak up the volume of $OSp(2|2,\mathbb{C})$. 
\vspace{0.2cm}\newline
Second, the localization methodology holds only under the assumption that contributions from the boundary of field configuration space are absent. This is known to fail in certain cases. Specifically, this phenomenon was observed to happen for $\mc{N}=2$ spacelike super-Liouville on flat space in \cite{Hori:2001ax}. The authors therein construct two gauged linear sigma models $A$ and $B$ related by mirror symmetry that flow, respectively, to the level $k$ $SL(2,\R)/U(1)$ Kazama--Suzuki supercoset model and to $\mathcal{N}=2$ spacelike super-Liouville theory. One then runs into an apparent paradox due to the fact that standard arguments imply that $D$-terms can neither affect the right-moving chiral algebra (on the $A$ side) nor the twisted superpotential (on the $B$ side). As in our case, this issue is resolved by realizing that these assumptions are affected by boundary terms in field space. Such boundary terms also appear in the sigma models above, for which the target space is non-compact and the $D$-term deformations do not decay fast enough at infinity.\footnote{A similar phenomenon occurs in localization computation of the elliptic genus of the Kazama--Suzuki supercoset model \cite{Murthy:2013mya, Ashok:2013pya}. Here the computation is done in the gauged linear sigma model $A$ introduced in \cite{Hori:2001ax}, but this also includes a chiral multiplet of `Stueckelberg-type' that is not minimally coupled to the Abelian gauge field (a feature that is dual to the non-linear transformation of the Liouville scalar under $R$-symmetry). Its action is $\mc{Q}$-exact but does not fall fast enough at infinity, and this leads to the presence of contribution to the partition function on the torus that is not holomorphic in the modular parameter, again in contrast to the canonical argument \cite{Witten:1986bf}. Of course, this is also reflected in the presence of a continuous spectrum of states on $S^1$ \cite{Troost:2010ud}.} 
\vspace{0.2cm}\newline
Relevant to the second point is, perhaps, Polchinski's observation \cite{Polchinski:1990ia} that the path integral of non-supersymmetric two-dimensional quantum gravity might follow entirely from such boundary terms. In that context, for instance, it was observed that the operator $e^{2\beta \varphi}$ is $\mc{Q}$-exact with respect to a BRST charge associated to the Weyl gauge choice. In the absence of boundary terms, this would lead to a path integral independent of the cosmological constant. The analogue in our context is (\ref{eq:N2_Exactness_Superpotential}), the $\mc{Q}$-exactness of the interacting part of $\mathcal{N}=2$ super-Liouville. From the perspective of the underlying $\mathcal{N}=2$ supergravity, the supercharge is also a BRST charge, now associated to the supe-Weyl gauge fixing.
\vspace{0.2cm}\newline
It seems likely, from the perspective of supersymmetric localization, that the $\mathcal{N}=2$ super-Liouville path integral has contributions from boundary terms. In such a situation, localization is still informing us that the $\mathcal{N}=2$ super-Liouville path integral can be arranged in such a way that the active sector of the path integral stems entirely from the boundary of configuration space --  a significantly reduced portion of the entire field space.

\section{Outlook}

We would like to end our discussion with a brief outlook regarding  the prospect of a quantum theory of dS$_2$ supergravity. 
\vspace{0.2cm}\newline
\textbf{Supersymmetric structure constants.} The structure constants of non-supersymmetric Liouville theory \cite{Dorn:1994xn,Zamolodchikov:1995aa}, and $\mathcal{N}=1$ super-Liouville \cite{Poghossian:1996agj, Rashkov:1996np} in the spacelike regime are well understood. Their construction provides a potential avenue to construct the structure constants of at least certain operators in the timelike regime  through conformal bootstrap methods as well as semiclassical techniques \cite{Teschner:2001rv,Belavin:2007gz,Harlow:2011ny,Giribet:2011zx,Anninos:2021ene}. The structure constants of $\mathcal{N}=2$ super-Liouville theory have been less explored but should be amenable to similar methods. In particular, the structure constant of the supersymmetric cosmological operator should serve as an independent computation of our semiclassical sphere path integral (\ref{ZtLN2}). 
\vspace{0.2cm}\newline
\textbf{Higher genus.} Throughout the paper, our analysis has been restricted to the $S^2$ topology. The supergravity path integrals (\ref{eq:ZSUGRA_General}) and (\ref{N2sugraPI}) invite us to consider higher topology also. In the non-supersymmetric case, the torus path integral displays divergent behaviour in the regime of vanishing cycle size \cite{Anninos:2022ujl}. It would be interesting to see whether the supersymmetric case affects this divergence. In the presence of fermions, the path integral includes a sum over spin structures \cite{Seiberg:1986by} that must be judiciously chosen to render the path integral invariant under large diffeomorphisms. Perhaps there exist permissible choices that exhibit fermionic zero modes leading to a vanishing non-spherical path integral as happens for instance in a variety of superstring partition functions  \cite{Seiberg:1986by,Martinec:1986wa}. 
\vspace{0.2cm}\newline
\textbf{Hartle--Hawking.} From the perspective of dS$_2$, a natural quantity to explore is the Hartle--Hawking path integral \cite{Hartle:1983ai}. This amounts to a disk path integral decorated with some boundary data \cite{Martinec:2003ka,Bautista:2021ogd}. From a semiclassical path integral perspective, there exists a contour for which the Hartle--Hawking saddle endows the disk with the round metric on the hemisphere continued  onto the Lorentzian dS$_2$ metric for larger spatial circles. The wrong sign kinetic term for $\varphi$ is reminiscent of the non-standard kinetic term of the Hamiltonian constraint in the Wheeler--de Witt equation \cite{DeWitt:1967yk}. In light of this, one may have to compute probability on a spatial slice of fixed $\varphi$. 
\vspace{0.2cm}\newline
\textbf{Gravitational entanglement.}  It is desirable to supplement our expressions for the Gibbons--Hawking entropy (\ref{eq:Z_grav_N1_final}) and (\ref{eq:Z_grav_N2_final}), as well as non-supersymmetric counterparts,  with a Lorentzian entanglement entropy picture.   For topological field theory  \cite{Dong:2008ft}, and to some extent for conformal field theory \cite{Casini:2011kv}, there is a fairly direct link between the sphere path integral and the Euclidean Rényi entropy picture of entanglement entropy, the latter encoding the Lorentzian entangling surface. In the case of topological quantum field theory, all the Rényi entropies are the same and consequently one finds a flat entanglement spectrum encoded by a maximally mixed edge mode theory \cite{Kitaev:2005dm,Levin:2006zz}. In de Sitter, considerations from the von Neumann algebra \cite{Chandrasekaran:2022cip} as well as Euclidean methods \cite{Dong:2018cuv,Anninos:2021ihe} also point to the static patch as a maximally mixed, potentially finite entropy  \cite{fischler,Banks:2006rx,Bousso:2000nf,Parikh:2004wh}, state. To leading order at large $c_m$, the expressions (\ref{eq:Z_grav_N1_final}) and (\ref{eq:Z_grav_N2_final}) take the form of an entanglement entropy of a two-dimensional conformal field theory. A clear target is to understand the entanglement content of the leading correction away from the large $c_m$ limit, which stems from coupling the conformal field theory to a fluctuating spacetime.
\vspace{0.2cm}\newline
\textbf{Lorentzian timelike super-Liouville theory.}  The fermionic part of the Lagrangian of $\cN=(2,2)$ timelike super-Liouville theory in Lorentzian signature in Minkowski space is
\begin{equation}\label{lorentzian}
	\mathcal{L}_{\rm fermi} = \ii \tpsi_- (\partial_0 + \partial_1) \psi_- + \ii \tpsi_+ (\partial_0 - \partial_1)\psi_+ - \ii \left( W''(\varphi) \psi_+ \psi_- + \widetilde{W}''(\tvarphi) \tpsi_- \tpsi_+ \right) \, .
\end{equation}
Here $W(\varphi) = \mu e^{\beta \varphi}/\beta$, while $\psi_\pm$ and $\tpsi_\pm$ are the components of the spinors in the chiral and anti-chiral multiplets, respectively. With the ``standard'' reality condition, $\tvarphi = \varphi^*$, and $\tpsi_\pm = \psi_\pm^*$, the kinetic term is real and the interaction term is pure imaginary. However, it is possible to impose a different reality condition on the spinors, namely $\tpsi_\pm = \psi_\mp^*$, such that the interaction term is real, at the cost of having a complex kinetic term. This suggests that it may be possible to obtain this by gauge-fixing a real-valued supergravity action, albeit with a non-standard reality condition. The interaction term can be related to a gravitino bilinear, and the complex kinetic term is entirely due to the gravitational path integral measure, since in two dimensions there is no kinetic term for the gravitino. The non-unitary nature of the kinetic term in (\ref{lorentzian}), which we recall stems from the gauge-fixed of the path integration measure, might be viewed as a fermionic counterpart to the wrong sign kinetic term of the Weyl factor.
\vspace{0.2cm}\newline
\textbf{Localization \& finiteness.}  As a final, more speculative remark, the prospect that the Gibbons--Hawking path integral exhibits some form of localization may be the Euclidean counterpart  of the hypothetical  finite-dimensional de Sitter Hilbert space  \cite{Anninos:2022ujl}. {Perhaps a template for this is the crystal melting picture of the path integral over K\"ahler geometries \cite{Iqbal:2003ds}.} The relation between localization methods, maximally mixed states, and topological quantum field theory may suggest that the origin of the Gibbons--Hawking entropy is of a more infrared nature than its black hole counterpart.

\subsection*{Acknowledgments}

We are grateful to Tarek Anous, Francesco Benini, Davide Cassani, Lorenz Eberhardt, Jackson Fliss, Sameer Murthy, Luigi Tizzano, David Tong, and Stathis Vitouladitis for helpful discussions. BM is supported in part by the Simons Foundation Grant No. 385602 and the Natural Sciences and Engineering Research Council of Canada (NSERC), funding reference number SAPIN/00047-2020. The work of PBG has been supported by the ERC Consolidator Grant N. 681908, “Quantum black holes: A macroscopic window into the microstructure of gravity”, by the STFC grant ST/P000258/1, and by the Royal Society Grant RSWF/R3/183010. D.A. is funded by the Royal Society under the grant The Atoms of a deSitter Universe. PBG gratefully acknowledges support from the Simons Center for Geometry and Physics, Stony Brook University at which some of the research for this paper was performed.

\appendix

\section{Spinor conventions}
\label{app:ConventionsEuclidean}

In this appendix we delineate our various choices of conventions for the spinors and their corresponding Euclidean reality conditions. 

\subsection*{Euclidean spinors}

The Clifford algebra Cliff$(2,0)$ is generated by the matrices $\gamma_a$ satisfying $\{\gamma_a, \gamma_b\} = 2\delta_{ab}$ in an orthonormal frame. Throughout the paper, we use $a$, $b$ for frame indices, and $\mu$, $\nu$ for coordinate indices. The $\gamma$ matrices are Hermitian, and there exists a unitary charge conjugation matrix $\cC$ satisfying
\begin{equation}
    \cC = - \cC^T ~, \qquad  \gamma_a^T = - \cC \gamma_a \cC^{-1} ~.
\end{equation}
The Majorana conjugate of a spinor is
\begin{equation}
\label{eq:MajoranaConjugate}
    \overline{\lambda} = \lambda^T \cC ~,
\end{equation}
which is such that $\overline{\lambda}\psi$ is a scalar under rotations.
\vspace{0.2cm}\newline
We indicate the antisymmetrization by $\gamma^{ab} \equiv \gamma^{[a}\gamma^{b]}$. The highest rank Clifford algebra element is $\gamma_* \equiv - \ii \gamma_1\gamma_2$, so that $\gamma_*^2 = \mathbb{I}$. We can use this to construct the chirality projectors 
\begin{equation}
    P_L = \frac{1}{2}(\mathbb{I} + \gamma_*) ~, \qquad P_R = \frac{1}{2}(\mathbb{I} - \gamma_*) ~.\end{equation}
In turn, we define a Weyl spinor as a spinor $\lambda_{L,R}$ satisfying
\begin{equation}
\label{eq:Weyl}
	P_{L,R}\lambda_{L,R} = \lambda_{L,R} ~,
\end{equation}
for either chirality.
\vspace{0.2cm}\newline
The charge conjugate of a spinor is given by
\begin{equation}
\label{eq:ChangeConjugateEuclidean}
    \lambda^C \equiv \cC^{-1}\lambda^* ~,
\end{equation}
and charge conjugation squares to minus the identity, that is $(\lambda^C)^C = -\lambda$. Therefore, we cannot define the reality condition in the standard way ``$\lambda = \lambda^C$''. Instead, we define a Euclidean Majorana spinor as a spinor satisfying the following reality condition
\begin{equation}
\label{eq:EuclideanReality}
    \lambda = \gamma_*\lambda^C ~.
\end{equation}
Notice that the chirality \eqref{eq:Weyl} and reality \eqref{eq:EuclideanReality} conditions are not compatible, so it is not possible to define a Majorana--Weyl spinor in Euclidean signature.
\vspace{0.2cm}\newline
Charge conjugation is more convenient than complex conjugation to study the reality properties of spinors and spinor bilinears. For scalars, charge conjugation is the same as complex conjugation, for a spinor it is defined as in \eqref{eq:ChangeConjugateEuclidean}, and for a matrix it is defined by $M^C \equiv \cC^{-1}M^* \cC$. Therefore, one finds that 
\begin{equation}
(\overline{\chi} M \lambda)^C = - \overline{\chi^C} M^C \lambda^C~.
\end{equation}
For two Majorana spinors $\chi_m, \lambda_m$, we find
\begin{equation}
\label{eq:ConjugationBilinearsMajorana}
(\overline{\chi_m}\lambda_m)^C = \overline{\chi_m}\lambda_m ~, \qquad (\overline{\chi_m}\gamma_a \lambda_m)^C = \overline{\chi_m} \gamma_a \lambda_m ~, \qquad (\overline{\chi_m}\slashed{\partial} \chi_m)^C = \overline{\chi_m}\slashed{\partial}\chi_m ~.
\end{equation}
All of these bilinears are real.
\vspace{0.2cm}\newline
If needed, we introduce indices writing $\lambda^\alpha$, $(\gamma_a)^\alpha_{\ph{\alpha}\beta}$, $\cC_{\alpha\beta}$, and use $\cC$ and $\cC^{-1}$ to, respectively, lower and raise indices, so that
\begin{equation}
    \overline{\lambda}\eta = \lambda^\alpha\cC_{\alpha\beta}\eta^\beta ~.
\end{equation}
Using this, it is easy to prove that the following relations hold for anti-commuting spinors
\begin{equation}
\label{eq:AntiCommutingRelations}
    \overline{\lambda}\eta = \overline{\eta}\lambda ~, \qquad \overline{\lambda}\gamma_a \eta = - \overline{\eta}\gamma_a\lambda ~, \qquad \overline{\lambda}\gamma_{ab}\eta = - \overline{\eta}\gamma_{ab}\lambda ~.
\end{equation}
The Fierz rearrangement in two dimensions is
\begin{equation}
\label{eq:Fierz}
\delta^\alpha_{\ph{\alpha}\delta} \delta^\gamma_{\ph{\gamma}\beta} = \frac{1}{2} \delta^\alpha_{\ph{\alpha}\beta} \delta^\gamma_{\ph{\gamma}\delta} + \frac{1}{2} (\gamma^a)^\alpha_{\ph{\alpha}\beta} (\gamma_a)^\gamma_{\ph{\gamma}\delta} + \frac{1}{2}(\gamma_*)^\alpha_{\ph{\alpha}\beta} (\gamma_*)^\gamma_{\ph{\gamma}\delta}~.
\end{equation}
To perform expansions and computations, we choose a concrete realization of the Clifford basis given by
\begin{equation}
\label{eq:ConcreteCliffordBasis}
\gamma_a = \sigma_a ~, \qquad \gamma_* = \sigma_3 ~,
\end{equation}
where $\sigma_m$ are the Pauli matrices. In terms of components $\cC=\sigma_2$. For instance, in this basis (\ref{eq:EuclideanReality}) reads
\begin{equation}
\lambda^1 = - i \lambda^{2 *}~ \quad \text{and} \quad \lambda^2 = - i \lambda^{1 *}~.
\end{equation}

\subsection*{Killing spinors}

A conformal Killing spinor on a $d$-dimensional manifold is a spinor satisfying the equation
\begin{equation}
\label{eq:ConformalKillingSpinor}
    D_\mu\epsilon = \frac{1}{d} \gamma_\mu \slashed{D}\epsilon ~.
\end{equation}
Here the covariant derivative acts on the spinor as
\begin{equation}\label{Dmu}
	D_\mu \epsilon = \partial_\mu \epsilon + \frac{1}{4} \omega_\mu^{ab}\gamma_{ab} \epsilon ~,
\end{equation}
and the Dirac operator is $\slashed{D}=\gamma^a D_a$. The conformal Killing spinor equation is so named because it is invariant under Weyl rescaling provided the spinor transforms with weight $-1/2$
\begin{equation}
    g_{\mu\nu}(x) \to \hat{g}_{\mu\nu}(x) = e^{2\sigma(x)}g_{\mu\nu}(x) ~, \qquad \epsilon(x) \to \hat{\epsilon}(x) e^{ \frac{1}{2} \sigma(x)} ~.
\end{equation}
A Killing spinor is a special case of the above satisfying the stronger relation
\begin{equation}
\label{eq:KillingSpinorEq}
    D_\mu \epsilon = \kappa \gamma_\mu \epsilon ~.
\end{equation}
The constant $\kappa$ is related to the curvature of the space
\begin{equation}
    \kappa^2 = - \frac{R}{4 d ( d - 1)} ~,
\end{equation}
as can be shown by taking the integrability condition for \eqref{eq:KillingSpinorEq}, and thus it is pure imaginary on a space of positive curvature. On such spaces, the bilinear $\epsilon'^\dagger \gamma^a\epsilon$ constructed from two Killing spinors $\epsilon$, $\epsilon'$ is a Killing vector.
\vspace{0.2cm}\newline
On the round two-sphere of radius $r$
\begin{equation}
\label{eq:MetricSphere}
    \rd s^2 = r^2 (\rd\theta^2 + \sin^2\theta \, \rd \phi^2) ~,
\end{equation}
with scalar curvature $R=\frac{2}{r^2}$, one can find ``positive'' or ``negative'' Killing spinors $\epsilon_\pm$ satisfying
\begin{equation}
	D_\mu \epsilon_\pm = \pm \frac{\ii}{2r} \gamma_\mu \epsilon_\pm ~.
\end{equation}
Notice that $\gamma_*\epsilon_\pm$ is a ``negative''/``positive'' Killing spinor.
Choosing the frame
\begin{equation}
\label{eq:zweibein}
 \rme^1 = r \, \rd\theta ~, \qquad \rme^2 = r \sin\theta \, \rd\phi ~, \qquad \omega^{12} = - \cos\theta \, \rd\phi ~, 
\end{equation}
and using the basis \eqref{eq:ConcreteCliffordBasis}, the most general ``positive'' Killing spinor is
\begin{equation}
\label{eq:filippo}
    \epsilon_+ = C_1 \rme^{-\ii \frac{\phi}{2}} \begin{pmatrix} \sin\frac{\theta}{2} \\ - \ii \cos\frac{\theta}{2} \end{pmatrix} + C_2 \rme^{\ii \frac{\phi}{2}} \begin{pmatrix} \cos\frac{\theta}{2} \\ \ii \sin\frac{\theta}{2} \end{pmatrix} ~,
\end{equation}
where $C_1, C_2\in \C$.

\section{\texorpdfstring{$S^2$}{S2} eigenfunctions}
\label{app:Fermions}

In this appendix we delineate our conventions of the various  harmonic expansions on the two-sphere.

\subsection*{Basis}

In order to expand the fields on the sphere, we choose a basis that simultaneously diagonalize the angular momentum and the kinetic operators. To construct it, we observe that using the frame \eqref{eq:zweibein} and the basis \eqref{eq:ConcreteCliffordBasis}, we can write the covariant derivative acting on scalars and spinors in a uniform way as
\begin{equation}
	D_\mu = \partial_\mu + \ii \hat{s} \, \cos\theta (\rd \phi)_\mu ~, 
\end{equation}
where $\hat{s}$ is a matrix that depends on the spin of the field: $\hat{s}\equiv 0$ for scalars, and $\hat{s} \equiv - \frac{\sigma_3}{2}$ for spinors. This allows us to compute the Laplacian
\begin{equation}
\label{eq:Laplacian}
	D^2 = \frac{1}{r^2} \left[ \frac{1}{\sin\theta} \partial_\theta ( \sin\theta \,  \partial_\theta ) + \frac{1}{\sin^2\theta} \left( \partial_\phi + \ii \hat{s} \, \cos\theta \right)^2 \right] ~.
\end{equation}
To define angular momentum on the two-sphere, we embed it into $\R^3$ and define the operator as
\begin{equation}
\label{eq:AngularMomentum}
\mb{L} = \mb{r} \wedge \mb{p} - \hat{s} \, \frac{\mb{r}}{r} ~, 
\end{equation}
The components
\begin{equation}
L_\pm = \rme^{\pm \ii \phi} \left( \pm \partial_\theta + \ii \cot\theta \, \partial_\phi - \hat{s} \, \frac{1}{\sin\theta} \right) ~, \qquad L_z = - \ii \partial_\phi
\end{equation}
generate a $\mf{su}(2)$ algebra
\begin{equation}
[L_z, L_\pm] = \pm L_\pm ~, \qquad [L_+, L_-] = 2L_z ~, 
\end{equation}
and their square is
\begin{equation}
\label{eq:Laplacian_AngularMomentum}
L^2 = \frac{1}{2}( L_+ L_- + L_-L_+) + L_z^2 = - r^2 D^2 + \hat{s}^2 ~, 
\end{equation}
where $D^2$ is the Laplacian \eqref{eq:Laplacian}.
\vspace{0.2cm}\newline
The kinetic operator on scalars is $D^2$, which is already diagonal in a basis of eigenfunctions of $L^2$ and $L_z$ thanks to \eqref{eq:Laplacian_AngularMomentum}.\footnote{It is more common to denote the scalar Laplacian by $\nabla^2$. We choose to use $D^2$ here to preserve the uniform treatment.} We work with the canonical spherical harmonics $\mathcal{Y}_{lm}(\Omega)$, which are defined for $l\in \Z_{\geq 0}$ and $m=-l, -l+1, \dots, l-1, l$, and satisfy
\begin{equation}
\label{eq:EigenvaluesSphericalHarmonics}
	L^2 \mathcal{Y}_{lm}(\Omega) = l(l+1) \mathcal{Y}_{lm}(\Omega) ~, \qquad L_z \mathcal{Y}_{lm}(\Omega) = m \mathcal{Y}_{lm}(\Omega) ~.
\end{equation}
Concretely, they have the expression
\begin{equation}
\label{eq:ConcreteSphericalHarmonics}
	\mathcal{Y}_{lm}(\Omega) = \sqrt{\frac{(2l+1)}{4\pi} \frac{(l-m)!}{(l+m)!} } P^m_l(\cos\theta) \, \rme^{\ii m \phi} ~, 
\end{equation}
where $P^m_l(x)$ is the associated Legendre polynomial. The coefficient is chosen to have the normalization
\begin{equation}
	\int_{S^2} \rd \Omega \, (\mathcal{Y}_{lm}(\Omega))^* \mathcal{Y}_{l'm'}(\Omega) = \delta_{ll'} \delta_{mm'} ~,
\end{equation}
where $\rd\Omega \equiv \sin\theta \, \rd\theta \rd\phi$. The real spherical harmonics $Y_{lm}(\Omega)$ used in the main text, are obtained from the $\mathcal{Y}_{lm}(\Omega)$ as follows
\begin{equation}\label{real_complex}
{Y}_{l m}(\Omega) \equiv \begin{cases}
\frac{i}{\sqrt{2}}\left(\mathcal{Y}_{l m}(\Omega)- (-1)^m \mathcal{Y}_{l, -m}(\Omega)\right)~,\quad \mathrm{if}~ m<0\\
\mathcal{Y}_{l 0}(\Omega)\\
\frac{1}{\sqrt{2}}\left(\mathcal{Y}_{l, -m}(\Omega)+ (-1)^m \mathcal{Y}_{l m}(\Omega)\right)~,\quad \mathrm{if}~ m>0~.
\end{cases}
\end{equation}
The real spherical harmonics satisfy the orthonormality conditions
\begin{equation}
	\int_{S^2} \rd \Omega \, {Y}_{lm}(\Omega) {Y}_{l'm'}(\Omega) = \delta_{ll'} \delta_{mm'} ~.
\end{equation}
The kinetic operator for spinors is the Dirac operator, which commutes with $L_\pm$, $L_z$, so it can be diagonalized simultaneously with angular momentum. In fact, more is true: the Lichnerowicz formula relates its square to the Laplacian, and thus to the angular momentum via \eqref{eq:Laplacian_AngularMomentum}
\begin{equation}
\label{eq:Lichnerowicz}
\slashed{D}^2 = D^2 - \frac{R}{4} = D^2 - \frac{1}{2r^2} = - \frac{1}{r^2}L^2 - \frac{1}{4r^2} ~.
\end{equation}
Therefore, the eigenvalues of $\slashed{D}$ are determined in terms of those of $L^2$ except for a sign, which is the additional label of the basis. We work with the basis $\psi^{(s)}_{\pm lm}(\Omega)$ where $s= \pm 1$, $l\in\Z_{\geq 0}$ and $m= 0, \dots, l$, which satisfies\footnote{To compare with the canonical eigenvalues used in \eqref{eq:EigenvaluesSphericalHarmonics} one can define $\mathtt{m}$ with $|\mathtt{m}|=m+ \frac{1}{2}$ and ${\rm sgn}(\mathtt{m}) = s$, and $\mathtt{l}=l+\frac{1}{2}$, in which case \eqref{eq:EigenvaluesSpinorHarmonics} becomes
\[
	L^2 \psi^{(s)}_{\pm \, lm} = \mathtt{l}(\mathtt{l}+1) \psi^{(s)}_{\pm \, lm} ~, \quad L_z \psi^{(s)}_{\pm \, lm} = \mathtt{m} \psi^{(s)}_{\pm \, lm} ~, \quad \slashed{D} \psi^{(s)}_{\pm \, lm} = \pm \ii \frac{\mathtt{l}+\frac{1}{2}}{r} \psi^{(s)}_{\pm \,  lm} ~.
\]}
\begin{equation}
\label{eq:EigenvaluesSpinorHarmonics}
	L^2 \psi^{(s)}_{\pm \, lm} (\Omega)= \left( l + \frac{1}{2} \right) \left( l + \frac{3}{2} \right) \psi^{(s)}_{\pm \, lm}(\Omega) ~, \qquad L_z \psi^{(s)}_{\pm \, lm}(\Omega) = s \left( m + \frac{1}{2} \right) \psi^{(s)}_{\pm \, lm} (\Omega)~.
\end{equation}
and 
\begin{equation}\label{snabla_app}
\slashed{D} \psi_{\pm\,lm}^{(s)}(\Omega) = \pm \frac{\ii}{r}(l+1)\psi_{\pm,lm}^{(s)}(\Omega)~,\quad s= \pm~.
\end{equation} 
Concretely, the eigenfunctions have the form \cite{Camporesi:1995fb}
\begin{equation}
\label{eq:ConcreteSpinorHarmonics}
	\psi^{(+)}_{\pm \, lm}(\Omega) = \frac{c_{lm}}{\sqrt{2}} \rme^{ \ii \left( m + \frac{1}{2} \right) \phi } \begin{pmatrix} \Phi_{lm}(\theta) \\ \pm \ii \Psi_{lm}(\theta) \end{pmatrix} ~, \qquad \psi^{(-)}_{\pm \, lm} (\Omega)= \frac{c_{lm}}{\sqrt{2}} \rme^{ - \ii \left( m + \frac{1}{2} \right) \phi } \begin{pmatrix} \pm \ii \Psi_{lm}(\theta) \\ \Phi_{lm}(\theta) \end{pmatrix} ~.
\end{equation}
Here
\begin{equation}
\begin{split}
	\Phi_{lm}(\theta) &= \left(\cos \frac{\theta}{2}\right)^{m+1} \left(\sin\frac{\theta}{2}\right)^m \, P_{l-m}^{(m,m+1)}(\cos\theta) ~, \\ 
	\Psi_{lm}(\theta) &= \left(\cos \frac{\theta}{2}\right)^{m} \left(\sin\frac{\theta}{2}\right)^{m+1} \, P_{l-m}^{(m+1,m)}(\cos\theta)  = (-1)^{l-m}\Phi_{lm}(\pi - \theta) ~,
\end{split}
\end{equation}
where $P_n^{(a,b)}(x)$ are the real-valued Jacobi polynomials. The coefficients
\begin{equation}
c_{lm}^2 = \frac{(l+m+1)! (l-m)!}{2\pi (l!)^2}
\end{equation}
Moreover, the eigenfunctions \eqref{eq:ConcreteSpinorHarmonics} satisfy
\begin{equation}
\label{ortho_appendix}
\int_{S^2} \dd \Omega\, \bpsi^{(+)}_{\pm \, lm}(\Omega) \psi^{(-)}_{\pm \, l'm'} (\Omega)= -  \ii \delta_{l,l'}\delta_{m,m'}~,\quad \int_{S^2} \dd \Omega\, \bpsi^{(-)}_{\pm \, lm}(\Omega) \psi^{(+)}_{\pm \, l'm'} (\Omega)=  +\ii \delta_{l,l'}\delta_{m,m'}~,
\end{equation}

\subsection*{Expansion of a spinor}

Given two spinors $\eta$, $\widetilde{\eta}$ on the two-sphere, we can expand them on the basis \eqref{eq:ConcreteSpinorHarmonics} with Grassmann-valued coefficients
\begin{equation}
\begin{split}
	\eta &=\frac{1}{\sqrt{r}} \sum_{l \geq 0} \sum_{0\leq m \leq l} \left( \alpha_{+\, lm} \, \psi^{(+)}_{+\, lm} + \alpha_{-\, lm} \, \psi^{(+)}_{-\, lm} + \beta_{+ \, lm} \, \psi^{(-)}_{+\, lm} + \beta_{- \, lm} \, \psi^{(-)}_{- \, lm} \right) ~, \\
	\widetilde{\eta} &=\frac{1}{\sqrt{r}} \sum_{l \geq  0} \sum_{0\leq m \leq l}\left( \widetilde{\alpha}_{+\, lm} \, \psi^{(+)}_{+\, lm} + \widetilde{\alpha}_{-\, lm} \, \psi^{(+)}_{-\, lm} + \widetilde{\beta}_{+ \, lm} \, \psi^{(-)}_{+\, lm} + \widetilde{\beta}_{- \, lm} \, \psi^{(-)}_{- \, lm} \right) ~.	
\end{split}
\end{equation}
Using \eqref{snabla_app} and \eqref{ortho_appendix}, we find
\begin{equation}
\begin{split}
	\int_{S^2} \rd\Omega \, \ii \, \overline{\widetilde{\eta}} \slashed{D} \eta &= \frac{1}{r} \sum_{l \geq 0} \sum_{0\leq m \leq l} ( {l+1} ) \left[ \tilde{\alpha}_{+\, lm} \beta_{+\, lm} - \tilde{\alpha}_{-\, lm} \beta_{-\, lm} - \tilde{\beta}_{+\, lm} \alpha_{+\, lm} + \tilde{\beta}_{-\, lm} \alpha_{-\, lm} \right] ~,\\
		\int_{S^2} \rd\Omega \, \overline{\widetilde{\eta}} \eta &= - \frac{\ii}{r}  \sum_{l \geq 0} \sum_{0\leq m \leq l}  \left[ \tilde{\alpha}_{+\, lm} \beta_{+\, lm} + \tilde{\alpha}_{-\, lm} \beta_{-\, lm} - \tilde{\beta}_{+\, lm} \alpha_{+\, lm} - \tilde{\beta}_{-\, lm} \alpha_{-\, lm} \right]~.
\end{split}
\end{equation}
In particular, for $\cN=2$ supersymmetry we are interested in imposing the condition $\widetilde{\eta} = \gamma_* \eta^C$, which implies
\begin{equation}
\label{eq:RelationsGrassmann_1}
	\tilde{\alpha}_{\pm \, lm} = - \ii \beta^*_{\mp \, lm} ~, \qquad \tilde{\beta}_{\pm \, lm} = - \ii \alpha^*_{\mp \, lm } ~,
\end{equation}
corresponding to \eqref{eq:psis_N2} in the text.
Substitution in the above expansions leads to
\begin{align}
	\int_{S^2} \rd\Omega \, \ii \, \overline{\widetilde{\eta}} \slashed{D} \eta &= \frac{1}{r} \sum_{l \geq 0} \sum_{0\leq m \leq l} ( {l+1} ) \left[ - \alpha^*_{-\, lm}\alpha_{+\, lm} - \alpha_{-\, lm}\alpha_{+\, lm}^* + \beta^*_{-\, lm}\beta_{+\, lm} + \beta_{-\, lm}\beta_{+\, lm}^* \right] ~, \nonumber \\
		\int_{S^2} \dd \Omega\,\overline{{\eta}}{\eta} &= -\frac{2\ii}{r}\sum_{l\geq 0}\sum_{0\leq m \leq l}\left(  \alpha_{+\,lm}   \beta_{+\,lm}  +  \alpha_{-\,lm}  \beta_{-\,lm}  \right)~, \\
\int_{S^2} \dd \Omega\,\overline{\widetilde{\eta}}\widetilde{\eta} &=-\frac{2\ii}{r}\sum_{l\geq 0}\sum_{0\leq m \leq l}\left(  \alpha_{+\,lm}^*   \beta_{+\,lm}^*  +  \alpha_{-\,lm}^*  \beta_{-\,lm}^*  \right)~. \nonumber
\end{align}
from which we see that the first bilinear is pure imaginary. In the main text, these expressions correspond to \eqref{N2fG}.
\vspace{0.2cm}\newline
A further specialization of this, used when discussing $\cN=1$ supersymmetry, is when the spinor $\eta$ is Majorana: imposing the reality condition \eqref{eq:EuclideanReality} corresponds to specialising the case above to $\widetilde{\eta}=\eta$, and \eqref{eq:RelationsGrassmann_1} gives
\begin{equation}
\label{eq:RelationsGrassmann_2}
	\beta_{\pm \, lm} = - \ii \alpha^*_{\mp \, lm} ~.
\end{equation}
The expansions of the bilinears further simplifies
\begin{equation}
\begin{split}
	\int_{S^2} \rd\Omega \, \ii \, \overline{{\eta}} \slashed{D} \eta &=  - \frac{2}{r} \sum_{l \geq 0} \sum_{0\leq m \leq l} ({l+1} ) \left( \alpha^*_{-\, lm}\alpha_{+\, lm} + \alpha_{-\, lm}\alpha_{+\, lm}^* \right)~, \\
		\int_{S^2} \rd\Omega \, \overline{{\eta}} \eta &= \frac{2}{r} \sum_{l \geq 0} \sum_{0\leq m \leq l} \left( \alpha^*_{-\, lm}\alpha_{+\, lm} - \alpha_{-\, lm}\alpha_{+\, lm}^* \right)~.
\end{split}
\end{equation}
The reality properties of these expansions is consistent with the observations around \eqref{eq:ConjugationBilinearsMajorana}. In the main text, these appear in \eqref{LF2}.

\section{Spacelike super-Liouville theory}\label{spacelikeL}

In this appendix we discuss some aspects of the spacelike $\mathcal{N}=1$ and $\mathcal{N}=2$ super-Liouville theory, which we can view as the super-Weyl gauge-fixed form of $\mathcal{N}=1$ and $\mathcal{N}=2$ supergravity coupled to a   matter SCFT with sufficiently small central charge. For the particular case of a minimal model SCFT, whose central charge is given by (\ref{cmMM}), a non-perturbative completion of the gravitational theory taking the form of a complex matrix integral has been conjectured \cite{Seiberg:2003nm}. 
\vspace{0.2cm}\newline
In contrast to $\mathcal{N}=1$ timelike super-Liouville theory, there is significant evidence that spacelike $\mathcal{N}=1$ super-Liouville is a unitary two-dimensional superconformal field theory with an explicit construction \cite{Rashkov:1996np,Poghossian:1996agj} for the OPE coefficients of all vertex operators. 

\subsection*{\texorpdfstring{$\mathcal{N}=1$}{N1} spacelike super-Liouville}
\label{sec:spacelikeN1}

The action for spacelike $\mathcal{N}=1$ super-Liouville theory  on the two-sphere is given by 
\begin{equation}
\label{eq:N1_SL_Action_appp}
\mathcal{S}^{\mathcal{N}=1}_{\text{L}}=\frac{1}{4\pi} \int_{S^2} \dd^2 x\, \tilde{\rme}\,\left(\frac{1}{2}\tilde{g}^{\mu\nu}\partial_\mu \varphi \partial_\nu {\varphi} - \frac{\ii}{2}  \bpsi \slashed{D} \psi -\frac{1}{2} F^2 + \frac{1}{2} Q \widetilde{R} \varphi + \mu e^{b\varphi} F - \frac{\ii}{2}  \mu b e^{b\varphi} \overline{\psi} \psi \right) ~,
\end{equation}
where $Q=b+b^{-1}$, $c_{\text{L}}^{\mathcal{N}=1}=3/2+6Q^2$, and $\mu \in \mathbb{R}$. Integrating out the auxiliary scalar $F$, the resulting undecorated path integral becomes
\begin{equation}
\label{eq:ZL_N1_IntegratedOut}
\mathcal{Z}_{{\text{L}}}^{\mathcal{N}=1}[\mu]  = \int \frac{[\mathcal{D} \varphi][\mathcal{D} \psi]}{{\rm vol}_{OSp(1|2;\C)}}\, e^{-\frac{1}{4\pi}\int_{S^2} \dd^2 x\, \tilde{\rme}\,\left(\frac{1}{2}\tilde{g}^{\mu\nu}\partial_\mu \varphi \partial_\nu {\varphi}  - \frac{\ii}{2}  \bpsi \slashed{D} \psi   + \frac{1}{2} Q \widetilde{R} \varphi  + \frac{1}{2}\mu^2  e^{2b\varphi}- \frac{\ii}{2}  \mu b e^{b\varphi} \overline{\psi} \psi \right)}~.
\end{equation}
{The Lagrangian obtained in this way matches \cite{Belavin:2007gz}.
We note that, similarly to the timelike case, the corresponding cosmological constant in the underlying gravitational theory goes as  $\mu^2/2>0$, and is hence positive.
\vspace{0.2cm}\newline
In the limit $b\to0^+$ one can employ the saddle point approximation. The saddle point solutions are given by the $OSp(1|2,\mathbb{C})$ orbit of $\psi_*=0$ and
\begin{equation}
\label{eq:N1_Spacelike_Saddles}
\varphi_*^{(n)} = \frac{1}{2b}\log \left(\frac{Q}{b \,\mu^2 r^2}\right) + \frac{\pi\ii}{2b}(2n+1)~,\qquad n \in \mathbb{Z}~.
\end{equation}
Therefore, this theory only admits complex solutions, which lie outside  the original path integration contour. This corresponds to a complexified metric on the sphere in the underlying $\mathcal{N}=1$ supergravity theory \eqref{eq:Z_grav_N1_0}. 
\vspace{0.2cm}\newline
Nonetheless, we can still  compute the path integral \eqref{eq:ZL_N1_IntegratedOut}, following the same prescription as in the non-supersymmetric case a subset of the saddle point solutions $\varphi_*^{(n)}$ may contribute to the  path integral. In \cite{Mahajan:2021nsd}, it was argued for the non-supersymmetric case that we should take those saddles with  $n \ge 0$.  This leads to
\begin{align}\label{ZLN1spacelike}
	\mathcal{Z}_{{\text{L}},\text{saddle}}^{\mathcal{N}=1}[\mu] &\approx -\ii\times \left(\frac{Q}{b \, \mu^2 r^2}\right)^{-\frac{Q}{2b}} \times e^{-\frac{Q}{2b}}e^{-\frac{\ii\pi}{2b^2}}\sum_{n\geq 0}(-1)^ne^{-\frac{\ii\pi}{b^2}n}~.
\end{align}
We note that the complex sphere saddles (\ref{eq:N1_Spacelike_Saddles}) are weighted by opposite signs when evaluated on the action (\ref{eq:ZL_N1_IntegratedOut}).  Performing the geometric sum over all non-negative saddles we obtain $\mathcal{Z}_{{\text{L}},\text{saddle}}^{\mathcal{N}=1}[\mu]  \propto (\cos(\frac{\pi}{2b^2}))^{-1}$. In the non-supersymmetric case \cite{Mahajan:2021nsd} all saddles are weighted by the same sign, replacing the cosine by a sine.
\vspace{0.2cm}\newline
The imaginary prefactor in (\ref{ZLN1spacelike}) is present because we kept only the leading part of the on-shell action. Indeed the bosonic fluctuations $\varphi= \varphi_*+\delta\varphi$ around the saddle (\ref{eq:N1_Spacelike_Saddles}) have a negative mass
\begin{equation}
\label{eq:ZL_N1_bosonFluctuations}
\mathcal{Z}_{{\text{L}},\text{bos.\,fluct.}}^{\mathcal{N}=1}  = \int[\mathcal{D}\delta\varphi] e^{-\frac{1}{8\pi}\int_{S^2} \dd^2 x\, \tilde{\rme}\,\delta \varphi\left(-\nabla^2 -\frac{2Qb}{r^2}\right)\delta\varphi}~,
\end{equation}
Expanding $\delta\varphi$ into eigenfunctions of the spherical Laplacian, we observe that the $l=0$ eigenfunction in (\ref{eq:ZL_N1_bosonFluctuations}) is Gaussian unsuppressed. We cure this following the approach of \cite{Gibbons:1978ac,Polchinski:1988ua} and  analytically continue $\delta \varphi_{00}\rightarrow \pm \ii \delta\varphi_{00}$. The corresponding Jacobian multiplies the one-loop contributions of (\ref{ZLN1spacelike}) by another imaginary factor, cancelling the imaginary unit in (\ref{ZLN1spacelike}) and rendering the sphere path integral real-valued. Of course to claim the latter one would need to investigate the theory to all-loop order. From a path integral perspective this is systematically tractable, though somewhat cumbersome \cite{Muhlmann:2022duj}. 
The one-loop contribution to the path integral from the fermionic fluctuations around the complex saddle  (\ref{eq:N1_Spacelike_Saddles}) reads
\begin{equation}
\mathcal{Z}_{{\text{L}},\text{fer.\,fluct.}}^{\mathcal{N}=1}  = \int[\mathcal{D}\delta\psi] e^{-\frac{\ii}{8\pi}\int_{S^2} \dd^2 x\,\tilde{\rme}\,\delta\bpsi \left(-\slashed{D} -\ii (-1)^n\frac{\sqrt{Qb}}{r}\right)\delta\psi}~.
\end{equation}
Combining the one-loop contributions with the saddle (\ref{ZLN1spacelike}) we obtain
\begin{equation}\label{ZLN1spacelike_final_app}
\mathcal{Z}_{\text{L}}^{\mathcal{N}=1} \approx    \text{const}\times \frac{1}{\cos\frac{\pi}{2b^2}}\times \left({\mu}{b}\right)^{\frac{1}{b^2}+1}\Lambda_{\text{uv}}^{\frac{5}{4}+b^2} \, r^{c_{\text{L}}^{\mathcal{N}=1}/3} \, e^{-\frac{1}{2b^2}+\frac{1}{b^2}\log b^2 } \times {b^{-1} } \times \left( 1 + \mathcal{O}(b^2) \right)~,
\end{equation}
where we also included the contribution from the volume of the super-Moebius group as discussed below (\ref{ZtLN1}).
\vspace{0.2cm}\newline
As was previously mentioned, $\mathcal{N}=1$ spacelike super-Liouville theory is a unitary superconformal field theory. The spectrum of Virasoro primary operators is continuous and can be further partitioned into Neveu--Schwarz or Ramond sectors, depending on the boundary conditions obeyed by the fermions. We denote the respective primary operators as $\mathcal{V}^{\text{NS}}_\alpha$ and $ \mathcal{R}_\alpha^{\text{R}}$. 
The respective conformal dimensions are given by 
\begin{equation}
\Delta_{\text{NS}} = \frac{\alpha(Q-\alpha)}{2} = \frac{c_{\text{L}}^{\mathcal{N}=1}}{24}-\frac{3}{48} +\frac{P^2}{2}~,\quad \Delta_{\text{R}} = \Delta_{\text{NS}} + \frac{1}{16}~,
\end{equation}
where we introduced the parameter $\alpha \equiv Q/2+ \ii P$, with $P\in \mathbb{R}$ \cite{Ahn:2002ev}. The spectrum is thus bounded from below by $(c_{\text{L}}^{\mathcal{N}=1}-3/2)/24$.
\vspace{0.2cm}\newline 
The OPE coefficients of $\mathcal{N}=1$ super-Liouville theory \eqref{eq:N1_SL_Action} have been obtained using various approaches \cite{Poghossian:1996agj, Rashkov:1996np,Belavin:2007gz}. By acting on $\mathcal{V}^{\text{NS}}_\alpha$ with the superconformal generators one can generate \cite{Belavin:2007gz} the on-shell operators 
\begin{equation}\label{eq:Win N1sL}
\mathcal{W}_\alpha = \alpha^2 \bpsi \psi \, e^{\alpha \varphi} - 2\pi \ii \mu \alpha  e^{(\alpha+b)\varphi}~.
\end{equation}
For the three-point function of the specific operator $\mathcal{W}_b$, the structure constant\footnote{The explicit form  is given by  
\begin{equation}
C(b,b,b) = \frac{\ii}{2}\left(\frac{\pi\mu}{2b} \gamma\left(\frac{1}{2}+\frac{b^2}{2}\right)b^{1-b^2}\right)^{\frac{1}{b^2}-2}\frac{\Upsilon_b\left(\frac{b}{2}\right)\Upsilon_b\left(\frac{1}{2b}\right)}{\Upsilon_b(b)}\times \frac{\gamma\left(\frac{1}{2}+\frac{b^2}{2}\right)^3  \gamma\left(\frac{1}{2}-\frac{1}{2b^2}\right)}{\gamma\left(\frac{b^2}{2}-\frac{1}{2}\right)}b^{-\frac{1}{b^2}-2-2b^2}~,
\end{equation}
with $\gamma(x) \equiv \Gamma(x)/\Gamma(1-x)$ and $\Upsilon_b(x)$ denotes the standard Liouville $\Upsilon$-function. 
} in the semiclassical limit is given by \cite{Belavin:2007gz}
\begin{align}\label{Cbbb}
C(b,b,b) 
&\overset{b\rightarrow 0}{\approx} \ii\, \left({\pi\mu b}\right)^{\frac{1}{b^2}-2} e^{\frac{1}{b^2}(1+\log 2+ \log b^2)}e^{-\frac{\ii\pi}{2b^2}} \sum_{n=0}^\infty (-1)^n e^{-\frac{\ii\pi}{b^2}n}{b^{-\tfrac{9}{2}}} \times \big(1-\frac{11}{6}b^2 + \ldots \big)~.
\end{align}
\vspace{0.2cm}\newline
One can compare the  path integral expression (\ref{ZLN1spacelike_final_app}) to (\ref{Cbbb}) obtained from the SCFT data. The $b$ dependence is different, which could stem from a different choice of normalization for the path integration measure. 

\subsection*{\texorpdfstring{$\mathcal{N}=2$}{N2} spacelike super-Liouville}

Spacelike super-Liouville theory with $\mathcal{N}=2$ supersymmetry is governed by the action
\begin{multline}
\label{eq:N22_SpacelikeLiouvilleSphere_app}
S_{\text{L}}^{\mathcal{N}=2} = \frac{1}{4\pi} \int_{S^2} \dd^2 x\, \tilde{\rme} \left(  \partial_\mu \tvarphi \partial^\mu \varphi - \ii \btpsi \slashed{D}\psi - F \tF + \frac{Q\widetilde{R}}{2} \left( \varphi + \tvarphi \right) \right. \\ \left.
   + \mu e^{b\varphi}  F - \frac{\ii}{2}\mu b e^{b\varphi} \overline{\psi}\psi +  \mu^* e^{b\tvarphi} \tF - \frac{\ii}{2}  \mu^* b e^{b\tvarphi} \btpsi\tpsi \right)~.
\end{multline}
The equations of motion obtained from (\ref{eq:N22_SpacelikeLiouvilleSphere_app}) for the bosonic fields $\varphi, F$ and their anti-chiral counterparts are 
\begin{equation}\label{N2_eom_spacelike}
\begin{aligned}[c]
0&= -\nabla^2\varphi + \frac{1}{b r^2} +\mu b e^{b\varphi}F  - \frac{\ii}{2} \mu b^2 \overline{\psi} \psi~, \cr
0&=\widetilde{F}-  \mu e^{b\varphi}~,
\end{aligned}
\quad
\begin{aligned}[c]
0&=-\nabla^2\tvarphi + \frac{1}{b r^2} +\mu^* b e^{b\tvarphi} \widetilde{F} -  \frac{\ii}{2}\mu^* b^2 \btpsi \tpsi~,\cr
0&=F - \mu^*  e^{b\tvarphi}~.
\end{aligned}
\end{equation}
Similarly to the $\mathcal{N}=1$ case $\mathcal{N}=2$ spacelike super-Liouville admits only complex saddles on $S^2$. Labeled by $n\in \mathbb{Z}$ the complex saddles of (\ref{eq:N22_SpacelikeLiouvilleSphere}) are given by
\begin{equation}
\label{eq:N2_Spacelike_Saddles}
(\varphi+ \tvarphi)_*^{(n)} = \frac{1}{b}\log \left(\frac{1}{b^2 |\mu|^2 r^2}\right) + \frac{\pi\ii}{b}(2n+1)~,\qquad n \in \mathbb{Z}~,
\end{equation}
while the fermions are vanishing. Assuming again that all $n\ge 0$ saddles contribute, we find
\begin{equation}\label{eq:Zsaddle_N2_spacelike}
\mathcal{Z}_{\text{L},\text{saddle}}^{\mathcal{N}=2}[\mu] = (|\mu^2|b^2 r^2)^{\frac{1}{b^2}}e^{\frac{1}{b^2}- \frac{\ii \pi}{b^2}} \sum_{n\geq 0}e^{-\frac{2\ii \pi}{b^2}n}=-\frac{\ii}{2} (e |\mu^2|b^2 r^2)^{\frac{1}{b^2}}\times \frac{1}{\sin(\frac{\pi}{b^2})}~.
\end{equation}
The sphere path integral of $\mathcal{N}=2$ Liouville theory thus resembles more closely the non-supersymmetric case \cite{Mahajan:2021nsd} than the $\mathcal{N}=1$ case. Note that the imaginary prefactor in (\ref{eq:Zsaddle_N2_spacelike}) is again just an artefact of (\ref{eq:Zsaddle_N2_spacelike}) only keeping track of the leading saddle. Upon including one-loop contributions,  the sphere partition function of $\mathcal{N}=2$ spacelike super-Liouville theory is rendered real-valued.

\bibliographystyle{JHEP}
\bibliography{bib}

\end{document}